\documentclass{iopart}
\usepackage{iopams}
\expandafter\let\csname equation*\endcsname\relax
\expandafter\let\csname endequation*\endcsname\relax
\usepackage{amsmath}
\usepackage{amssymb}
\usepackage{caption}
\usepackage{amscd}
\usepackage{dsfont}
\usepackage{url}
\usepackage{graphicx}

\newcommand{\ue}{\mathrm{e}}
\newcommand{\ui}{\mathrm{i}}

\newcommand{\T}{\mathcal{T}}

\begin{document}
\title{A non-Hermitian $PT$-symmetric kicked top}
\author{Steve Mudute-Ndumbe and Eva-Maria Graefe}
\address{Department of Mathematics, Imperial College London, London SW7 2AZ, United Kingdom}
\begin{abstract}
A non-Hermitian $PT$-symmetric version of the kicked top is introduced to study the interplay of quantum chaos with balanced loss and gain. The classical dynamics arising from the quantum dynamics of the angular momentum expectation values are derived. It is demonstrated that the presence of $PT$-symmetry can lead to ``stable'' mixed regular chaotic behaviour without sinks or sources for subcritical values of the gain-loss parameter. This is an example of what is known in classical dynamical systems as \textit{reversible} dynamical systems. For large values of the kicking strength a strange attractor is observed that also persists if $PT$-symmetry is broken. The intensity dynamics of the classical map is investigated, and found to provide the main structure for the Husimi distributions of the subspaces of the quantum system belonging to certain ranges of the imaginary parts of the quasienergies. Classical structures are also identified in the quantum dynamics. Finally, the statistics of the eigenvalues of the quantum system are analysed and it is shown that if most of the eigenvalues are complex (which is the case already for fairly small non-Hermiticity parameters) the nearest-neighbour distances of the (unfolded) quasienergies follow a two-dimensional Posisson distribution when the classical dynamics is regular. In the chaotic regime, on the other hand they are in line with recently identified universal complex level spacing distributions for non-Hermitian systems, with transpose symmetry $\hat A^T=\hat A$. It is demonstrated how breaking this symmetry (by introducing an extra term in the Hamiltonian) recovers the more familiar universality class for non-Hermitian systems given by the complex Ginibre ensemble. Both universality classes display cubic level repulsion. The $PT$-symmetry of the system does not seem to influence the complex level spacings. Similar behaviour is also observed for the spectrum of a $PT$-symmetric extension of the triadic Baker map.  \end{abstract}

\section{Introduction}

Non-Hermitian and in particular $PT$-symmetric quantum systems, that can have either complex conjugate or purely real eigenvalues, have been the subject of extensive investigations over the past two decades. The interest reaches from their spectral theory to technological applications \cite{Bend18_book,Chris18_book}. Yet, apart from a few exceptions \cite{Bend09b,West10,Long17,Yusu19}, what has been little investigated hitherto is the interplay between non-Hermiticity and chaos. Since mixed regular-chaotic systems are the norm rather than the exception, it is clearly of great importance to study this interplay to gain deeper insights into the \textit{typical} behaviour of $PT$-symmetric quantum systems and their classical counterparts. 

Many of the implementations of $PT$-symmetric systems have used analogous classical wave systems, such as microwaves or classical optics (see e.g. \cite{Chris18_book} for some examples). Such systems are exceptionally well suited for the investigation of quantum (or more precisely \textit{wave}) chaos. Recently, however, also quantum mechanical realisations of $PT$-symmetric systems have been experimentally implemented \cite{Xiao17,Li19,Klau19}. Here we study the interplay of $PT$-symmetry and quantum chaos for a toy model based on one of the paradigm models of Hermitian/unitary quantum chaos. While we believe that it is entirely possible to implement this model in a suitable experimental setup, the emphasis of the present study is on the identification of typical behaviour that may be observed in a more readily available experimental system, in the theoretical comfort of a simple model system that is numerically and analytically more accessible. 
 
For this purpose we introduce a $PT$-symmetric extension of one of the most prominent example systems of traditional quantum chaos: the kicked top. In dependence on the parameters it can show regular, mixed or (almost) entirely chaotic behaviour, and is faithful to random matrix theory \cite{Frah85,Haak86,Haak10}. More than twenty years after its theoretical inception, it has been experimentally realised using cold atoms \cite{Chau09}.  A dissipative version of the kicked top modelling decoherence via Lindblad dynamics has been studied in the literature before \cite{Brau01}, and a $PT$-symmetric version of the non-kicked system has been studied in \cite{Grae08,Grae08b,Grae10b}. Here we present the first study of the $PT$-symmetric kicked system showing a wealth of new behaviour on both the quantum and the classical side, compared to both the non-Hermitian system without $PT$-symmetry and the Hermitian system.
In the classical limit, the dynamics shows a mixture of what may be termed \textit{quasi-Hamiltonian} chaos and dissipative chaos. This is in fact an example of what has been studied under the name of reversible dynamical systems \cite{Robe92} in classical dynamics. 

The paper is organised as follows. We introduce the model of a $PT$-symmetric kicked top, adding a linear gain-loss profile to the non-kicked part, in section \ref{sec-model}, and briefly discuss a possible experimental realisation. In section \ref{sec-classical} we derive the classical dynamics using the dynamics of the expectation values of the angular momentum operators, and analyse it in some detail.  We provide a very brief introduction to the concept of reversibility in classical maps, which is the classical counterpart of the quantum $PT$-symmetry. This $PT$-symmetric/reversible classical dynamics shows \textit{stable} Poincare dynamics without any sinks or sources, if the gain-loss parameter is not too large. There is no counterpart of this behaviour in non-Hermitian or other open systems without $PT$-symmetry/reversibility. Furthermore, in our system, we find that the presence of gain and loss leads to an early onset of chaos. Compared to the unitary system, the non-Hermitian $PT$-symmetric kicked top has an additional classical variable resulting from the norm of the quantum wave function, that can be viewed as a classical intensity. The behaviour of this intensity variable is analysed as a function of the initial state. In section \ref{sec-quantum} we then study some of the features of the quantum system, starting with the phase-space structure arising from the Husimi representation of the subspaces corresponding to certain lifetimes, which we find to be closely related to the classical intensity as a function of the initial state. Furthermore, the statistics of the nearest-neighbour spacings of the complex quasi energies are analysed to gain an insight into potentially universal features. Recently the spectral statistics of open quantum chaotic systems have attracted renewed interest (see e.g. \cite{Deni19,Sa20,Akem19,Hama20}), and new universality classes for non-Hermitian random matrices have been identified in \cite{Hama20}. Here we demonstrate that the nearest-neighbour distributions for our $PT$-symmetric kicked top in the chaotic regime indeed follow the corresponding universality class of \cite{Hama20}. This universality class is dictated by a symmetry of our model under transposition, that is independent of the $PT$-symmetry. We show that  we can switch to another universality class, by introducing an additional nonlinear kicking term in the Hamiltonian, similar to the switch from GOE to GUE statistics for the standard kicked top analysed in \cite{Kus88}. The resulting behaviour is described by the complex Ginibre ensemble in the large $N$ limit. In the regular regime the (unfolded) nearest-neighbour spacings follow the two-dimensional Poisson distribution. This behaviour is in line with what has been identified over two decades ago by Haake and coworkers for regular and chaotic systems with complex eigenvalues \cite{Grob88,Grob89,Haak92c} and which has recently been reinforced by detailed analysis in \cite{Akem19}. We find that when we exclude eigenvalues too close to the boundary or the symmetry line, the nearest-neighbour spacing statistics is in fact independent of the $PT$-symmetry, which can be explained due to the latter introducing a global correlation in the spectrum, while the nearest-neighbour correlation is clearly local as has been argued in \cite{Akem19} and \cite{Hama20}.  
We finally consider the eigenvalues statistics of a $PT$-symmetric generalisation of a triadic Baker map. This system exhibits the same level-spacing characteristics as observed in the $PT$-symmetric kicked top. We finish with a brief summary in section \ref{sec_conclusion}. 

\section{The model}
\label{sec-model}
We introduce a non-Hermitian kicked top, described by the time-dependent Hamiltonian
\begin{equation}
\label{eqn_Ham}
\hat{H}(t) =p \hat{L}_x + (\epsilon + \ui \gamma) \hat{L}_z + \frac{k}{L}\hat{L}_z^2 \sum_{n=-\infty}^{\infty} \delta (t - n \tau),
\end{equation}
where $\hat L_j$ are angular momentum operators, fulfilling the $SU(2)$ commutation relations 
\begin{equation}
[\hat L_z,\hat L_{\pm}]=\pm\hat L_{\pm},\quad {\rm and} \quad [\hat L_+,\hat L_-]=\hat L_z,
\end{equation}
with $\hat L_{\pm}=\hat L_x\pm\rmi \hat L_y$. The system has five real-valued parameters $p$, $\epsilon$, $\gamma$, $k$ and $\tau$. For $\gamma=0$ we recover the familiar Hermitian kicked top \cite{Frah85,Haak86,Haak10}. The first two terms describe a linear precession of a quantum angular momentum; the quadratic delta-kicked term leads to the onset of chaos in the classical dynamics. The total angular momentum $\hat L^2 = \hat L_x^2+\hat L_y^2+\hat L_z^2$ is a conserved quantity. Thus, one can restrict the analysis to a $(2L+1)$-dimensional sector of the Hilbert space belonging to the eigenvalue $L(L+1)$ of $\hat L^2$. In the classical picture this restricts the system to a spherical phase space. The kicked top has been experimentally realised using the spin angular momentum of a single $^{133}$Cs atom as reported in \cite{Chau09}. Another interpretation of the system is an ensemble of bosonic particles populating two modes (either in an external potential or using internal degrees of freedom) where the interaction between the particles is periodically switched on for infinitesimally short times \cite{Strz08}, which could be experimentally realised in cold atoms similar to the experiments reported in \cite{Tomk17}.

For non-vanishing $\gamma$ the Hamiltonian is no longer Hermitian, that is, the dynamics is not unitary, and the norm is not conserved. The $\gamma$ term describes a linear gain/loss profile along the $L_z$ direction. In the bosonic two-mode context this could be realised (up to an imaginary  energy shift) by a loss of particles from one of the modes. A related model without explicit time-dependence, where the quadratic part is constantly switched on, has been studied theoretically in detail in \cite{Grae08b,Grae10b}.  Experimentally atom losses from lattices can be controlled with high precision \cite{Wurt09,Ott16}. They are naturally described by Lindblad dynamics, and an effective non-Hermitian dynamics can be achieved by post selection (see, e.g. \cite{Dale14}), as has been the case in the recent experimental realisations of quantum $PT$-symmetric systems \cite{Klau19,Grae19}. In this spirit, the model introduced here could be realised with atoms in a double well trap, with induced losses from one of the sites, and a time-dependent interaction strength, or equivalently, time-dependent barrier height and loss rate \cite{Strz08}.

There are various definitions of $PT$-symmetry in use throughout the literature, that differ in small details. Here we use the definition of \cite{Jone10b}: A non-Hermitian Hamiltonian is called $PT$-symmetric if it satisfies $[PT,H]=0$ where $T$ is an antilinear operator and $P$ is a linear involution that commutes with $T$. $T$ is usually set to be the time-reversal operator, which, in our case fulfils $T^2 = +1$ (sometimes referred to as even $PT$-symmetry). A $PT$-symmetric Hamiltonian $\hat H$ with even $PT$-symmetry thus satisfies $\hat H = P\hat H^*P$, where the asterisk denotes complex conjugation. As a consequence $\hat H$ has either real eigenvalues when the corresponding eigenstates are also eigenstates of the operator $PT$, or otherwise the eigenvalues come in complex conjugate pairs. For $\epsilon = 0$ for any given time the Hamiltonian (\ref{eqn_Ham}) is $PT$-symmetric, with 
\begin{equation}
PT:\quad \hat{L}_x \mapsto \hat{L}_x, \quad \hat{L}_y \mapsto \hat{L}_y, \quad\hat{L}_z \mapsto -\hat{L}_z,\quad {\rm and}\quad {\rmi\to-\rmi}.
\end{equation}

The investigation of kicked systems is most easily facilitated by the study of the Floquet operator, describing the time-evolution of a state over one time period, $\hat F=\hat U(t_0,t_0+\tau)$. Here, we choose a time-interval symmetric around the kick, that is, we consider a free evolution with $\hat H_0=p \hat{L}_x +(\epsilon+ \ui \gamma) \hat{L}_z$ for time length $\frac{\tau}{2}$, followed by an instantaneous kick with $\hat H_1=\frac{k}{L}\hat{L}_z^2$, and then another free evolution for time $\frac{\tau}{2}$. That is, the Floquet operator is given by
\begin{equation}
\label{eqn_Floq}
\hat F =  \ue^{-\ui(p \hat{L}_x +(\epsilon+ \ui \gamma) \hat{L}_z)\frac{\tau}{2}} \ue^{-\ui \frac{k}{L}\hat{L}_z^2} \ue^{-\ui(p \hat{L}_x +(\epsilon+ \ui \gamma) \hat{L}_z)\frac{\tau}{2}}.
\end{equation}
The choice of the time interval ensures that the Floquet operator is indeed $PT$-symmetric with respect to the $PT$ operator above for $\epsilon=0$. Here we follow \cite{Moch16} and define a quantum map as $PT$-symmetric if it fulfils the relation $\hat F = P(\hat F^*)^{-1}P$. Denoting the eigenvalues of a $PT$-symmetric quantum map by $\lambda =  r\ue^{-\ui \theta}$, the eigenvalues are either unimodular, when the corresponding eigenstates are also eigenstates of the $PT$ operator, or come in pairs $\lambda_1=(\lambda_2^*)^{-1}$, i.e., $r_1=\frac{1}{r_2}$ and $\theta_1=\theta_2$, in which case the $PT$ operator acting on one of the eigenvectors yields the other. That is, the quasienergies $E=\rmi \ln(\lambda)$ are either real or come in complex conjugate pairs, just as the eigenvalues of time-independent $PT$-symmetric Hamiltonians. Note that for another choice of the time interval the Floquet operator would be isospectral to the one considered here and still be $PT$-symmetric, however, with a non-trivial $P$ operator.

In what follows we shall study the occurrence of chaos in the classical dynamics and its fingerprints in the quantum system for the $PT$-symmetric quantum map (\ref{eqn_Floq}). For this purpose, we shall first introduce the classical map, arising in the limit of large Hilbert space dimension $L\to\infty$, and analyse the resulting dynamics, before we return to its quantum features.

\section{Classical dynamics}
\label{sec-classical}
To derive the classical dynamics we consider the quantum evolution of the expectation values of the three angular momentum operators in the limit $L \to \infty$. We will deduce the mapping of the expectation values of the angular momentum operators over one period, by expressing 
\begin{equation}
\langle \hat L_j \rangle_{n+1} = \frac{\langle \psi^{(n+1)}|\hat L_j|\psi^{(n+1)}\rangle}{\langle \psi^{(n+1)}|\psi^{(n+1)}\rangle}
\end{equation}
as functions of the $\langle\hat  L_j \rangle_{n}$, where
\begin{equation}
|\psi^{(n+1)}\rangle= \hat F|\psi^{(n)}\rangle=\hat F^{(n+1)}|\psi^{(0)}\rangle,
\end{equation}
is the quantum state after $n+1$ evolution periods. 

Let us first consider the free evolution with the Hamiltonian $\hat H_0=p \hat{L}_x +(\epsilon+ \ui \gamma) \hat{L}_z$. The equations of motion for the expectation values of the angular momentum operators can for example be deduced from the $2\times 2$ representation of the system, which, due to the linearity in the $\hat L_j$ provides the full dynamics for all values of the total angular momentum. The time evolution of the expectation values of the rescaled angular momentum expectation values $s_j= \langle\hat L_j\rangle/L$ for the free evolution is given by \cite{Grae08b,Grae10b}
\begin{eqnarray}\label{eqn_free_eom}
\dot s_x=&-\epsilon s_y- \gamma \,s_z s_x, \nonumber\\
\dot s_y=& \epsilon s_x-p  s_z - \gamma \,s_z s_y,\\
\dot s_z=&  p  s_y+\gamma\, (1-s_z^2)\,.\nonumber
\end{eqnarray}
In the Hermitian case $\gamma=0$ this describes a rigid rotation around the axis $\left(p,\, 0,\, \epsilon\right)^T$.

\begin{figure}[htb]
\centering
\includegraphics[width=0.32\textwidth]{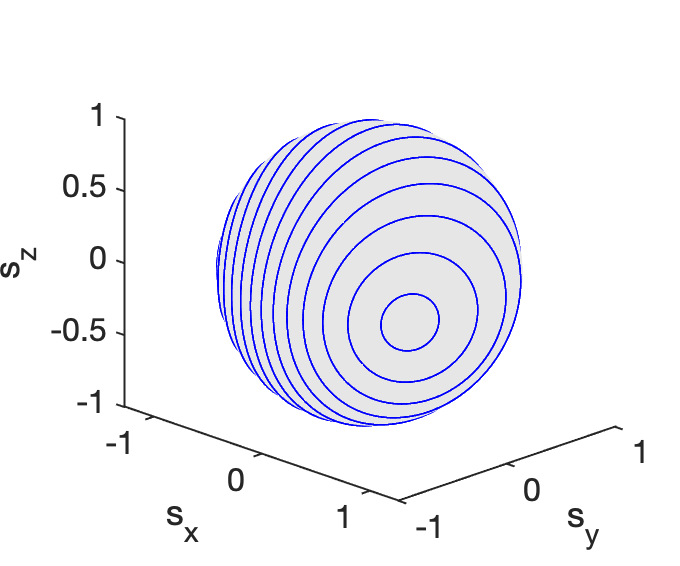}
\includegraphics[width=0.32\textwidth]{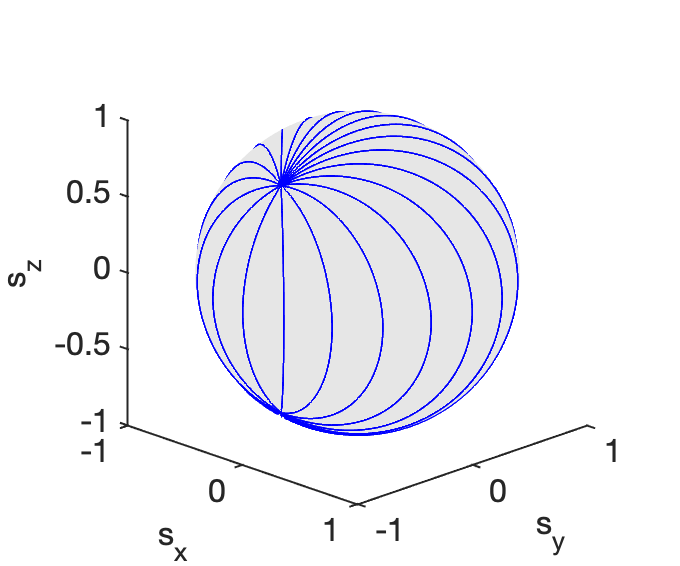}
\includegraphics[width=0.32\textwidth]{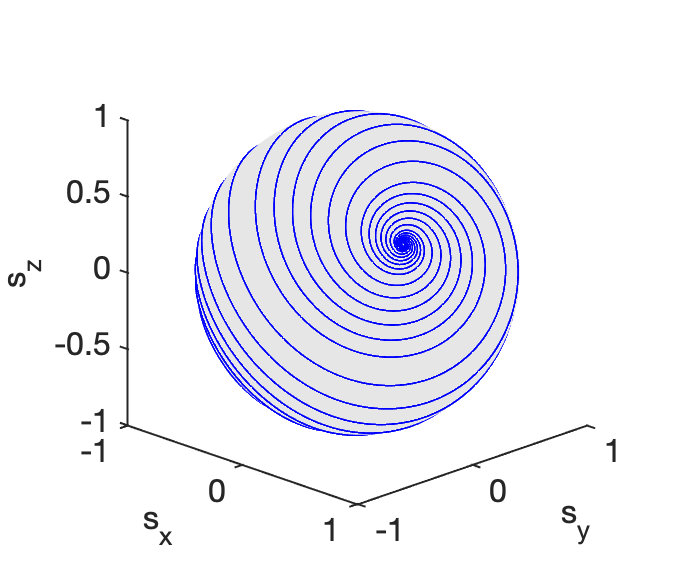}\\
\includegraphics[width=0.32\textwidth]{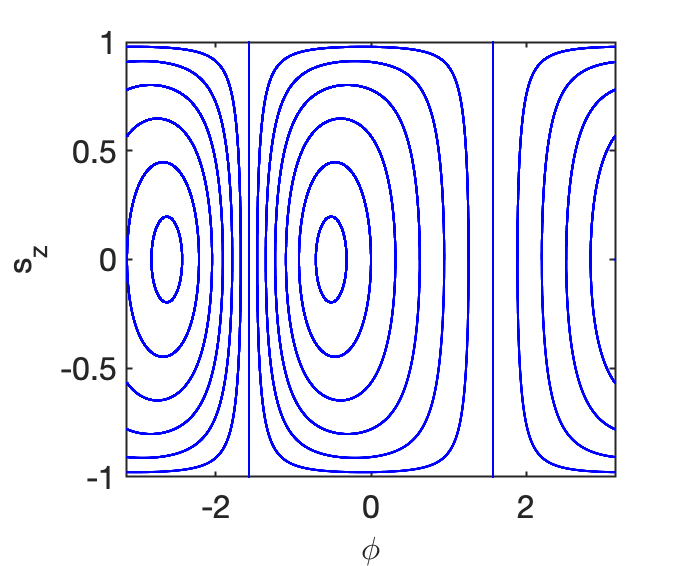}
\includegraphics[width=0.32\textwidth]{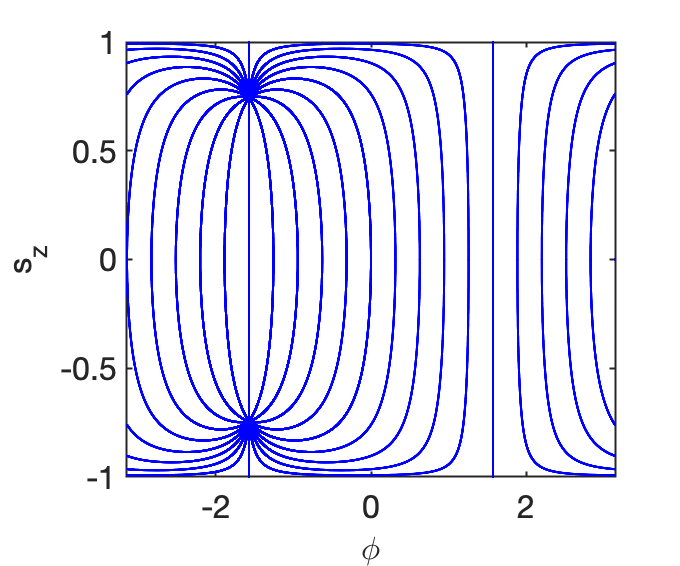}
\includegraphics[width=0.32\textwidth]{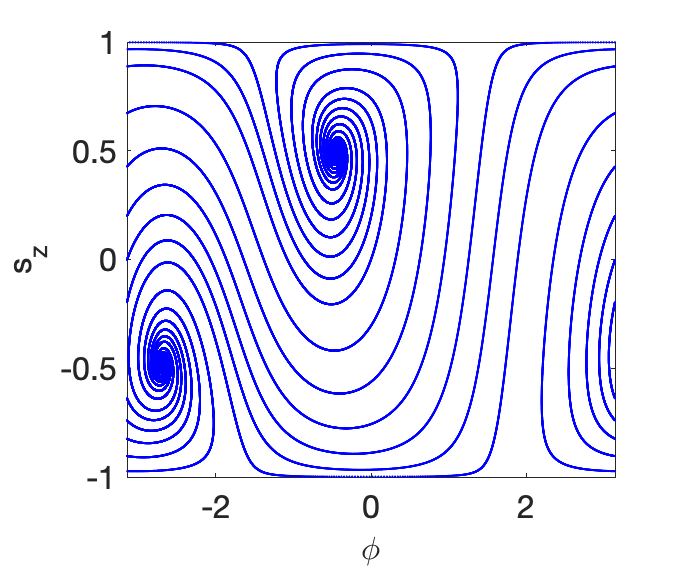}
\caption{Trajectories of the free motion for $p=1$ and different values of $\epsilon$ and $\gamma$, $\gamma=0.5$ and $\epsilon=0$ (left), $\gamma=1.5$ and $\epsilon=0$ (middle), and $\gamma=0.5$ and $\epsilon=0.5$ (right).}
\label{fig_freedyn}
\end{figure} 

Some examples of the phase-space trajectories for non-vanishing $\gamma$ are depicted in figure \ref{fig_freedyn}. For more details on the dynamics see, e.g., \cite{Grae10b}. Figure \ref{fig_freedyn} depicts both the dynamics on the sphere (in the top row) as well as the projection onto a flat phase space in (canonical) cylindrical coordinates $s_z$ and $\phi={\rm atan}\left(\frac{s_y}{s_x}\right)$ (in the bottom row). In the $PT$-symmetric case, $\gamma\neq 0,\, \epsilon=0$, there are two distinct phases of motion for subcritical and supercritical values of $\gamma$, respectively, where the critical value is given by $\gamma_{crit}=p$. For $\gamma<p$, an example of which is shown in the left panel of figure \ref{fig_freedyn}, the motion is a combination of a rotation around the x-axis and a Lorentz boost around the z-axis, with two stable elliptic fixed points that, for increasing values of $\gamma$ move from the opposite poles at $s_x=\pm 1$ along the equator, approaching $s_x=0$, where they meet at the critical value. Note that while there are no sinks and sources in this evolution the dynamics is not area preserving, but has balanced amounts of contraction and expansion over the phase sphere. For supercritical increasing values of $\gamma>\gamma_{crit}$ the fixed points separate and travel along the great circle with $s_x=0$ towards the north and south pole. The southern fixed point is a source and the northern fixed point is a sink of the dynamics. An example of this motion is depicted in the middle panel of figure \ref{fig_freedyn}. This classical behaviour is directly related to the $PT$-symmetry and its breaking in the quantum system \cite{Grae10b}. 
Without the $PT$-symmetry, for $\epsilon\neq 0$ the elliptic fixed points present for the Hermitian system turn into a sink and a source for arbitrarily small values of $\gamma$, and the rigid rotation around the $\left(p,\, 0,\, \epsilon\right)^T$-axis turns into a spiralling phase-space flow from source to sink, where again the sink and the source move closer to the poles of the sphere for increasing values of $\gamma$. An example of the resulting trajectories is depicted in the right panel of figure \ref{fig_freedyn}. 

To deduce the phase-space map generated by the free motion over half a period $\tau/2$ we need to integrate the dynamical equations (\ref{eqn_free_eom}). For the $PT$-symmetric case $\epsilon=0$, that explicitly yields
\begin{align} 
\tilde s_x=&\,\, \frac{1}{\Gamma} s_x\\
 \tilde s_y =&\,\, \frac{1}{\Gamma}\left( \left(\frac{p^2}{\omega^2} \mathrm{cos}\left(\frac{\omega \tau}{2}\right) - \frac{\gamma^2}{\omega^2} \right)s_y-\frac{p}{\omega}\mathrm{sin}\left(\frac{\omega \tau}{2}\right) s_z  +\frac{p\gamma}{\omega^2}\cos\left(\frac{\omega \tau}{2}\right) - \frac{p\gamma}{\omega^2}\right)\\
 \tilde s_z=&\, \, \frac{1}{\Gamma} \left(\ \frac{p}{\omega}\mathrm{sin}\left(\frac{\omega \tau}{2}\right) s_y+ \mathrm{cos}\left(\frac{\omega \tau}{2}\right)s_z + \frac{\gamma}{\omega}\sin\left(\frac{\omega \tau}{2}\right) \right),
 \end{align}
with 
\begin{equation}
\label{eqn_normfactor}
\Gamma=\frac{p^2}{\omega^2} - \frac{\gamma^2}{\omega^2}\mathrm{cos}\left(\frac{\omega \tau}{2}\right) + \frac{p\gamma}{\omega^2}\big(1 - \mathrm{cos}\left(\frac{\omega \tau}{2}\right)\big) s_y  + \frac{\gamma}{\omega}\mathrm{sin}\left(\frac{\omega \tau}{2}\right)s_z ,
\end{equation}
and  $\omega = \sqrt{p^2 - \gamma^2}$. It is straightforward to verify that this indeed reduces to a rigid rotation by an angle $\frac{p\tau}{2}$ around the x-axis, for $\gamma=0$. 

The full kicked top map is composed of the mapping resulting from the free motion followed by the mapping arising from the kick, which is not altered by the non-Hermiticity of the free evolution,  and then another free motion of half a period. The kicked part of the Floquet operator alone yields the map \cite{Haak86}
\begin{align} 
\tilde{s}_x &= \mathrm{cos}(2k s_z)s_x - \mathrm{sin}(2k s_z)s_y\nonumber\\
\tilde{s}_y &= \mathrm{sin}(2k s_z)s_x + \mathrm{cos}(2k s_z)s_y\label{eqn_map_k}\\
\tilde{s}_z &= s_z,\nonumber
 \end{align}
that is, a torsion around the $z$-axis. In what follows we shall set the period $\tau$ to unity. 

The $PT$ symmetry of the quantum system results in the classical map $F=F_{free}F_{kick}F_{free}$ being \textit{reversible} \cite{Robe92}. A map is called \textit{reversible} if there exists an involution $G$ (i.e. applying $G$ twice yields the identity) such that 
\begin{equation}
F\circ G\circ F=G
\end{equation}
which is equivalent to 
\begin{equation}
F^{-1}=G\circ F\circ G,
\end{equation}
i.e. under the transformation of the phase space by $G$ the map (that is the time evolution) is inversed. Not surprisingly, the involution $G$ for our system is the classical counterpart of the $P$ operator,  that is the reversing of the z-direction $G=P: s_z\mapsto -s_z$. 

There are many consequences  of reversibility, regarding symmetry lines,  the occurrence of \textit{pseudo}  conservative features such as periodic orbits and KAM tori, as well as dissipative features  such as attractors, and universal scaling behaviours \cite{Robe92}. An exhaustive analysis of the classical dynamics goes well beyond of the scope of the present paper, however, and will be reserved for another study. Here we shall content ourselves with gaining a first overview of the dynamical features of our classical map. 

\begin{figure}[htb]
\centering
\includegraphics[width=0.32\textwidth]{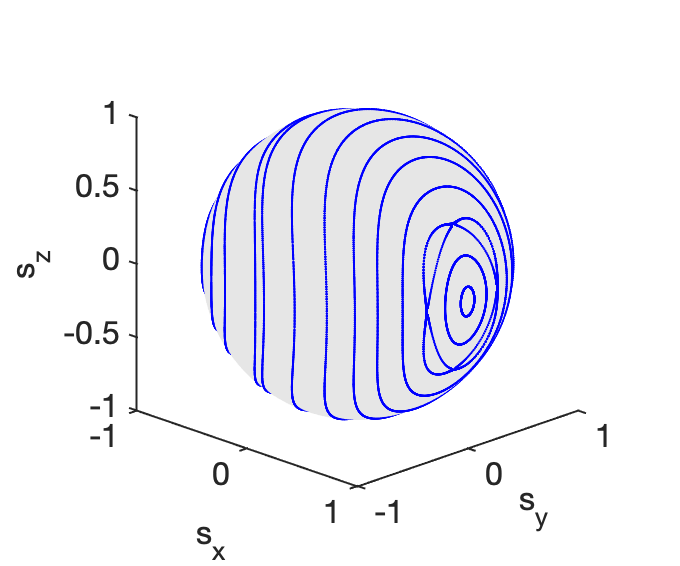}
\includegraphics[width=0.32\textwidth]{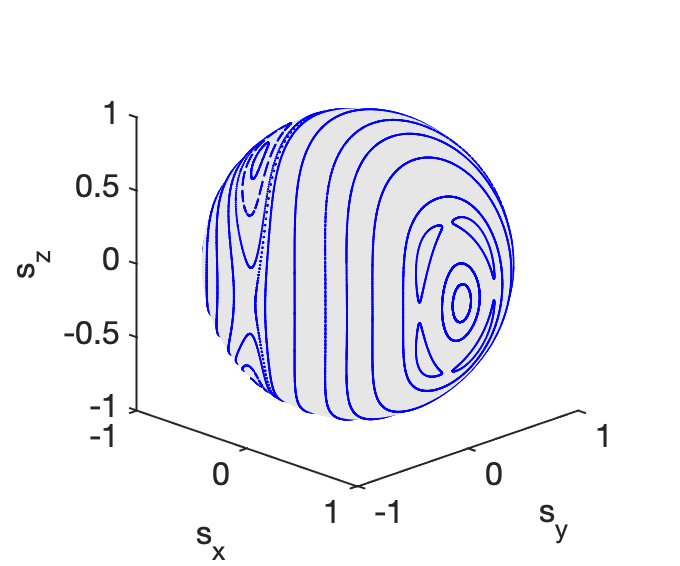}
\includegraphics[width=0.32\textwidth]{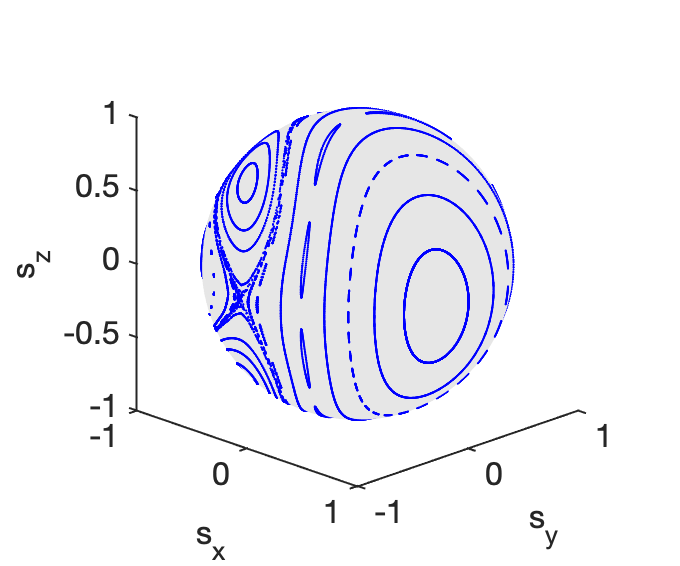}
\includegraphics[width=0.32\textwidth]{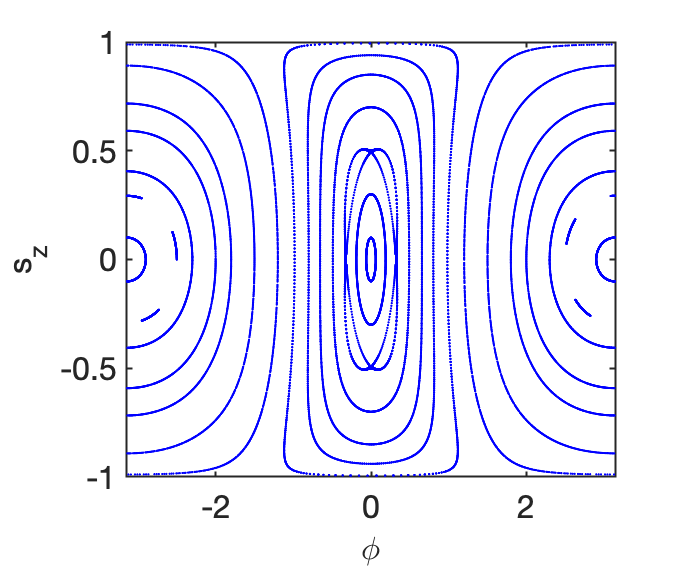}
\includegraphics[width=0.32\textwidth]{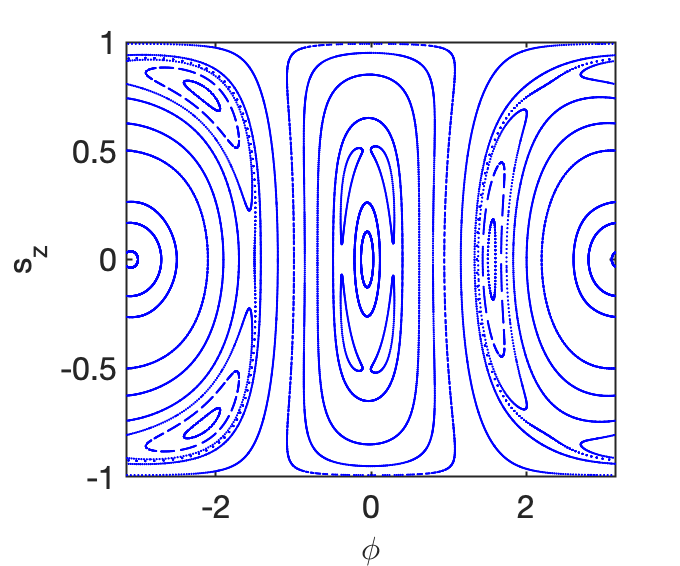}
\includegraphics[width=0.32\textwidth]{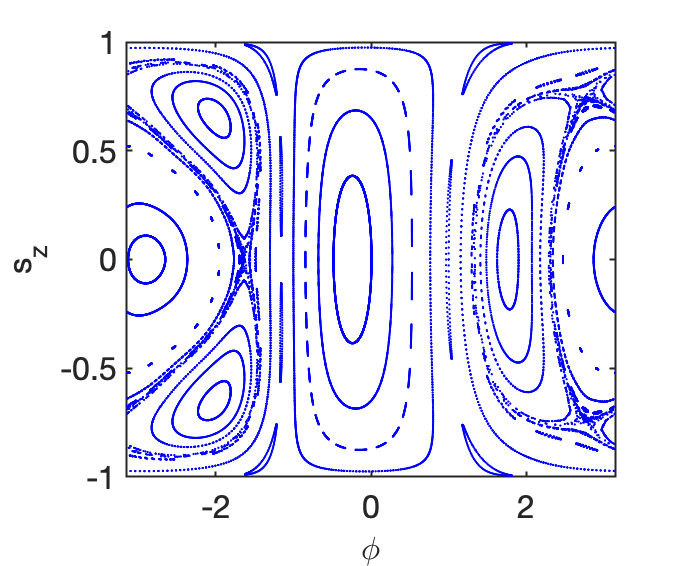}
\includegraphics[width=0.32\textwidth]{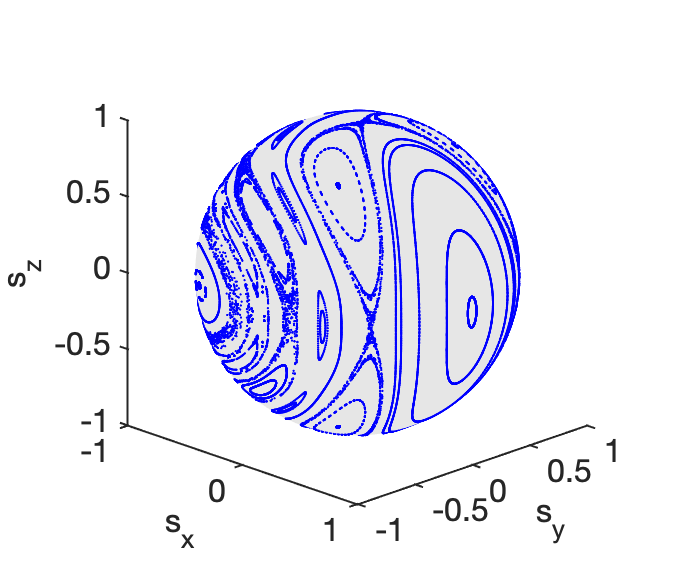}
\includegraphics[width=0.32\textwidth]{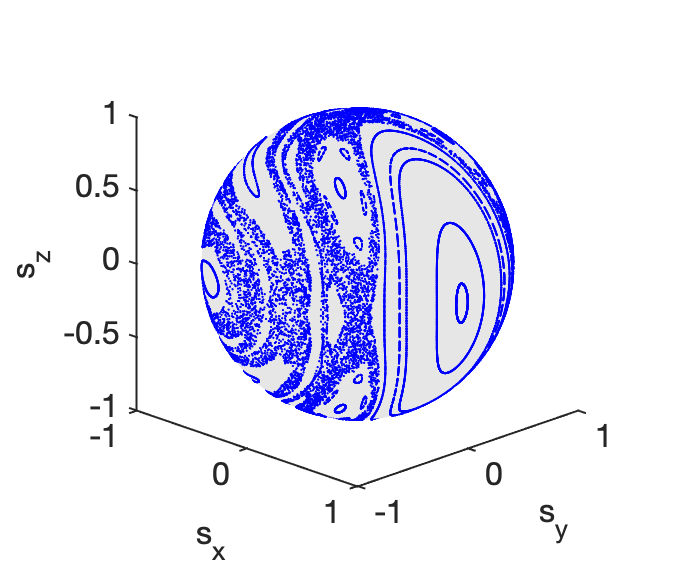}
\includegraphics[width=0.32\textwidth]{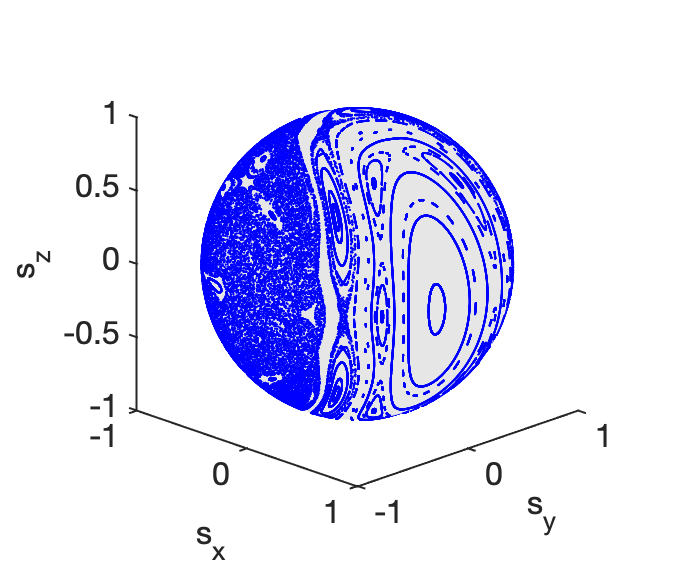}
\includegraphics[width=0.32\textwidth]{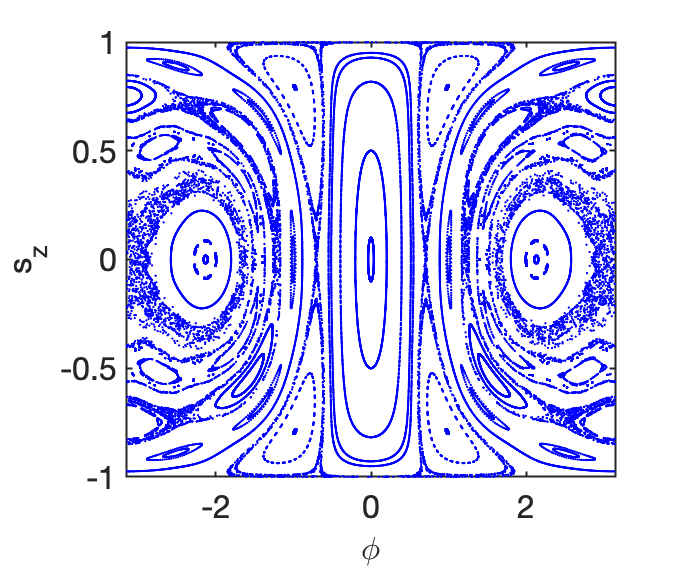}
\includegraphics[width=0.32\textwidth]{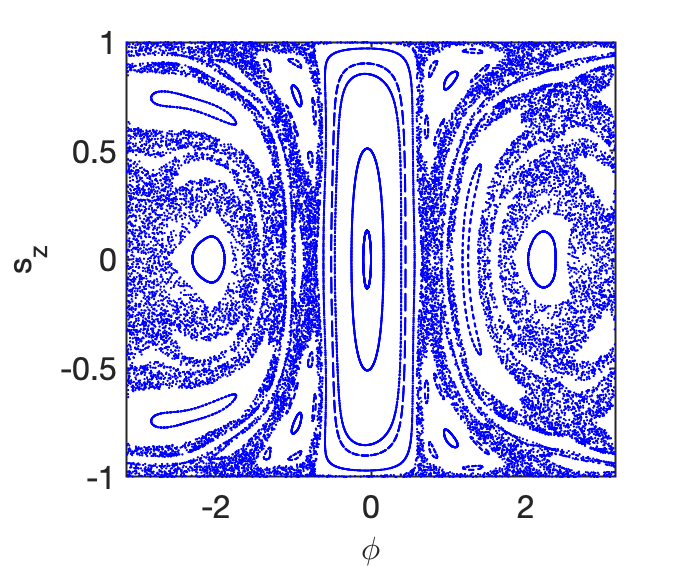}
\includegraphics[width=0.32\textwidth]{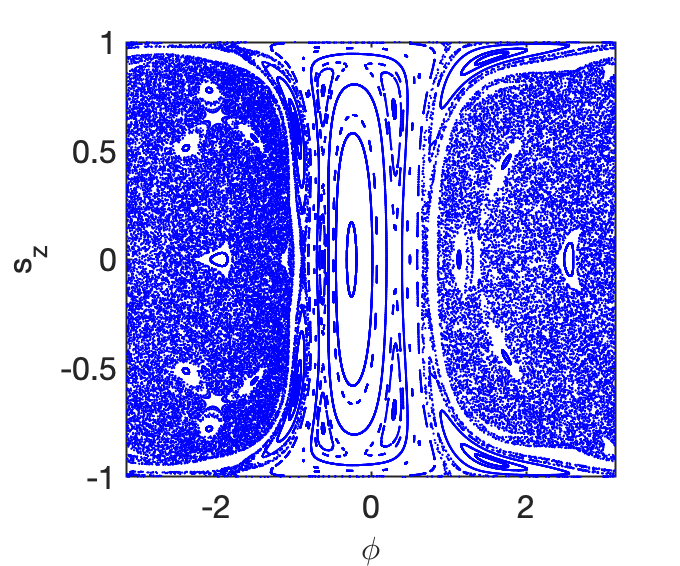}
\caption{Poincare plots of the classical map for $\epsilon=0$ and different values of $k$ and $\gamma$. The top two rows show the Poincare dynamics on the sphere (top) and the projection to the (canonical) cylindrical variables (bottom) for $k=0.5$ and $\gamma=0,\, 0.1,\, 0.5$ (from left to right). The bottom two rows show the same for $k=1$.}
\label{fig_dyn1}
\end{figure}

In the Hermitian case $\gamma=0$ an increasing kicking strength $k$ leads to the onset of chaos and for large kicking strength $k>>p$ the dynamics is almost entirely chaotic. Here we shall investigate how the non-Hermiticity changes this behaviour, in particular in the $PT$-symmetric case $\epsilon=0$. 

Note that even for infinitesimal values of $\gamma$ the dynamics is no longer area preserving. Since the kicking part of the dynamics remains Hamiltonian, the rate of contraction and expansion does not depend on the value of the kicking strength. Some examples of Poincare plots of the phase-space dynamics for $\epsilon=0$, $p=2$, and relatively small $\gamma$  and $k$ are depicted in figure \ref{fig_dyn1}. We observe that the non-Hermiticity seems to enhance the chaos, but does not lead to the appearance of sinks and sources in the dynamics for the small values depicted here.  This behaviour is due to the $PT$-symmetry/reversibility of the system, and an inclusion of a small symmetry breaking parameter $\epsilon$ would lead to a fully contracting phase-space flow with a single point sink. 

\begin{figure}[htb]
\centering
\includegraphics[width=0.32\textwidth]{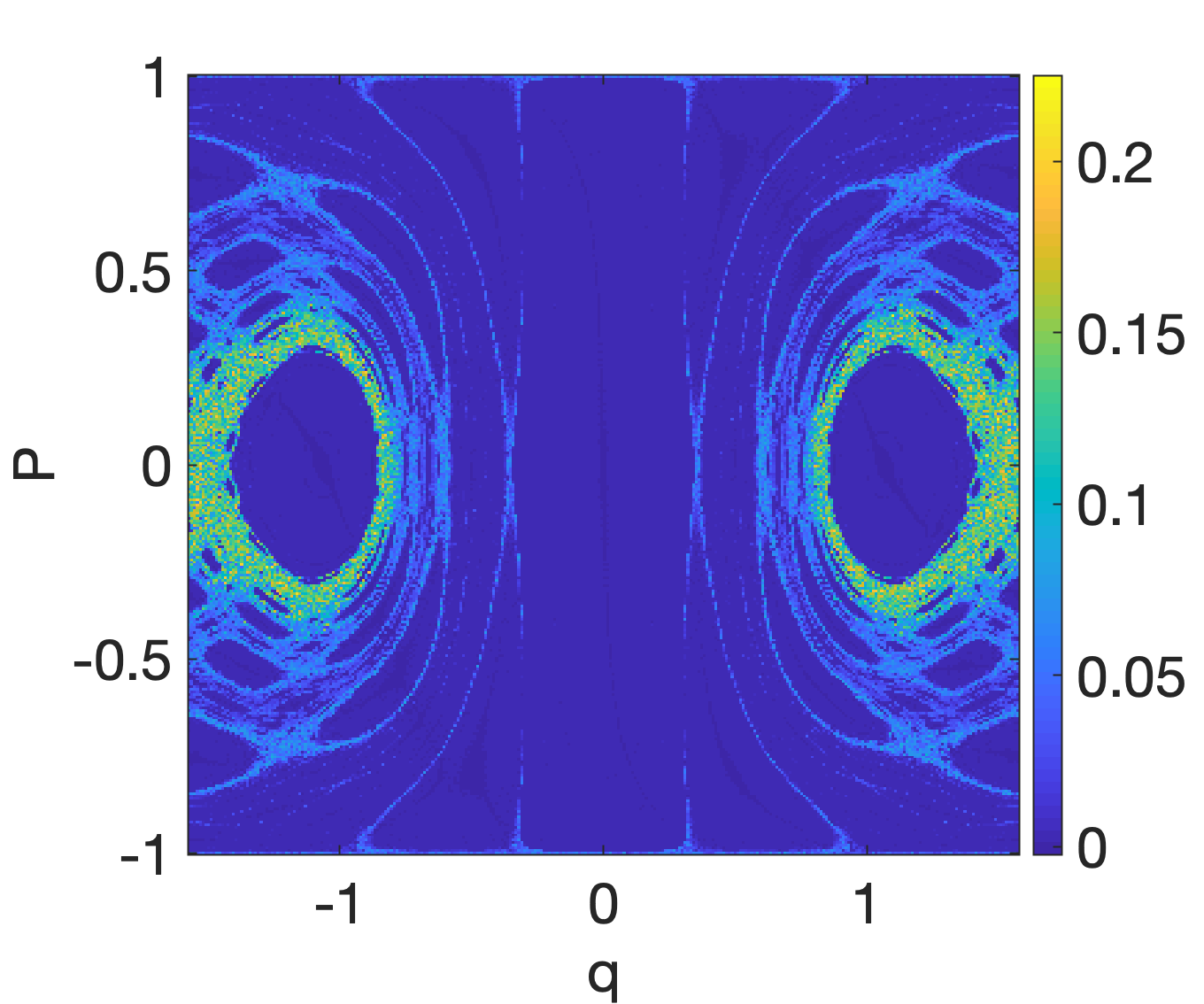}
\includegraphics[width=0.32\textwidth]{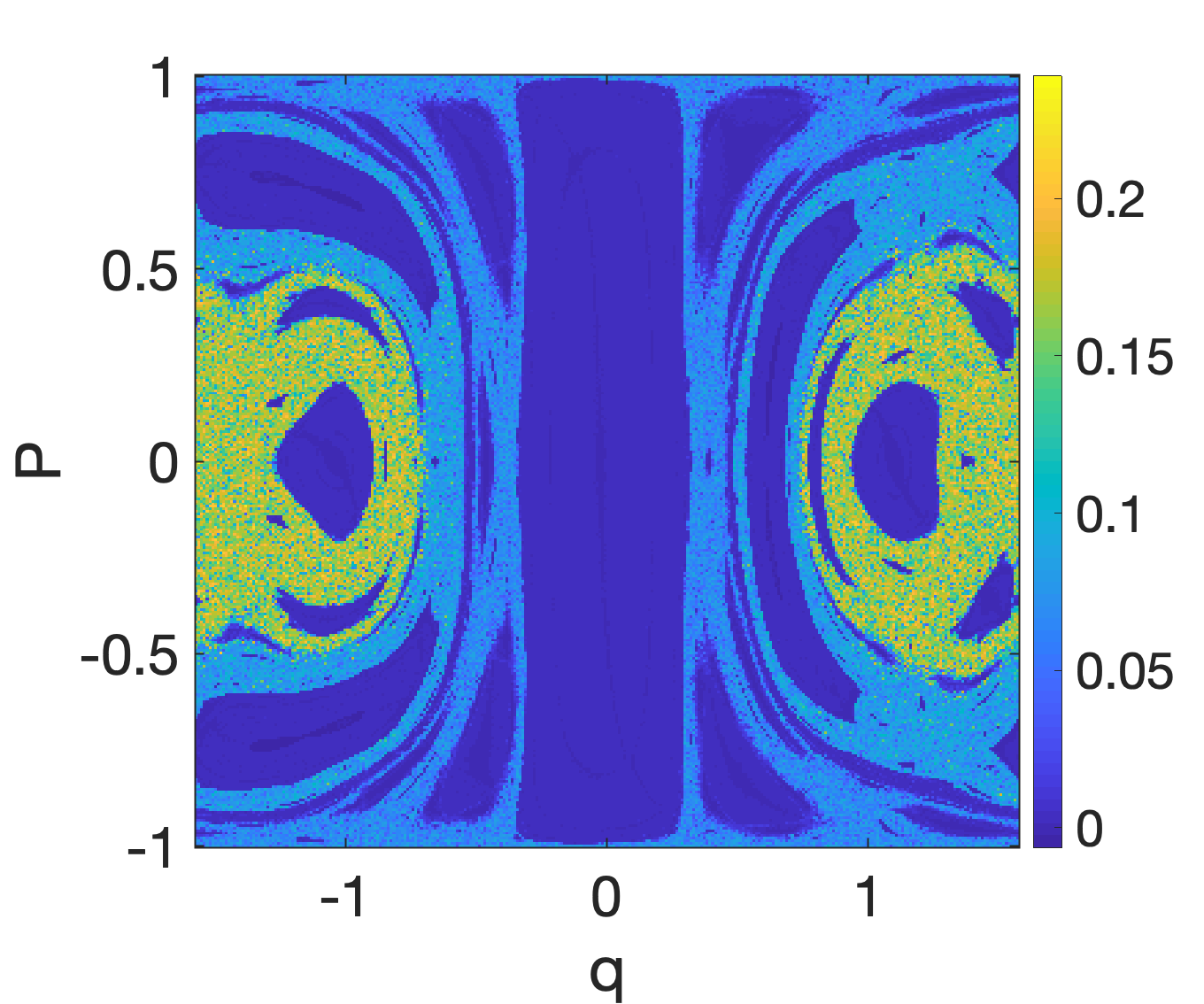}
\includegraphics[width=0.32\textwidth]{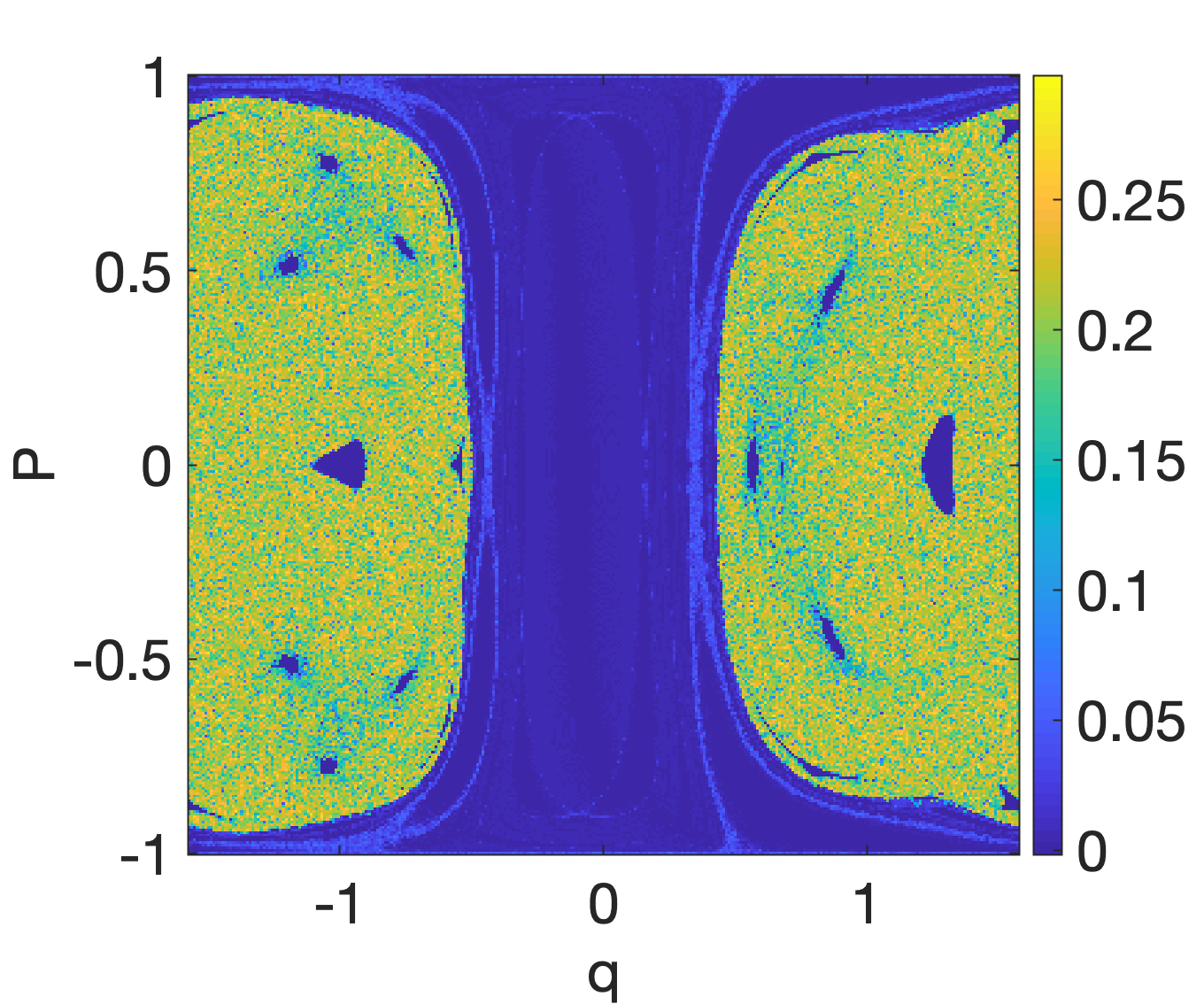}
\caption{Lyapunov exponents on the projection of the phase space for $k=1$, $p=2$ and three values of $\gamma$ ($\gamma=0,\, 0.1,\, 0.5$, from left to right). The Lyapunov exponent is estimated numerically using 2000 iterations of the map and a variation of $10^{-7}$ of the initial phase-space coordinates.}
\label{fig_lyap1}
\end{figure}  
 
The enhancement of chaos in the classical phase-space dynamics observed in the Poincare sections is confirmed by the analysis of the Lyapunov exponent of the dynamics, depicted in false colours for $p=2$, $k=1$ and three different values of $\gamma$ in Figure \ref{fig_lyap1}. A positive Lyapunov exponent indicates exponential sensitivity to small variations in the initial conditions, that is, chaos. In the examples depicted, while most of the phase space is associated to a vanishing Lyapunov exponent, corresponding to regular motion, for vanishing $\gamma$, the region of positive Lyapunov exponent grows significantly with increasing $\gamma$ and for $\gamma=0.5$ already a large proportion of the phase space is covered by a chaotic sea.

\begin{figure}[htb]
\centering
\includegraphics[width=0.32\textwidth]{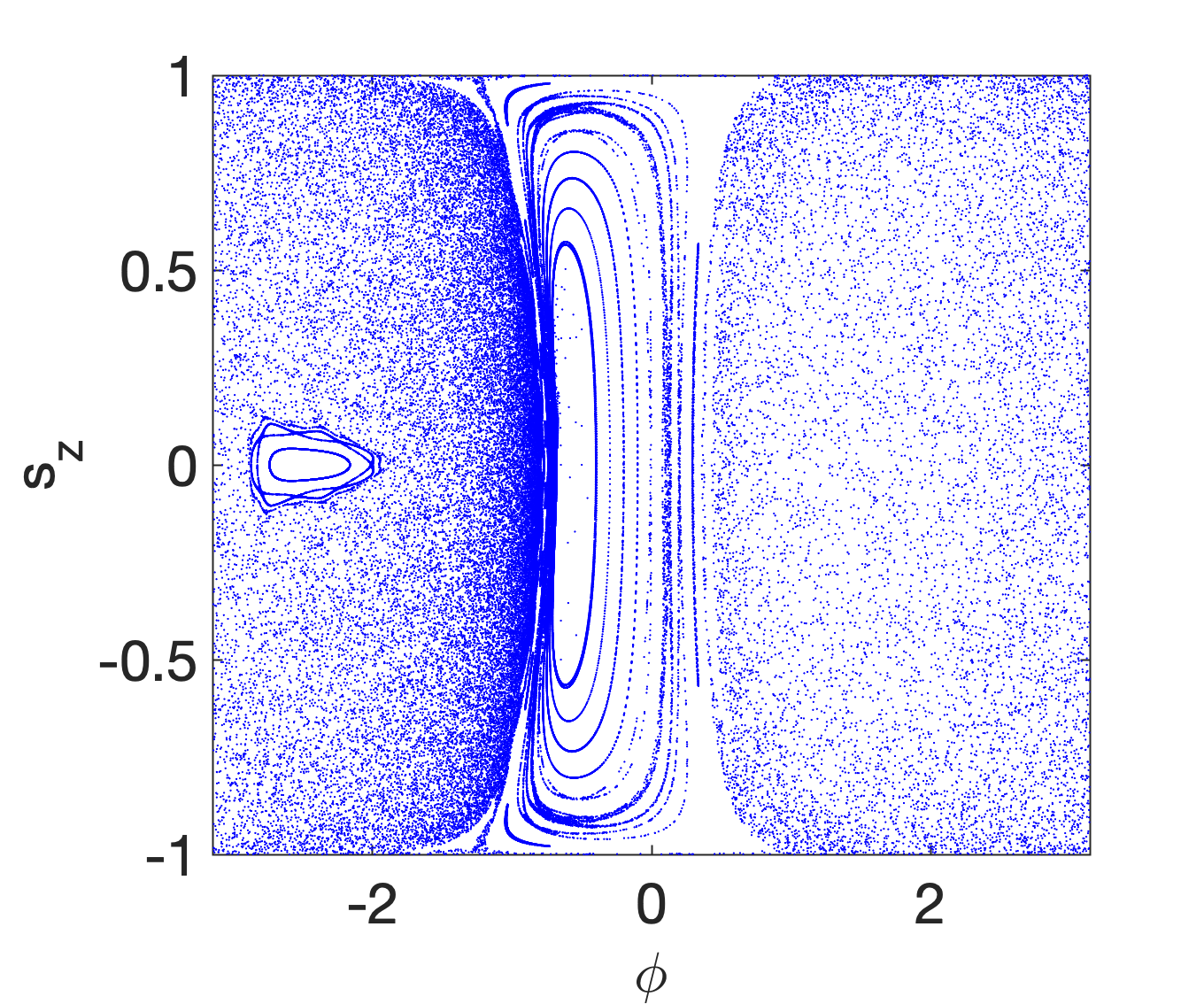}
\includegraphics[width=0.32\textwidth]{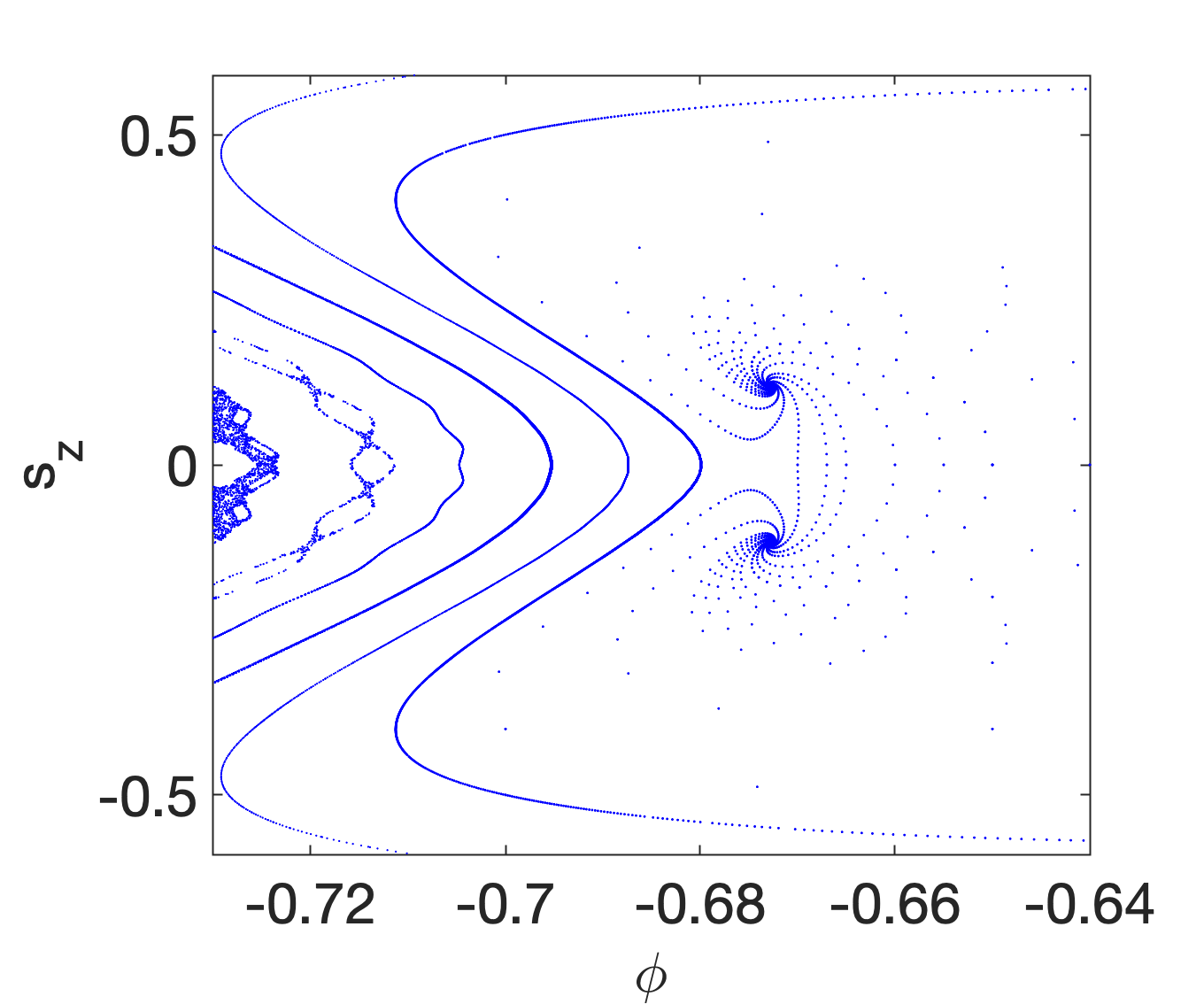}
\includegraphics[width=0.32\textwidth]{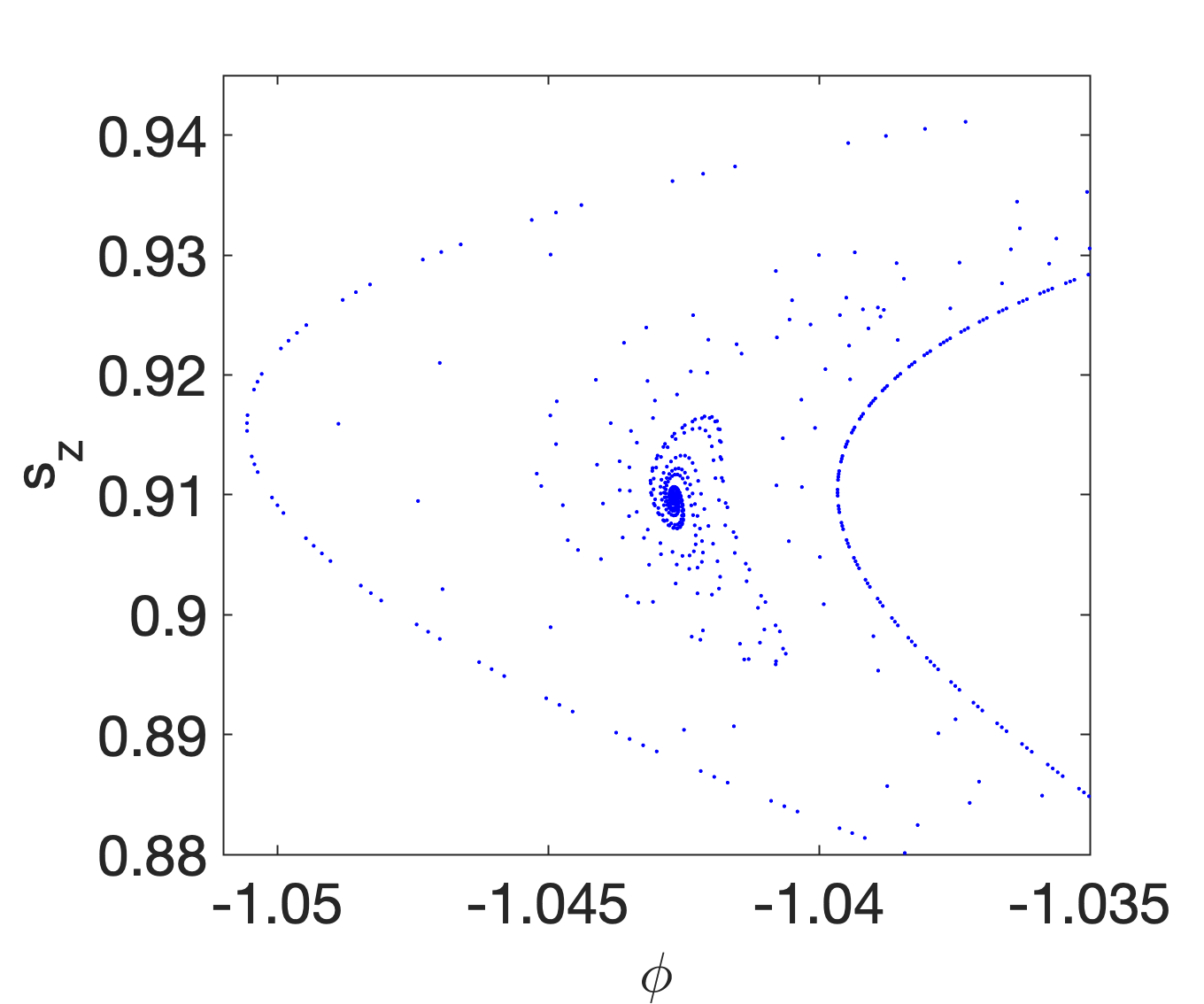}
\caption{Poincare plot of the classical map for $\epsilon=0$, $p=2$, $k=1$, and $\gamma=1.25$ (left). The middle and right panels depict magnifications of smaller regions of phase space.}
\label{fig_dyn1x}
\end{figure}

The most prominent fixed point in the Poincare sections in figure  \ref{fig_dyn1} is the one that originates from the elliptic fixed point at $\phi=0$ for vanishing $\gamma$ and $k$. It moves along the equator with increasing $\gamma$, much as it would have done for $k=0$. In fact, its position is unaffected by the value of $k$, provided $k$ is not so large that the fixed point  has bifurcated, which can be seen as follows. It is a consequence of the reversibility of the system that any possible elliptic fixed point has to lie on the line $s_z=0$. In addition, at an elliptic fixed point the linearisation of the map must be area preserving, i.e. the Jacobian determinant must be one. Now that Jacobian determinant, as mentioned  above, is independent of the (conservative) kicking  and thus the parameter $k$. Therefore, the only possible elliptic  fixed points of the system are the two fixed points of the free evolution, that are situated at $s_z=0$ and $\phi_1=-\arctan\left(\frac{\gamma}{p}\right)$ and $\phi_2=-\pi-\arctan\left(\frac{\gamma}{p}\right)$, respectively. However, with increasing $k$ the fixed point on the left side in our phase-space plots bifurcates into a two-cylce and a hyperbolic fixed point in a chaotic band. For not too large $k$, as $\gamma$ increases further, the bifurcation is reversed, and the stable fixed point on the left reappears, as can be seen for example in figure \ref{fig_dyn1x}. The regions around the two fixed points are separated from each other by increasing numbers of layers of chains of small regular islands and chaotic bands, in the typical KAM fashion.

\begin{figure}[htb]
\centering
\includegraphics[width=0.32\textwidth]{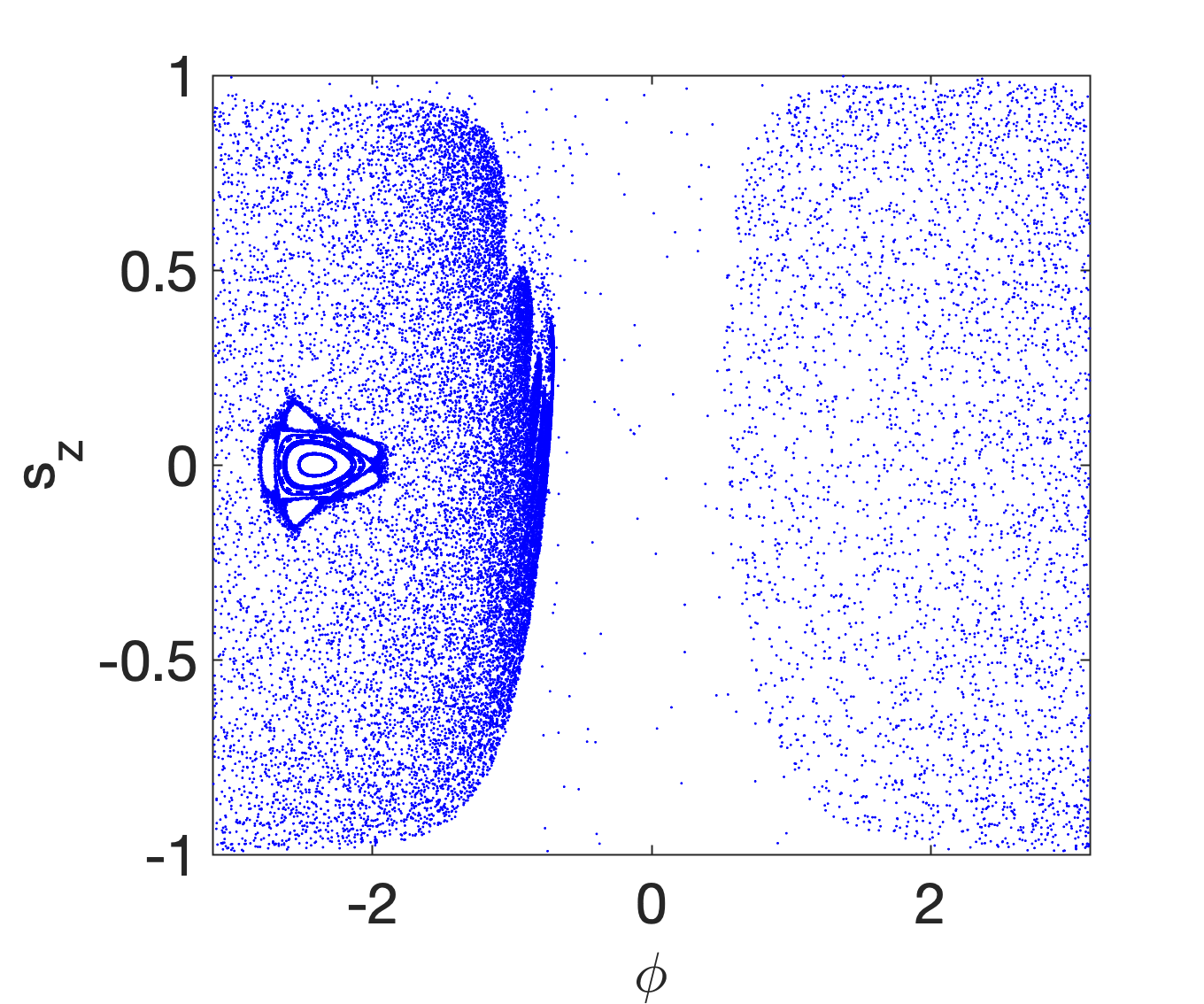}
\includegraphics[width=0.32\textwidth]{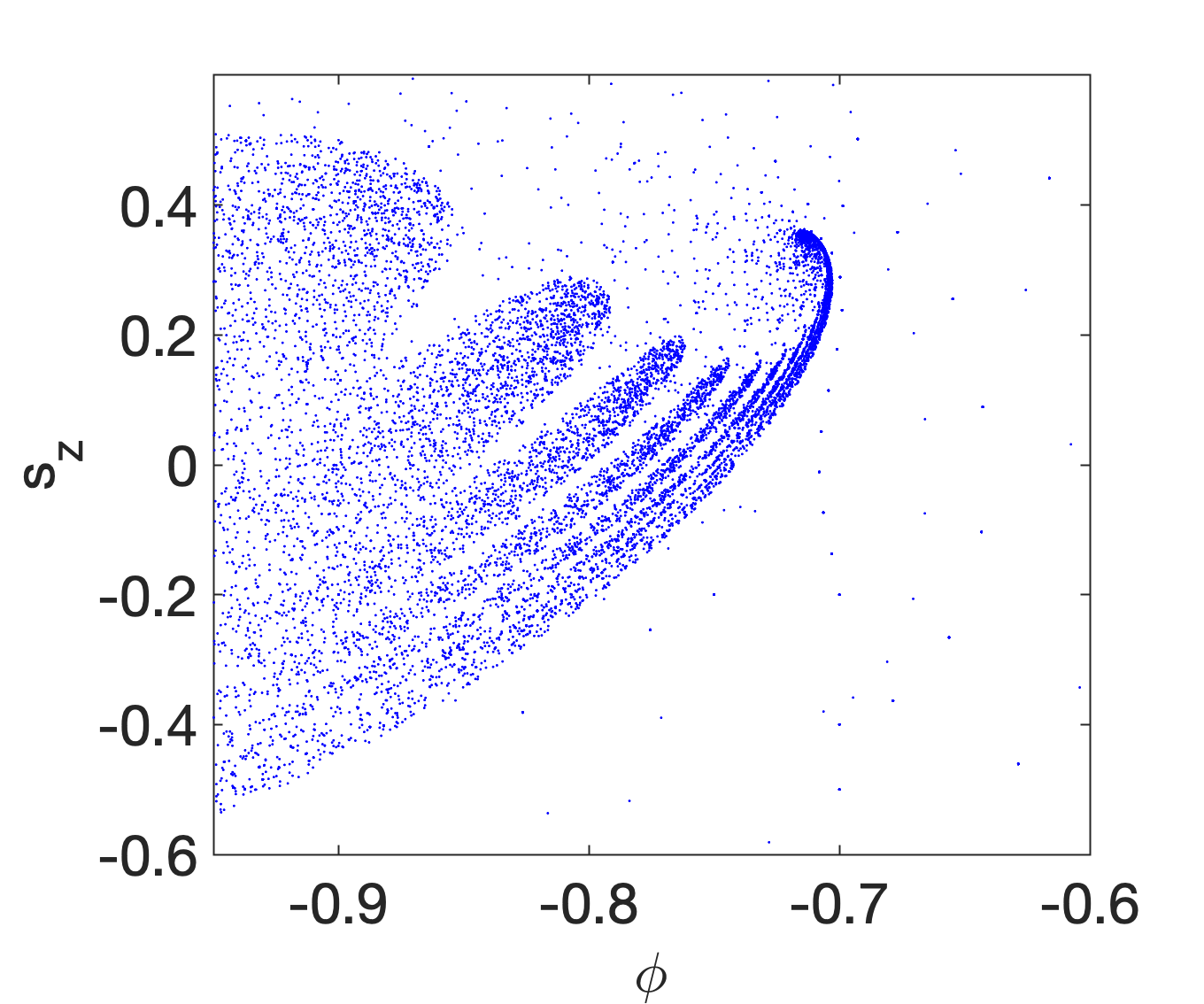}
\includegraphics[width=0.32\textwidth]{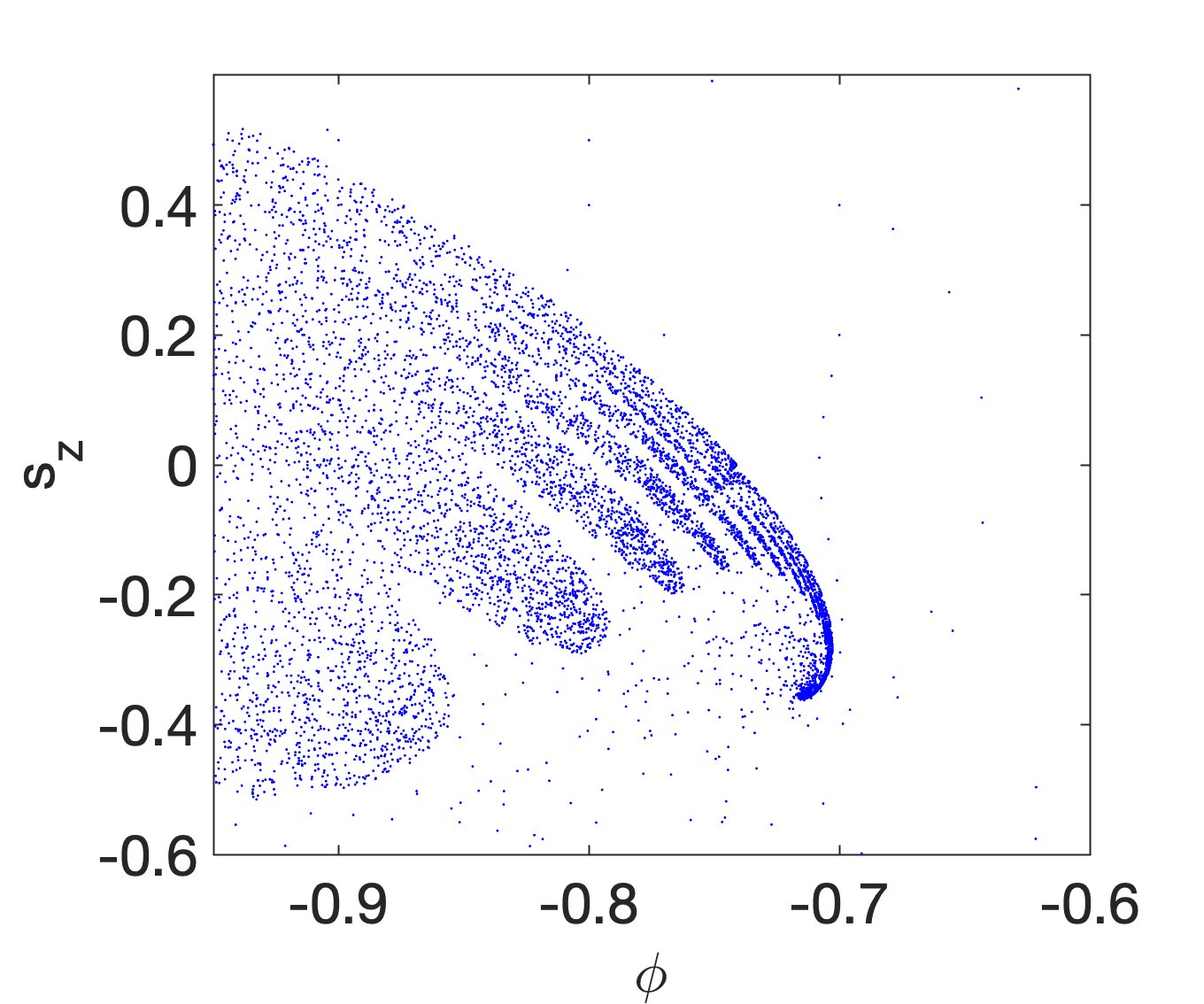}
\includegraphics[width=0.32\textwidth]{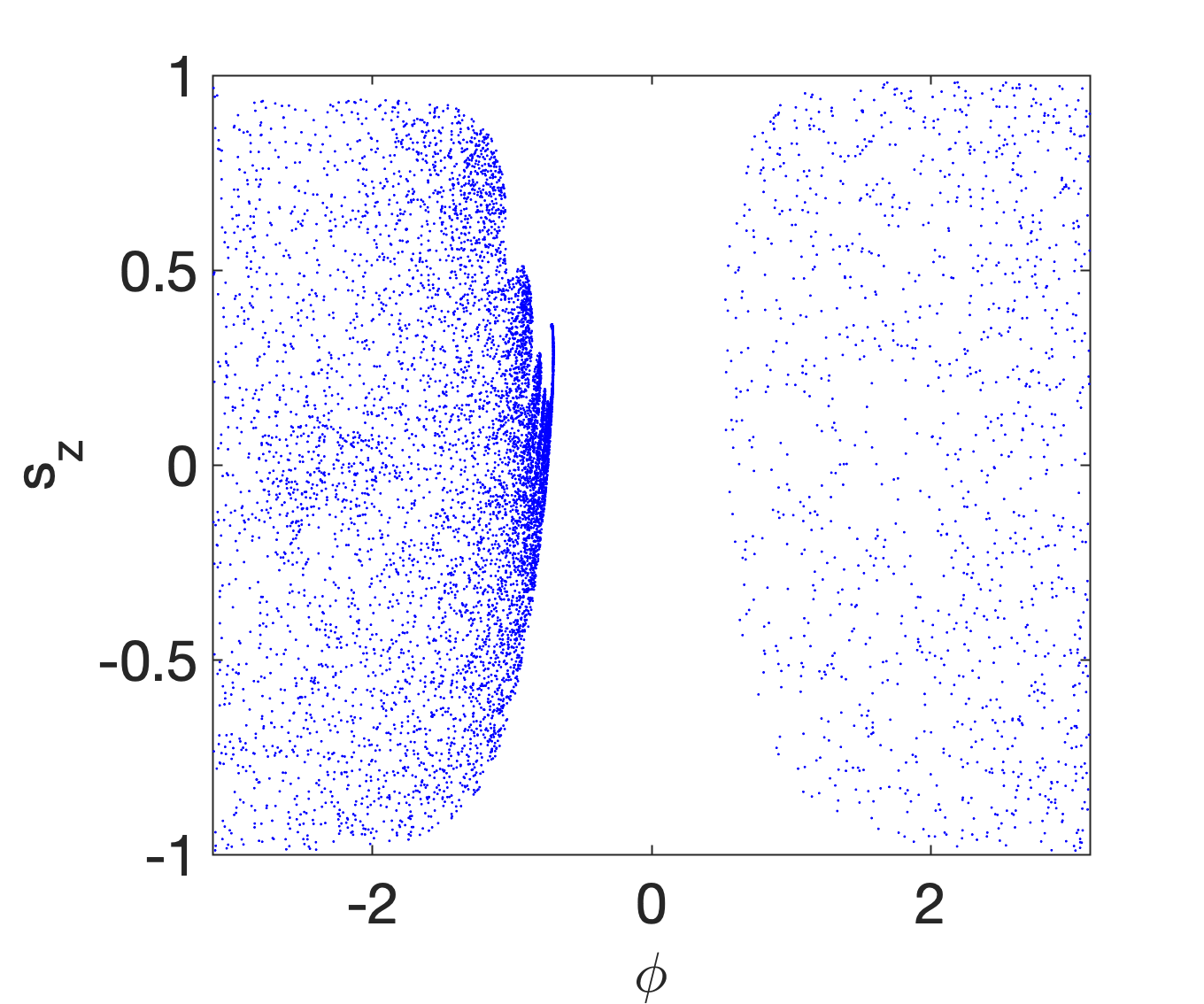}
\includegraphics[width=0.32\textwidth]{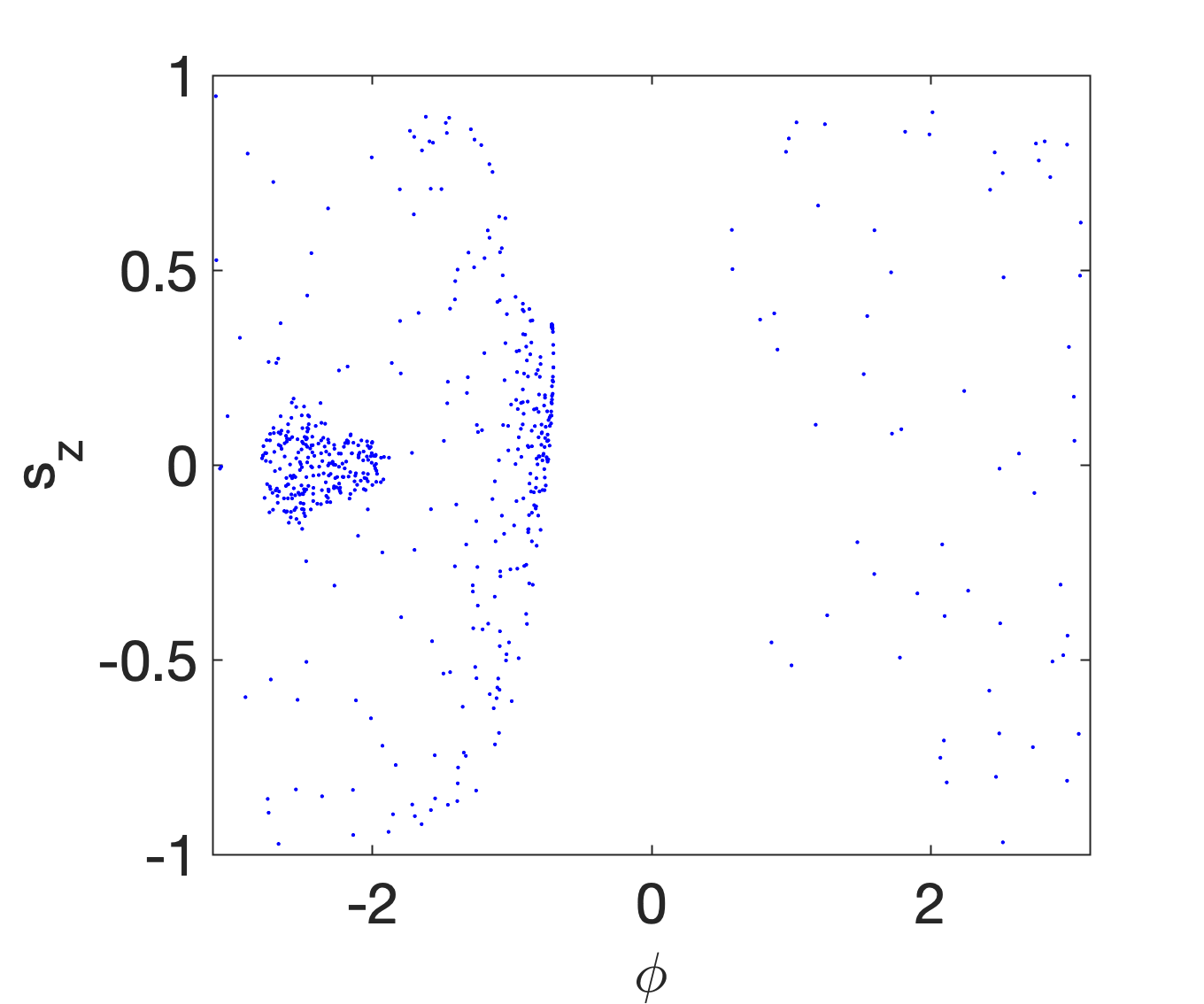}
\includegraphics[width=0.32\textwidth]{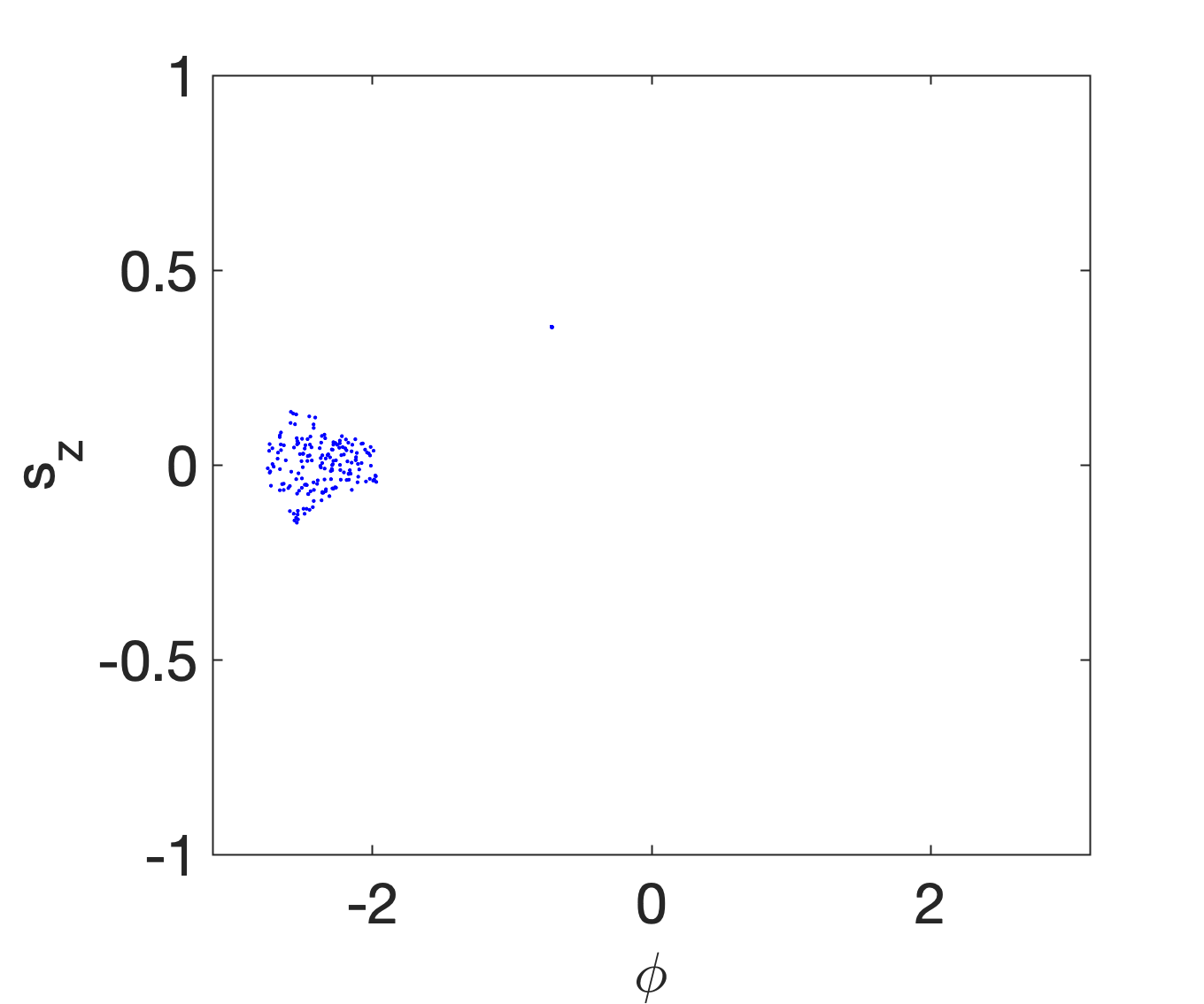}
\caption{Dynamics of the classical map for $\epsilon=0$, $p=2$, $k=1$, and $\gamma=1.35$. The left plot on the top shows a Poincare section of the dynamics and the middle plot on the top shows a magnification of the region around the sink and the source. The right panel on the top shows a magnification of the same area for the backwards propagation. The bottom row shows the distribution of $20000$ phase-space points that were initially chosen from a uniform distribution on phase space after propagation for $100$ (left), $500$ (middle), and $10,000$ (right) iterations.}
\label{fig_dyn1xx}
\end{figure} 

For larger, but still subcritical values of $\gamma<p$ the prominent central fixed point moves further along the equator and finally bifurcates into a sink and a source for a small volume surrounding it. An example of the Poincare section just above this bifurcation, and a magnification of the area around the central fixed point, are depicted in figure \ref{fig_dyn1x} in the left and middle panels. At this value of $\gamma$ also other of the countless fixed points of the system have turned into sinks for tiny phase-space volumes, as is revealed by a zoom into the Poincare plot, depicted in the right panel of the same figure. To generate these figures both the forwards and backwards time-evolution have been plotted, to get a more complete picture of the phase-space flow.

For further increasing $\gamma$ the basin of attraction of the sink in the central region grows. At another critical value of $\gamma$, finally this region of attraction reaches the chaotic sea, and now the volume of the latter is also attracted by the sink, which by now is the only sink of the flow. An example of the resulting Poincare map and a magnification of the area around the sink and source are depicted in the top left and middle panels of figure \ref{fig_dyn1xx}. The right panel on the top of the same figure shows the same magnified area as the figure in the middle, however, for the backwards propagation, revealing the source corresponding to the sink. The sink attracts all of the phase-space volume but the small volume in the remaining regular island at $s_z=0$ and $\phi=-\pi-\arctan\left(\frac{\gamma}{p}\right)$. This can be seen in the bottom panels of the same figure, that depict the distribution of $20000$ phase-space points that were initially chosen from a uniform distribution on phase space and propagated for $100$, $500$, and $10,000$ iterations, respectively. It can be clearly seen that while the points that started out in the regular island remain within this volume, all other points are eventually mapped into the sink.

\begin{figure}[htb]
\centering
\includegraphics[width=0.32\textwidth]{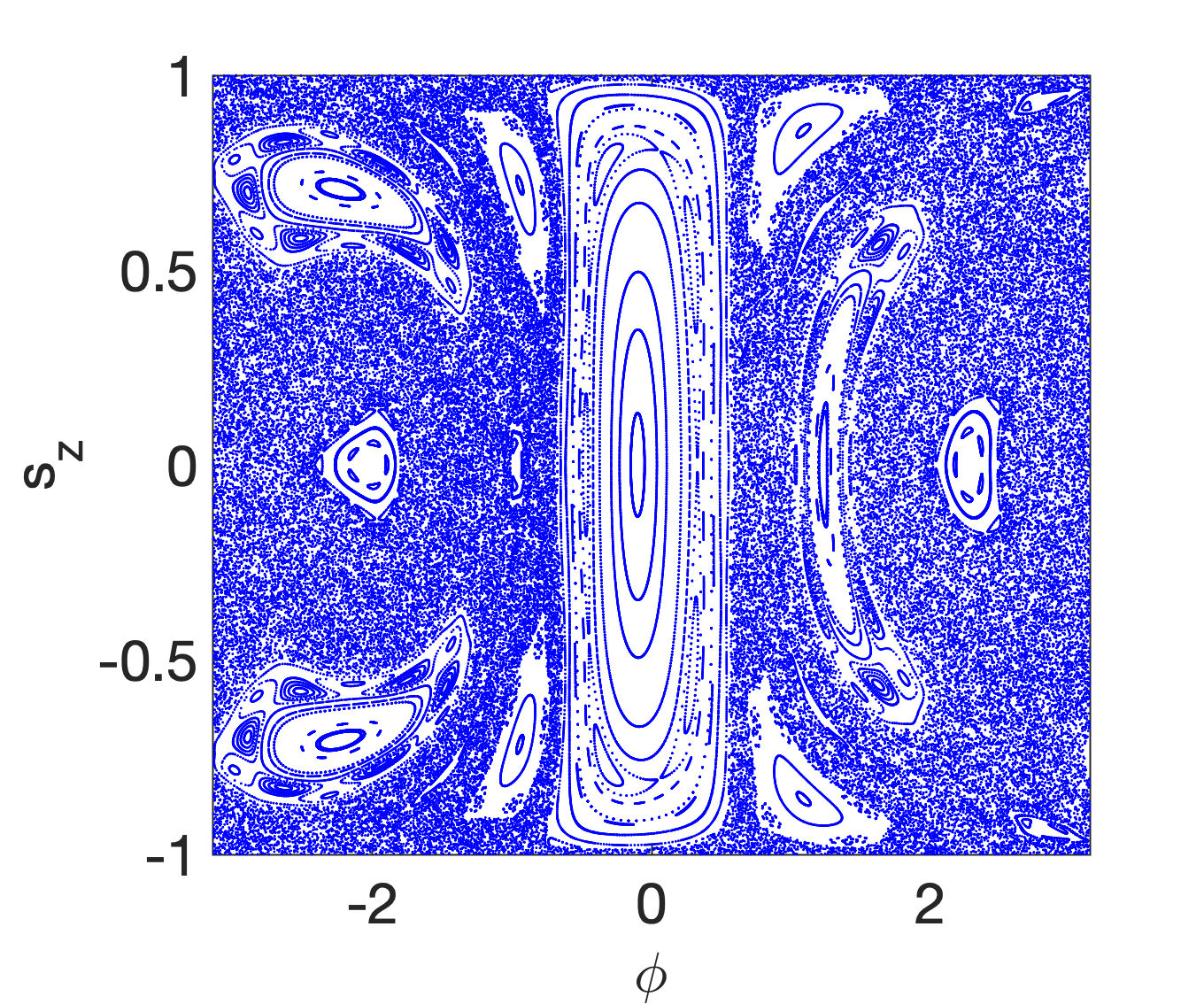}
\includegraphics[width=0.32\textwidth]{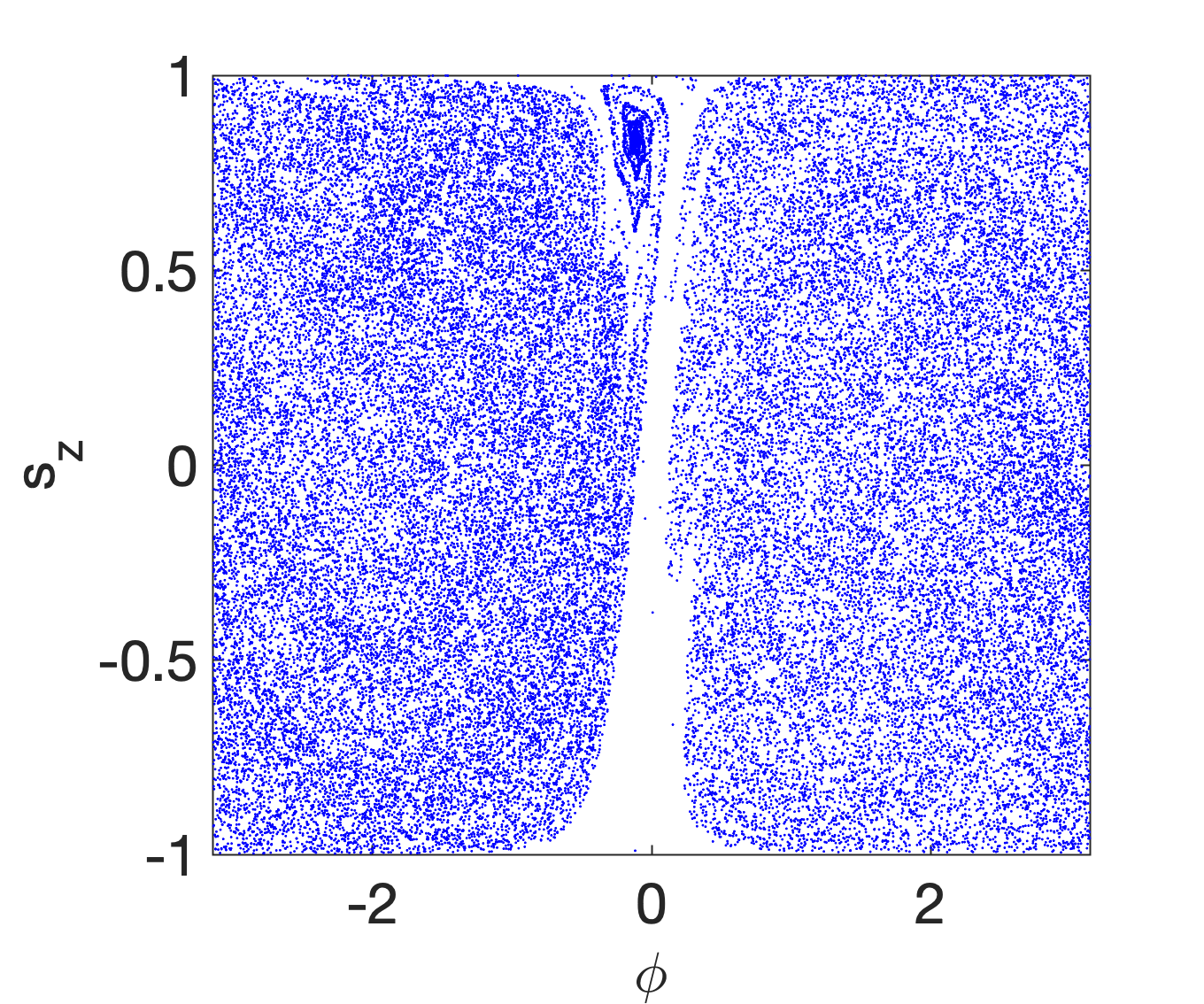}
\includegraphics[width=0.32\textwidth]{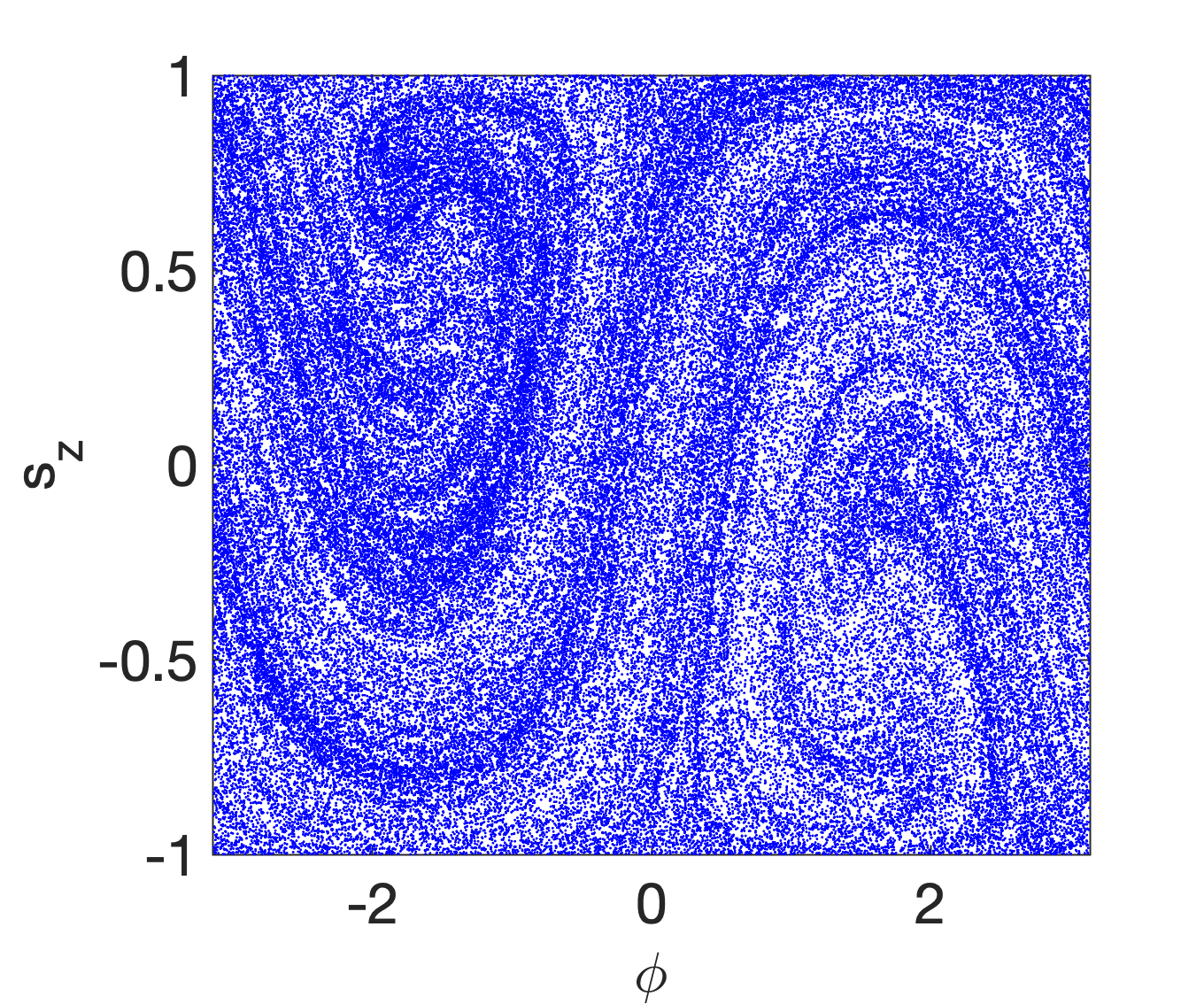}
\includegraphics[width=0.32\textwidth]{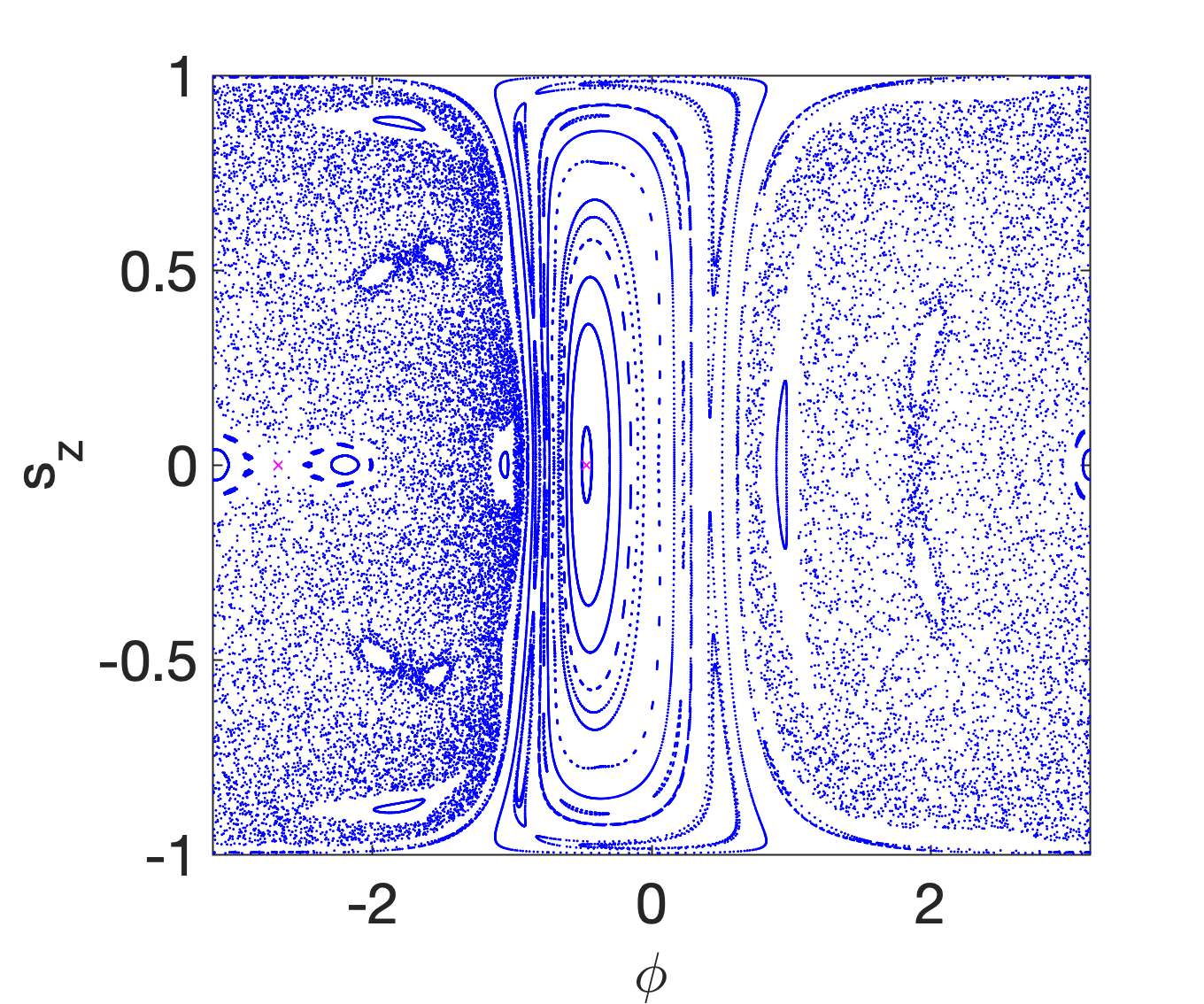}
\includegraphics[width=0.32\textwidth]{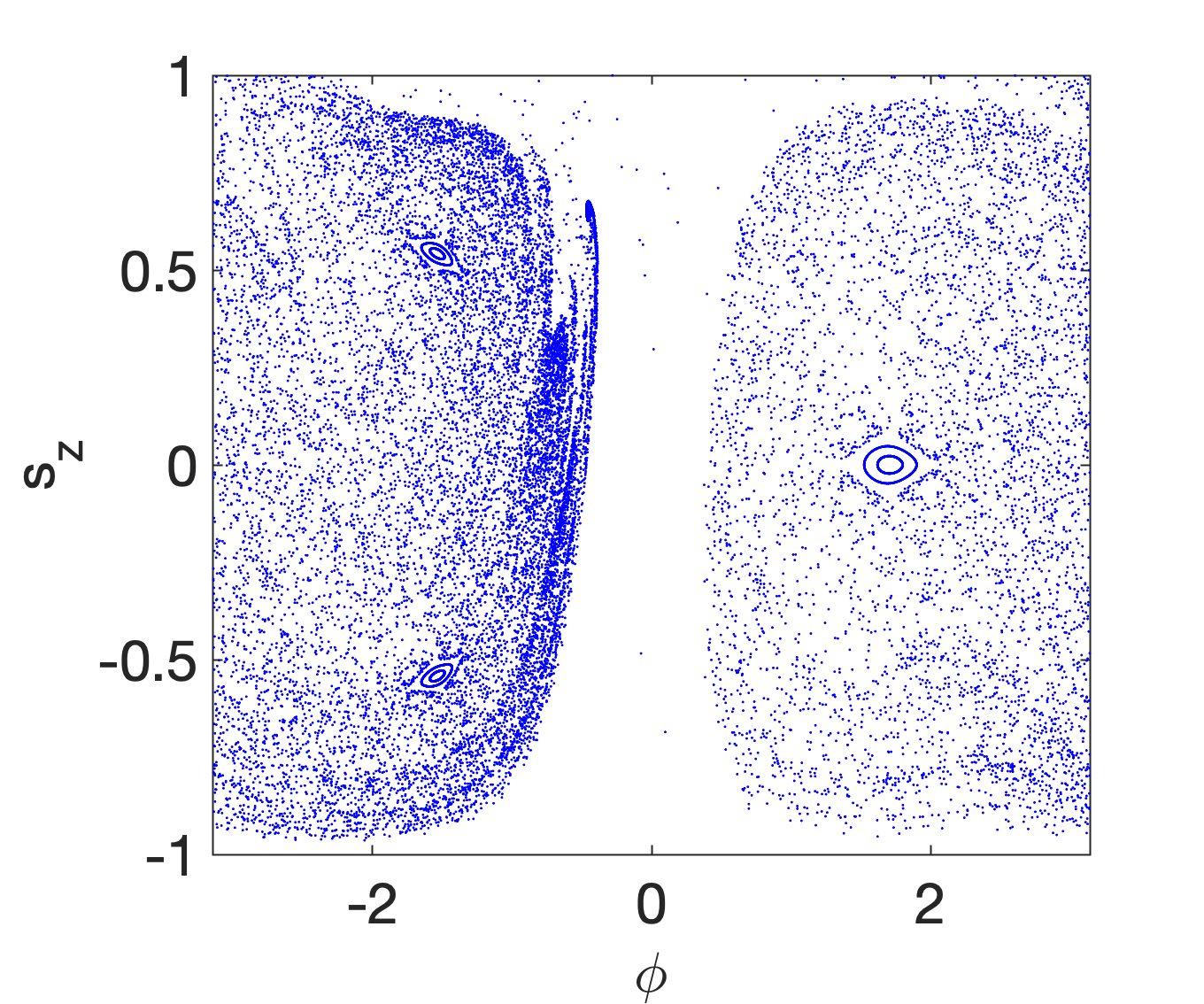}
\includegraphics[width=0.32\textwidth]{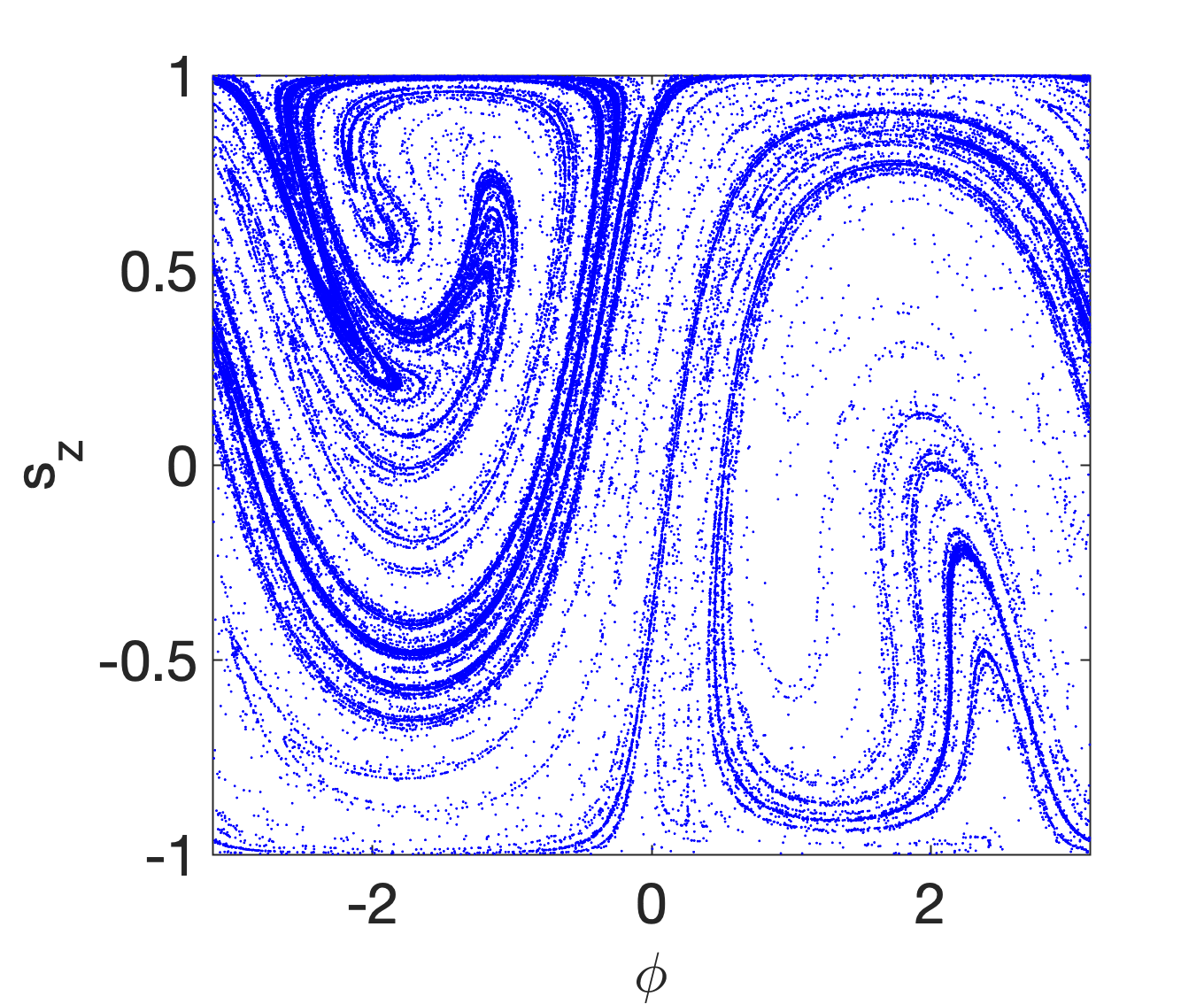}
\caption{Poincare plots of the classical map for $\epsilon=0$, $p=2$, $\gamma=0.2$ (top) and $\gamma=0.9$ (bottom) and different values of $k$ (From left to right top row: $k=1,\, 2,\, 5$, bottom row: $k=1,\, 1.5,\, 4$ ). }
\label{fig_dyn2b}
\end{figure}

For even larger values of $\gamma$ the remaining regular island shrinks further and moves further along the equator approaching the hyperbolic fixed point between the sink and the source that moves towards it along the equator from the other side. At the critical value $\gamma=p$ the stable island finally vanishes when it meets the hyperbolic fixed point, and the whole phase-space volume is attracted to the single remaining sink. While the details of the dynamics in dependence on $\gamma$ depend strongly on $k$, independently of the value of $k$ there are no conservative features in the dynamics for $\gamma>p$.  This is related to the fact that the dynamics does no longer allow for elliptic fixed points above this critical value. Of course, depending on $k$ the transition away from pseudo conservative phase-space portraits can happen for much smaller $\gamma$ already.

To provide a better idea of typical phase-space portraits we depict  a number of further examples in figure \ref{fig_dyn2b} for two different values of $\gamma$ and increasing values of $k$. Note that for intermediate values of $k$ as in the middle column, most of the pahse space volume is attracted to the sink on the northern hemisphere, much as in figure \ref{fig_dyn1xx}. For larger values of $k$ we clearly see a spiral shaped fractal structure emerging. Depending on the exact values of $\gamma$ and $k$ this structure can be transient while for long times the phase-space volume is attracted to a smaller sink, or it can be a fully developed strange attractor.

\begin{figure}[htb]
\centering
\includegraphics[width=0.32\textwidth]{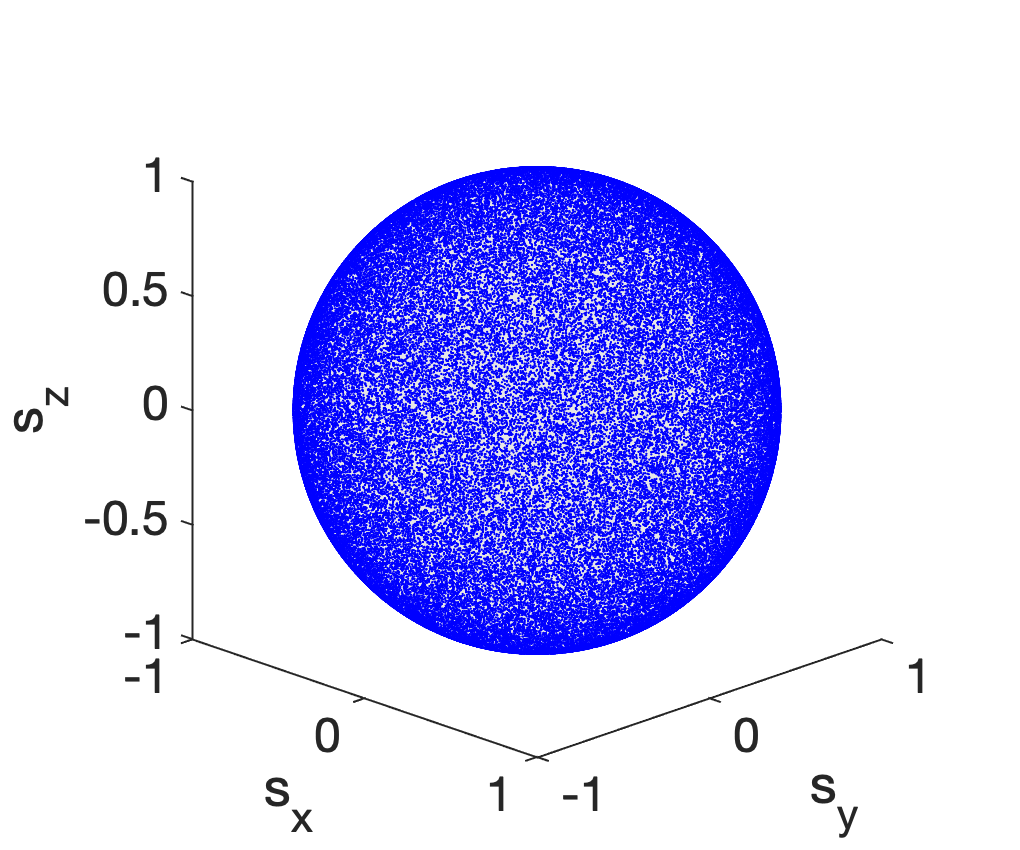}
\includegraphics[width=0.32\textwidth]{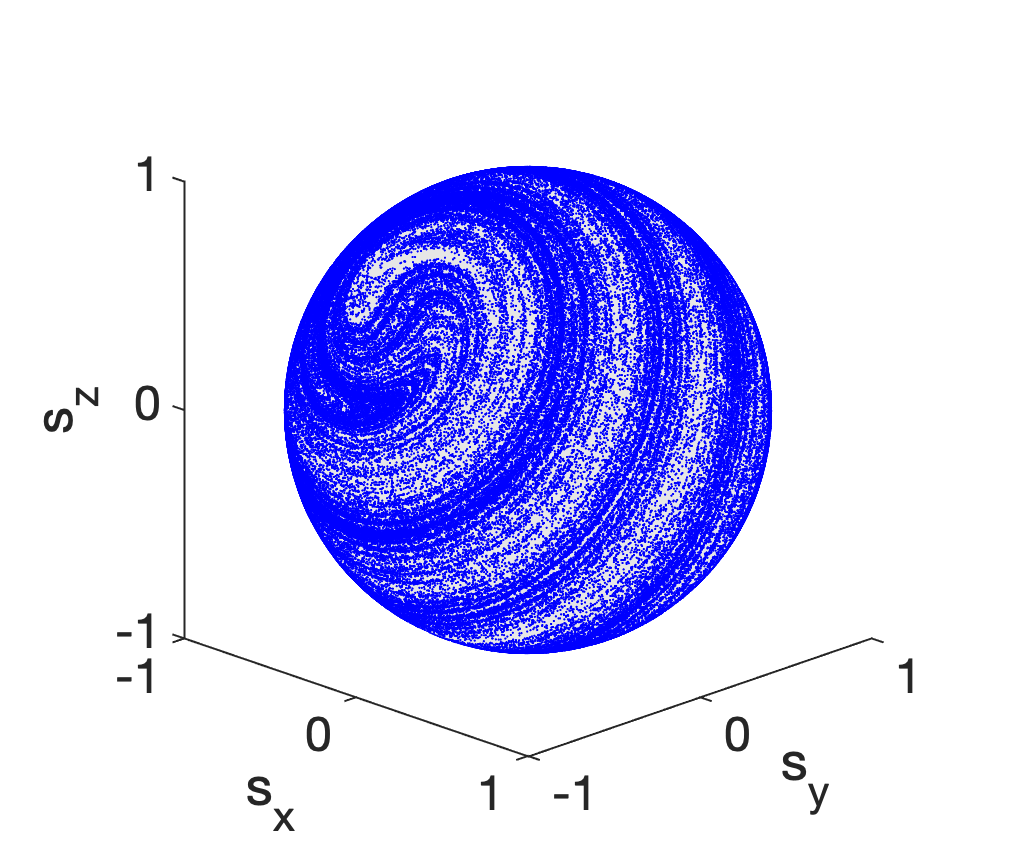}
\includegraphics[width=0.32\textwidth]{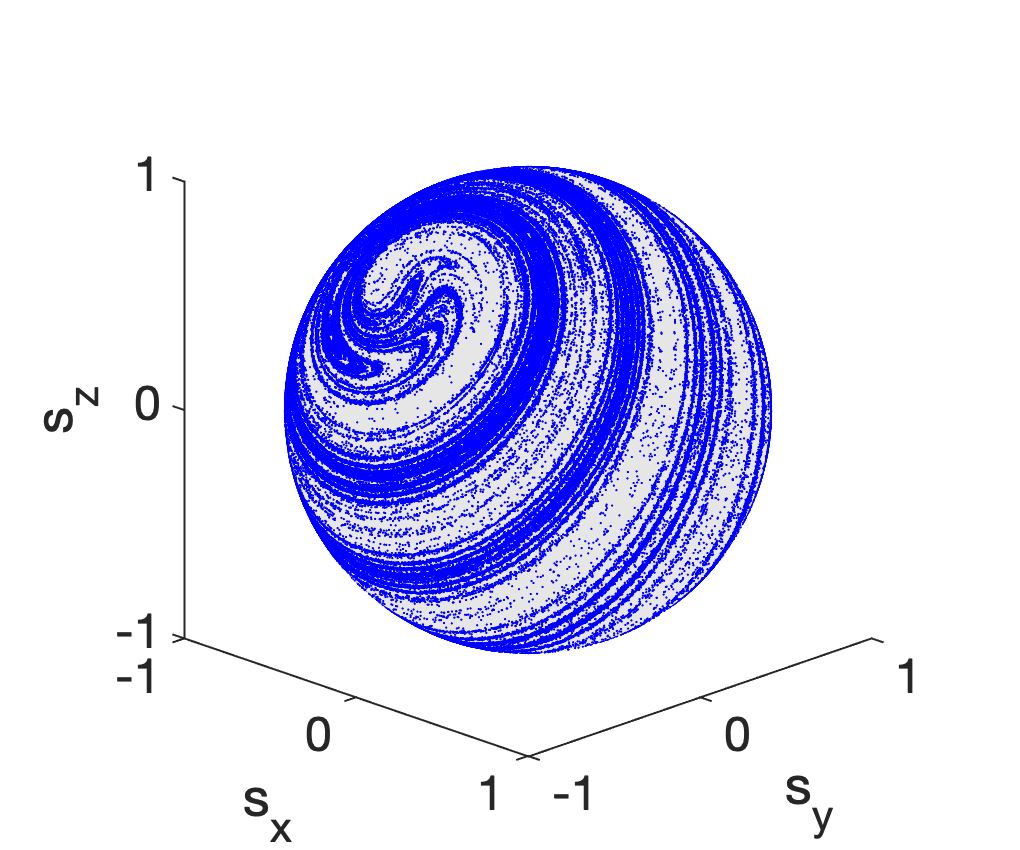}
\includegraphics[width=0.32\textwidth]{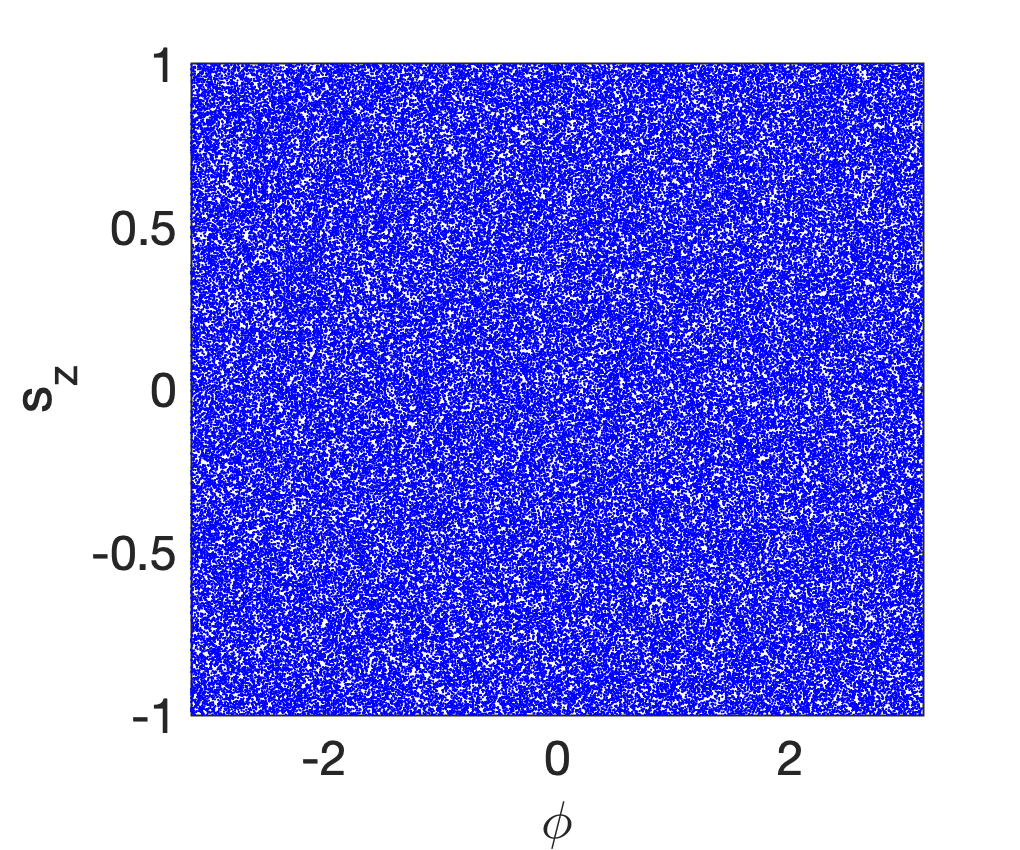}
\includegraphics[width=0.32\textwidth]{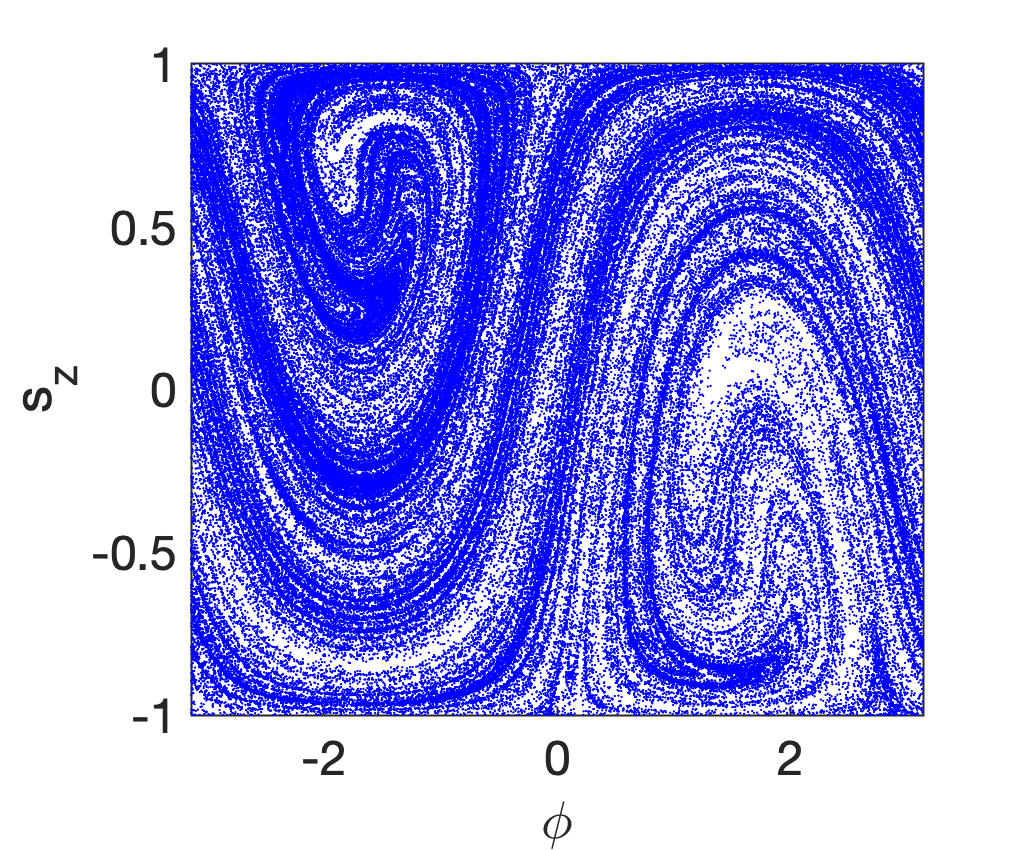}
\includegraphics[width=0.32\textwidth]{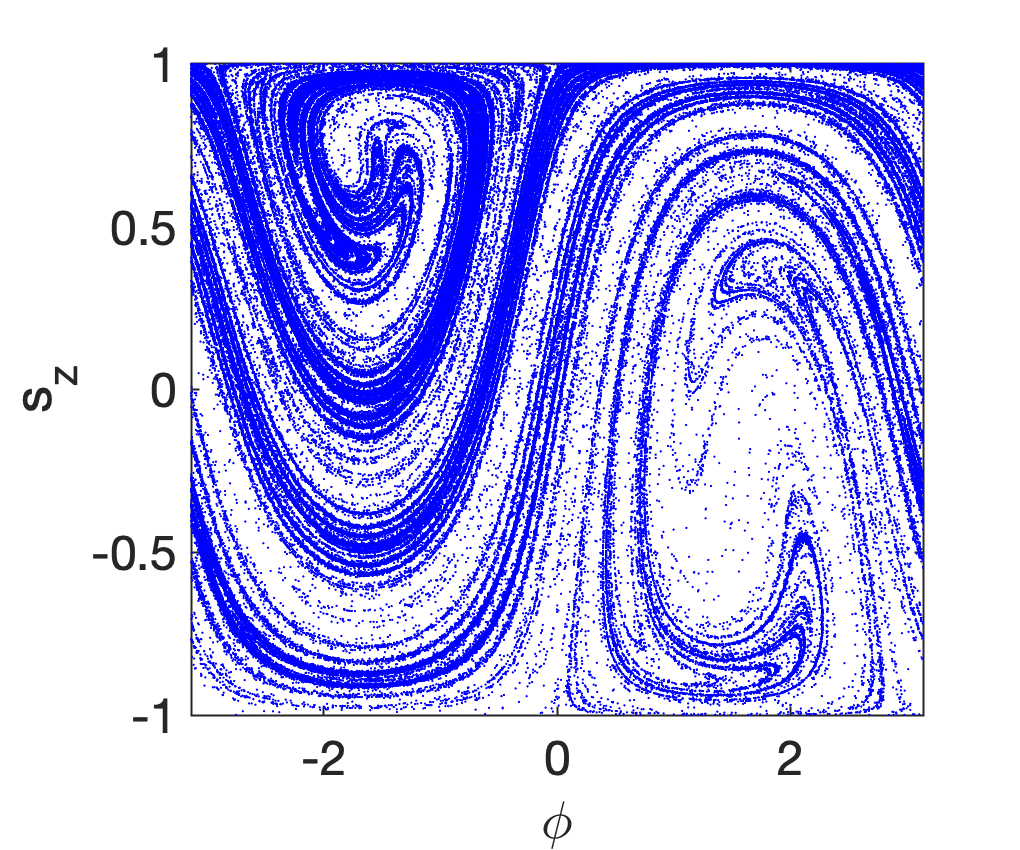}
\caption{Poincare plots of the classical map for $\epsilon=0$, $p=2$, $k=7$, and different values of $\gamma$ (From left to right:  $\gamma=0,\, 0.5,\, 1$). }
\label{fig_dyn3}
\end{figure} 

\begin{figure}[htb]
\centering
\includegraphics[width=0.3\textwidth]{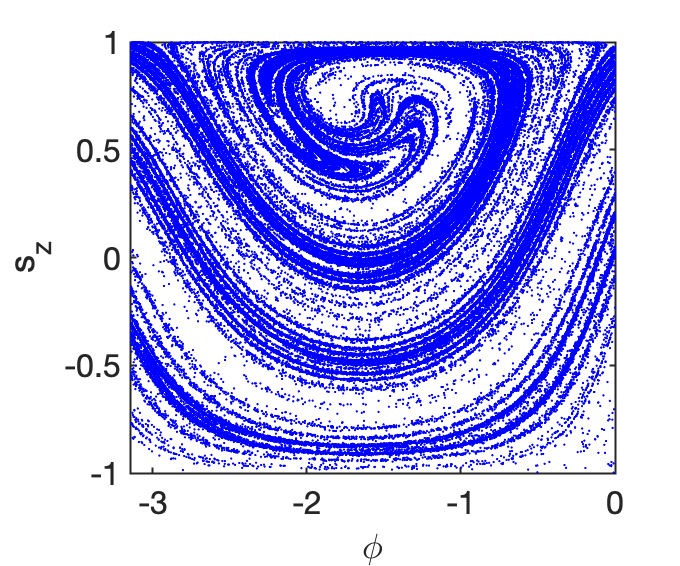}
\includegraphics[width=0.3\textwidth]{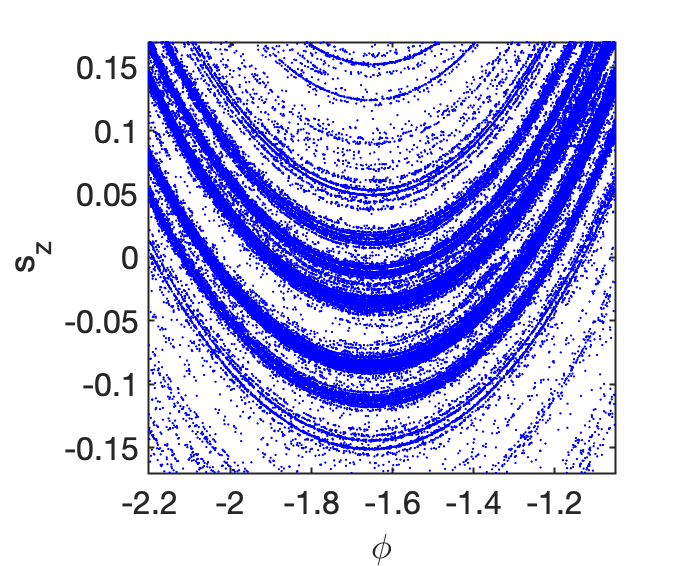}
\includegraphics[width=0.3\textwidth]{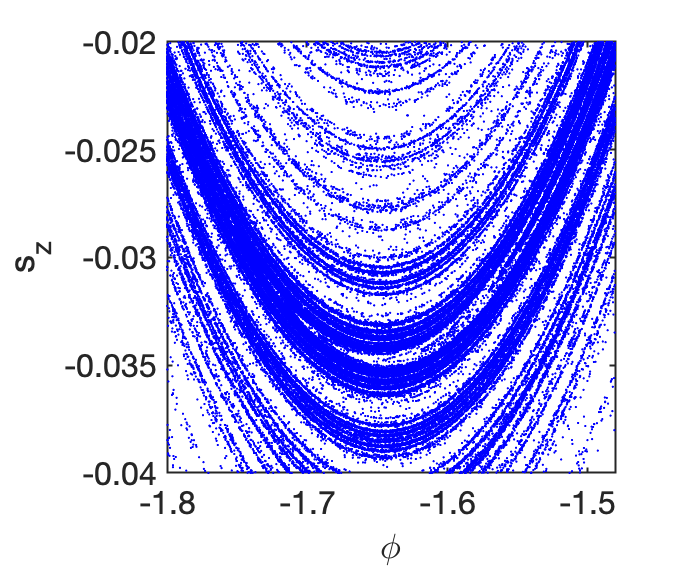}
\caption{Consecutive zooms into the Poincare phase-space plot for $\gamma=1$, $p=2$, and $k=7$.}
\label{fig_strangeat}
\end{figure}

In the Hermitian case $\gamma=0$ increasing values of the kicking strength $k$ lead to enlarged chaotic seas spreading over the phase sphere that eventually almost cover the sphere for sufficiently large values of $k$. An example of the Poincare section for $p=2$ and $k=7$ is depicted on the left in figure \ref{fig_dyn3}. Introducing gain and loss, $\gamma\neq 0$, in this case leads to the appearance of spiral-shaped attractors in the phase space, as can be seen in the middle and right panels of the same figure. These spirals appear to be strange attractors. That is, over time the dynamics continues to move around the spiral in a random manner, and zooms into the Poincare plots, as depicted for example in figure \ref{fig_strangeat}, indeed reveal a fractal structure of the attractor. 

\begin{figure}[htb]
\centering
\includegraphics[width=0.32\textwidth]{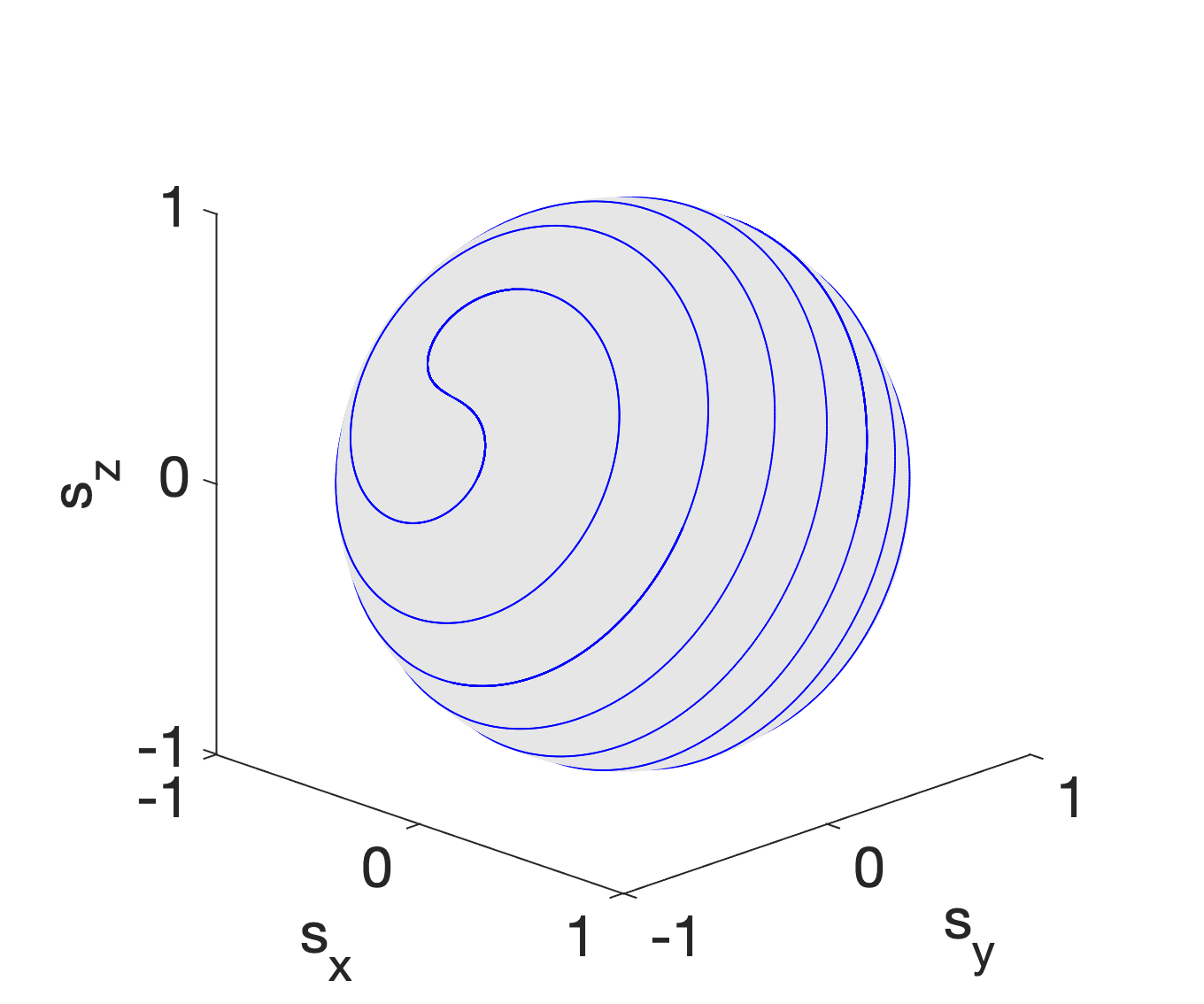}
\includegraphics[width=0.32\textwidth]{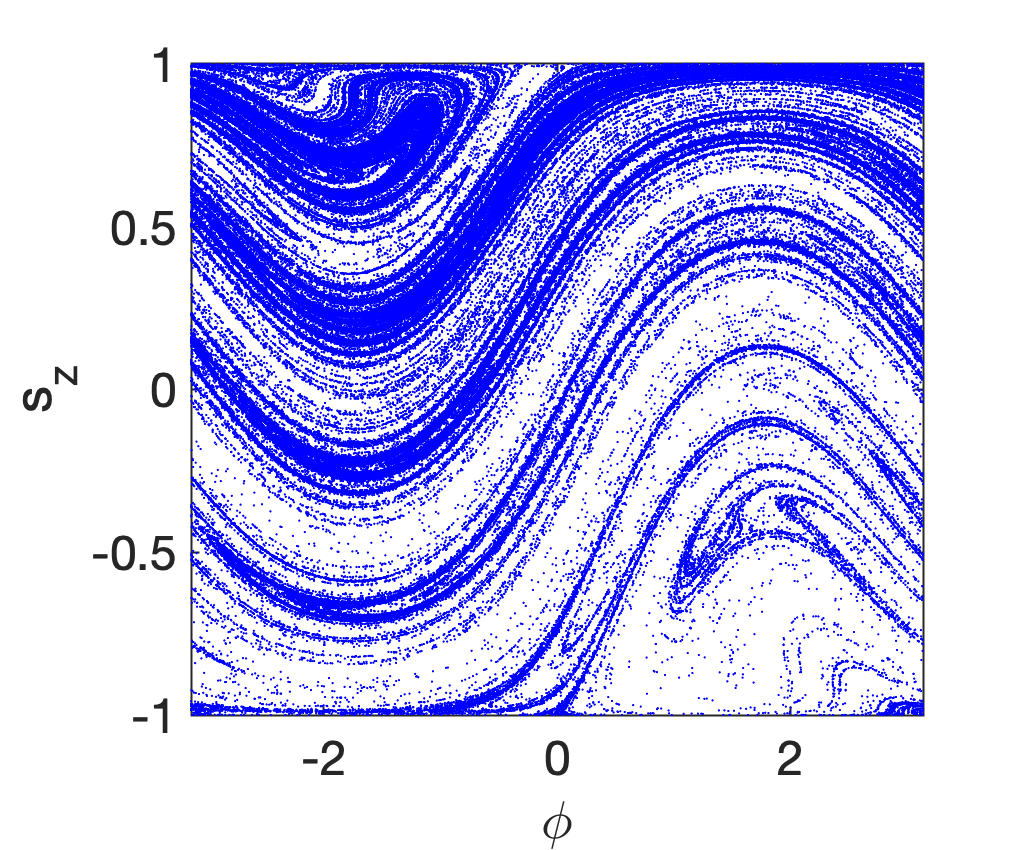}
\includegraphics[width=0.32\textwidth]{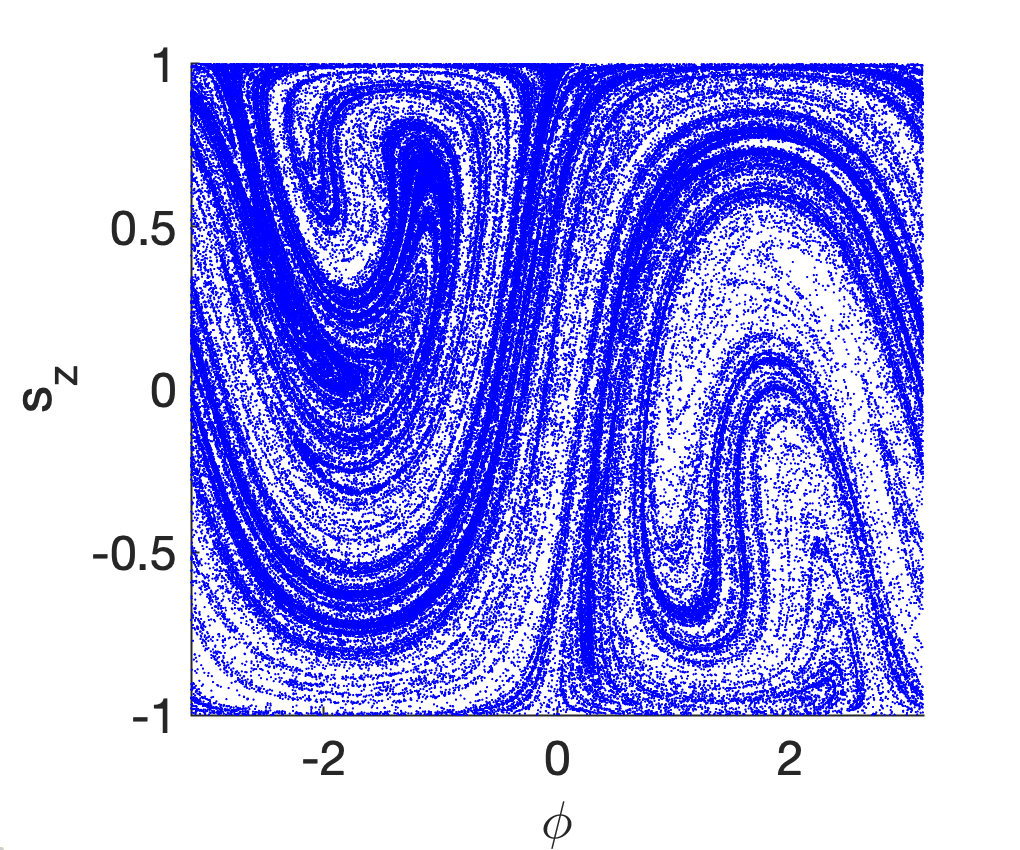}
\includegraphics[width=0.32\textwidth]{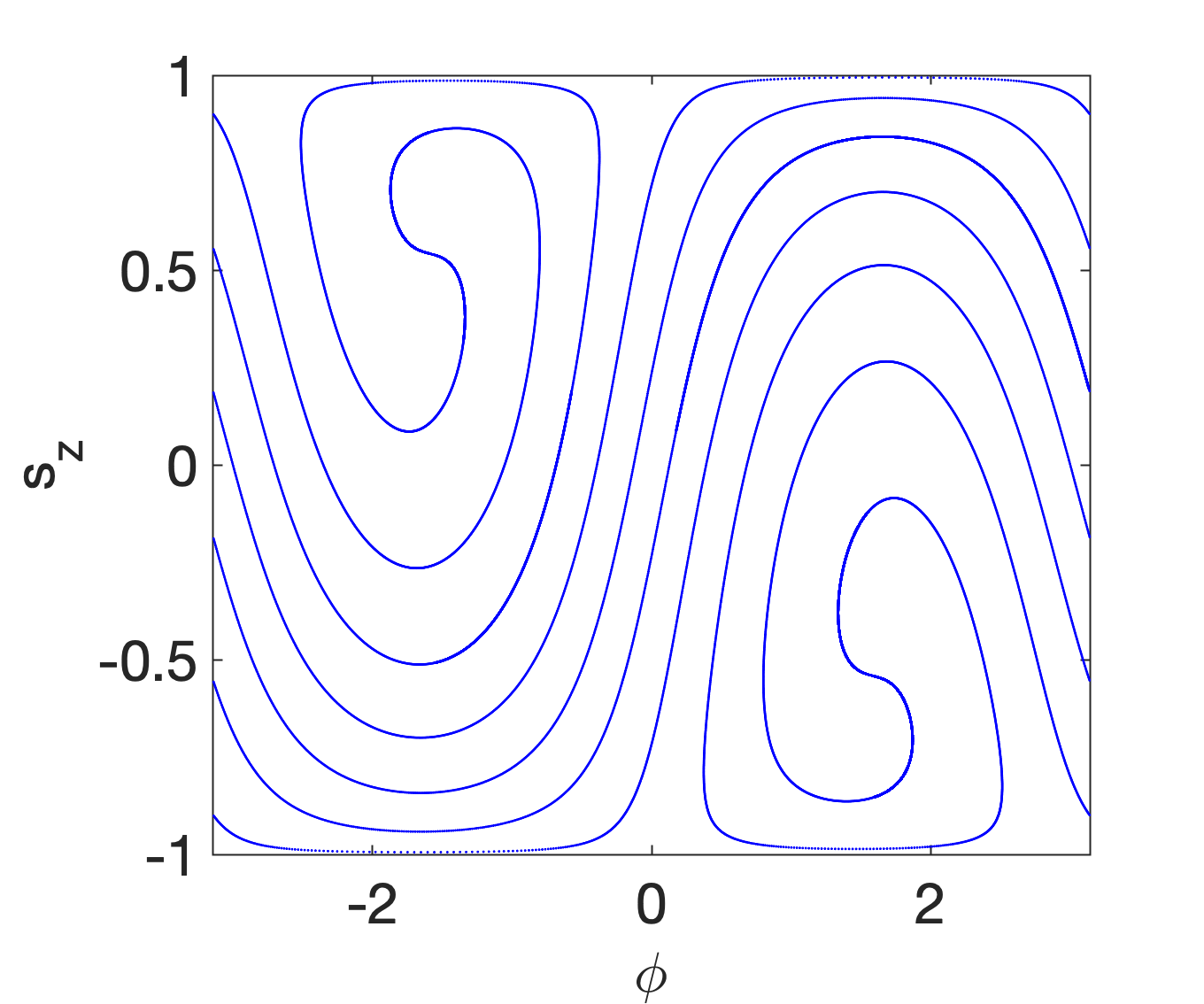}
\includegraphics[width=0.3\textwidth]{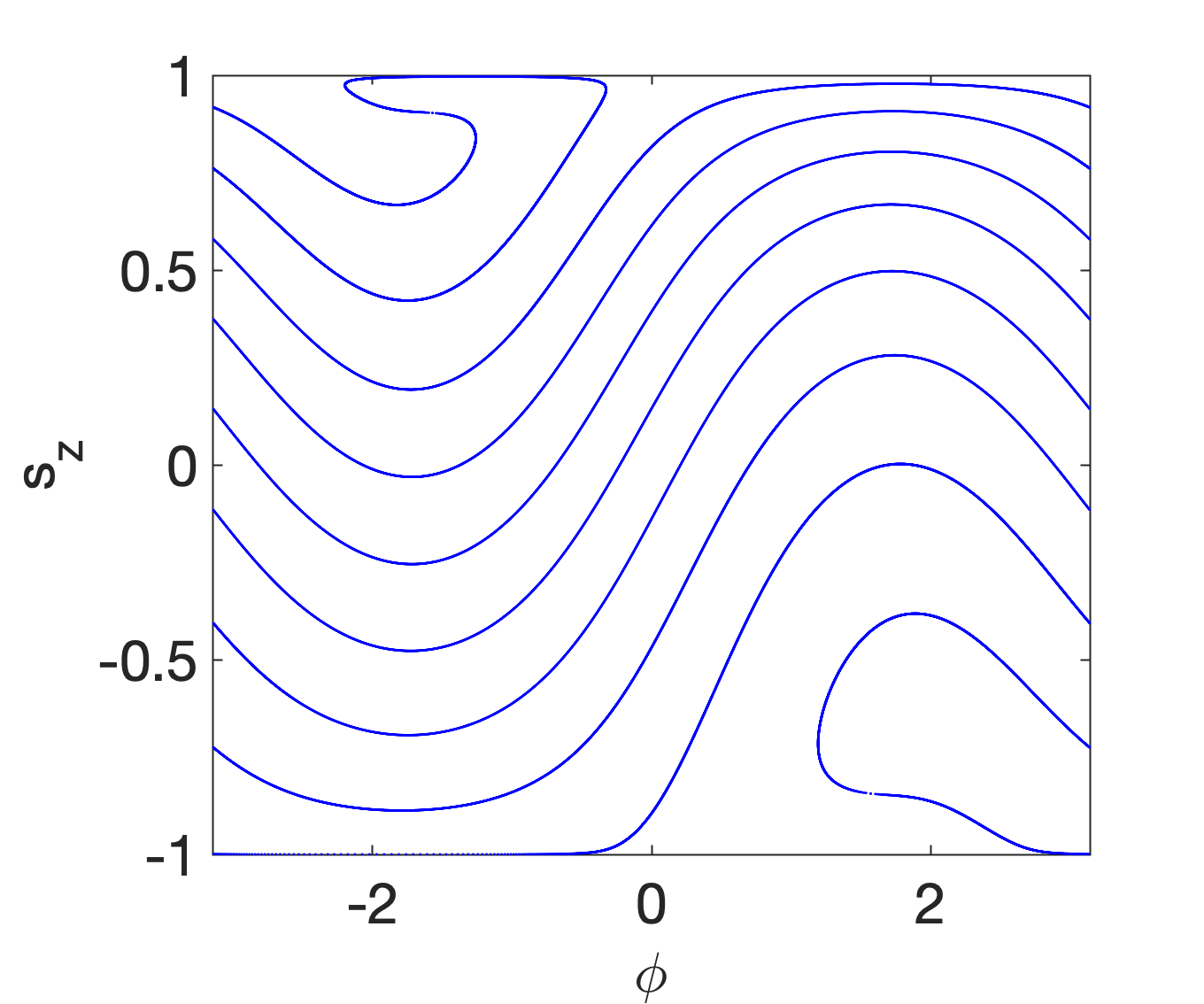}
\includegraphics[width=0.3\textwidth]{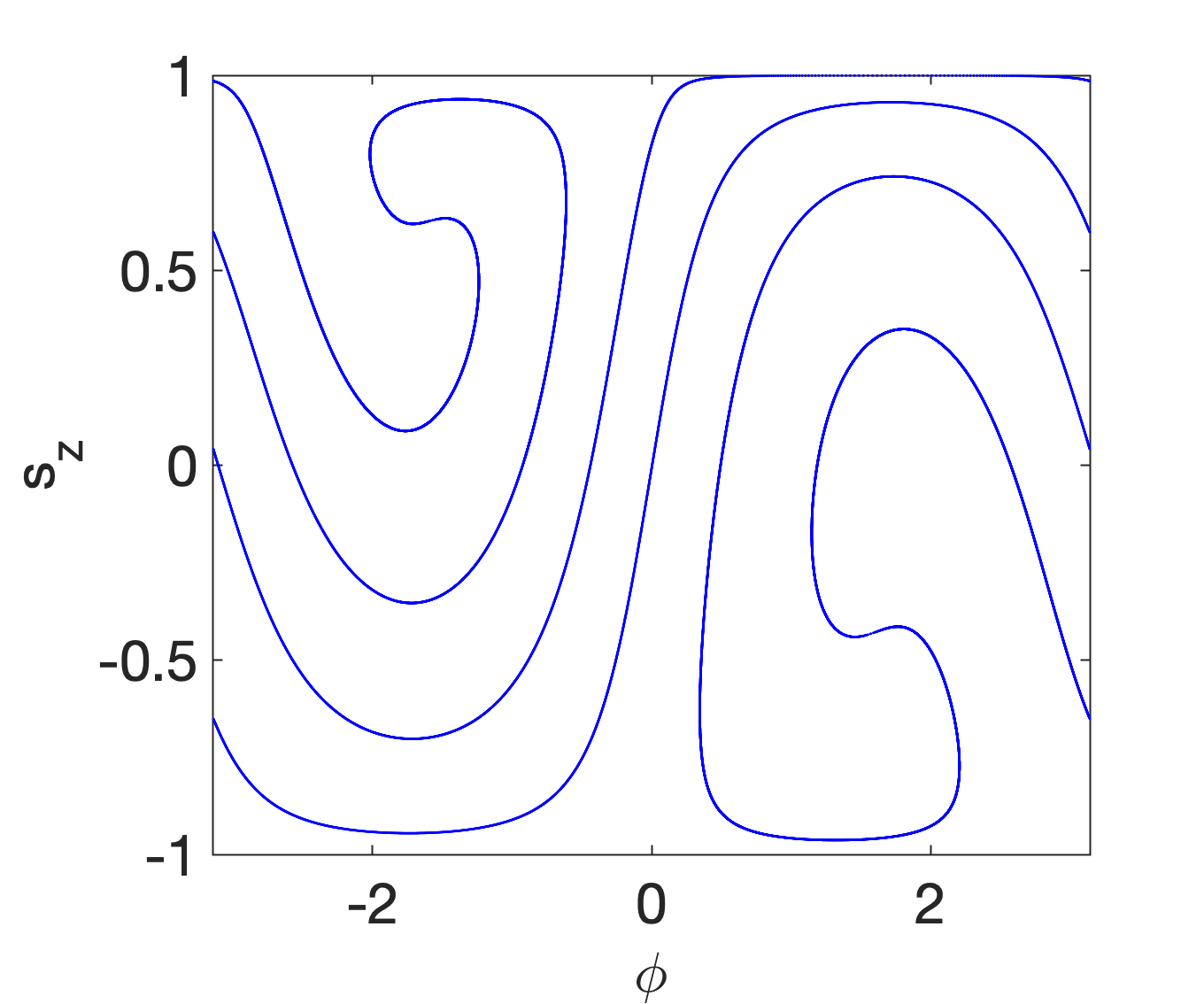}
\caption{Structure of the strange attractor. The left panel shows the image of the great circle under the torsion around the z-axis with $k=7$ followed by the rotation around the x-axis by an angle $p/2$, for $p=2$ (corresponding to $\gamma=0$). The top shows the image on the sphere, the bottom on the flat projection. The left panel should be compared to the strange attractors in figure \ref{fig_dyn3}. The middle and right panels show the comparison between the Poincare section of the classical map (top) with $p=1$, $k=7$, and $\gamma=0.5$ (middle) and $p=2$, $k=4$, and $\gamma=0.5$ (right), and the attractor skeleton arising from the image of the great circle for the same parameters (bottom).}
\label{fig_spiral_reg}
\end{figure} 

The shape of these strange attractors can be easily understood as being structured around the skeleton of the orbits arising from the single period map, that is, the free motion distorted by the torsion of the kicking dynamics and then followed by another free motion. The left panels in figure \ref{fig_spiral_reg} depict the image of the great circle under this transformation on the sphere as well as on the projection for the same values of $p$ and $k$ as in figure \ref{fig_dyn3}. We have chosen $\gamma=0$ for this purpose, since the deformation of the spiral for small values of $\gamma$ is negligible. This demonstrates clearly, that the seed of the strange attractor actually lies within the Hermitian system, but is made visible only by the non-Hermiticity. Using the image of the great circle under this transformation to describe the strange attractor, we can easily deduce that the winding number of the spiral is given by $\frac{4 k}{\pi}$, since the north pole is rotated by $2k\pi$ in the positive direction, and the south pole is rotated by the same angle in the negative direction by the kicking torsion. We can further deduce that the ``centre" of the spiral, is located at $\phi=-\pi$ (i.e. $s_x=0$), and $s_z=\cos(\frac{p}{2})$. In the middle and right panels of figure \ref{fig_spiral_reg} this dependence of the winding number and orientation of the attractor is demonstrated using two different sets of parameters. The top row shows the Poincare sections of the map, while the bottom row shows the image of the great circle under the transformation of the kicking torsion followed by the free motion.  

Note that in contrast to the behaviour for smaller $k$, where an arbitrarily small value of the $PT$-symmetry breaking parameter $\epsilon$ leads to appearance of a single sink in the phase-space dynamics, these strange attractors are stable and also appear in the non-Hermitian system without $PT$-symmetry. We have verified for a large range of parameters that the motion indeed continues to erratically move around the attractor rather than going into a sink for large time scales with or without $PT$-symmetry.  

\begin{figure}[htb]
\centering
\includegraphics[width=0.32\textwidth]{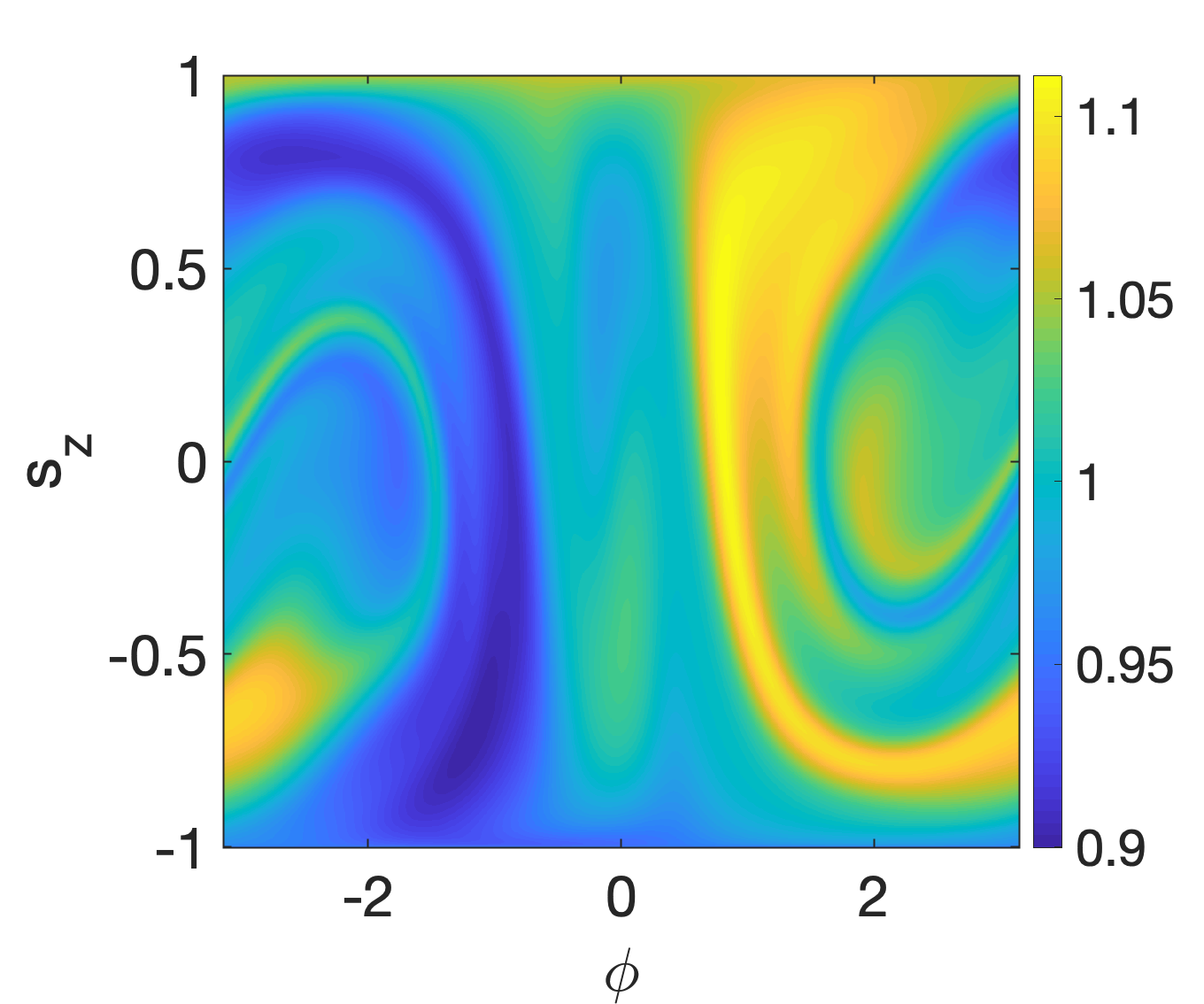}
\includegraphics[width=0.32\textwidth]{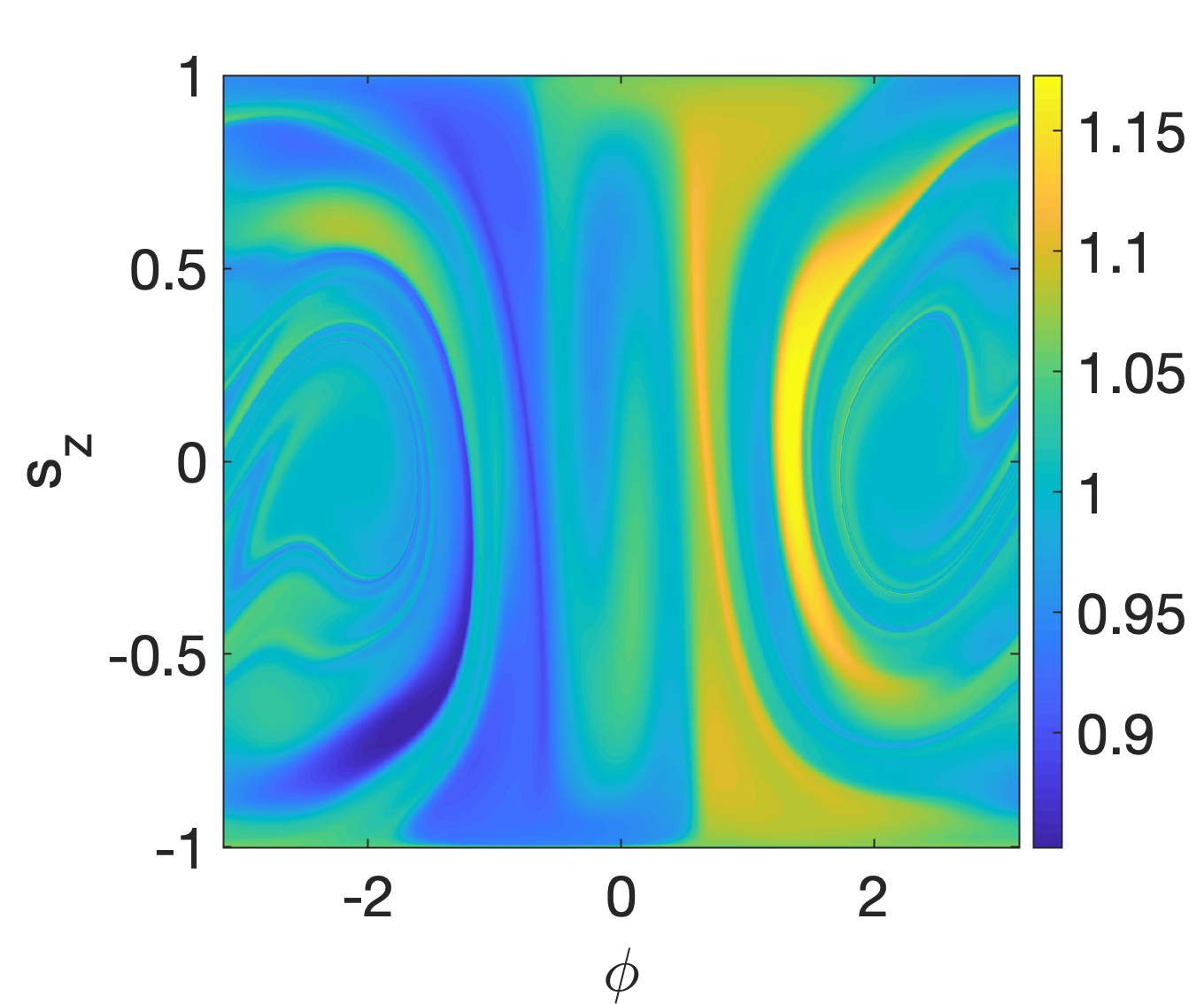}
\includegraphics[width=0.32\textwidth]{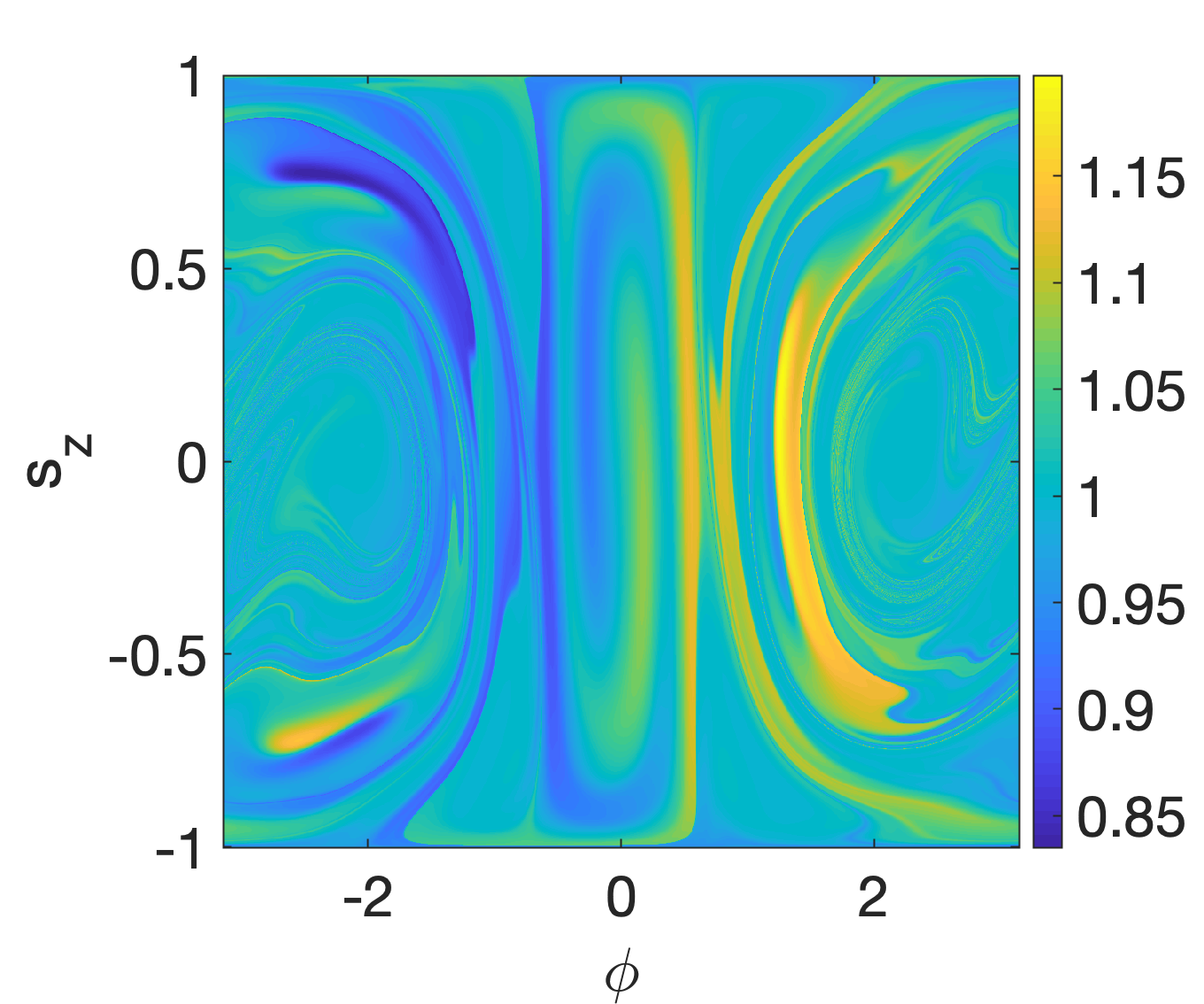}
\includegraphics[width=0.32\textwidth]{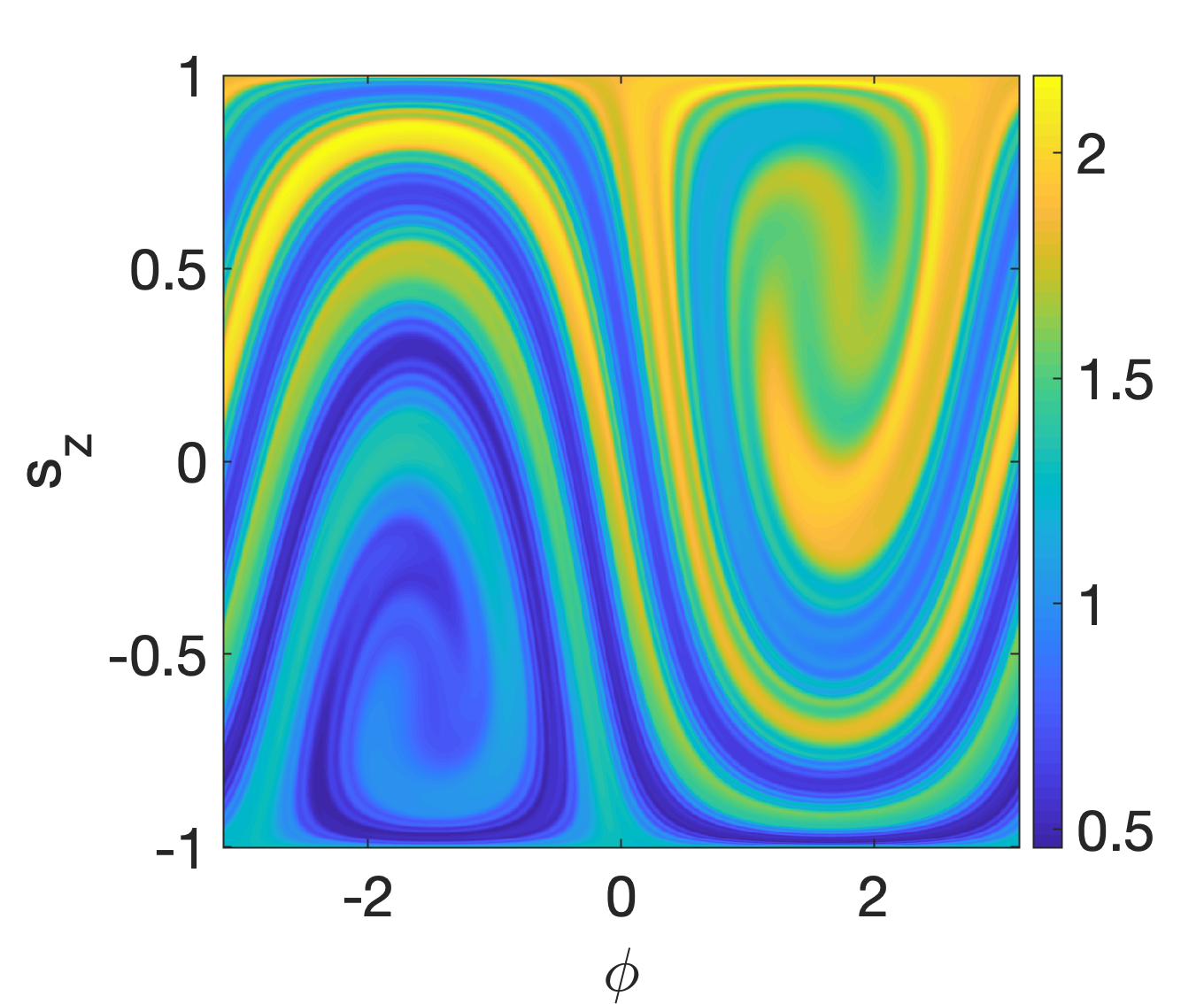}
\includegraphics[width=0.32\textwidth]{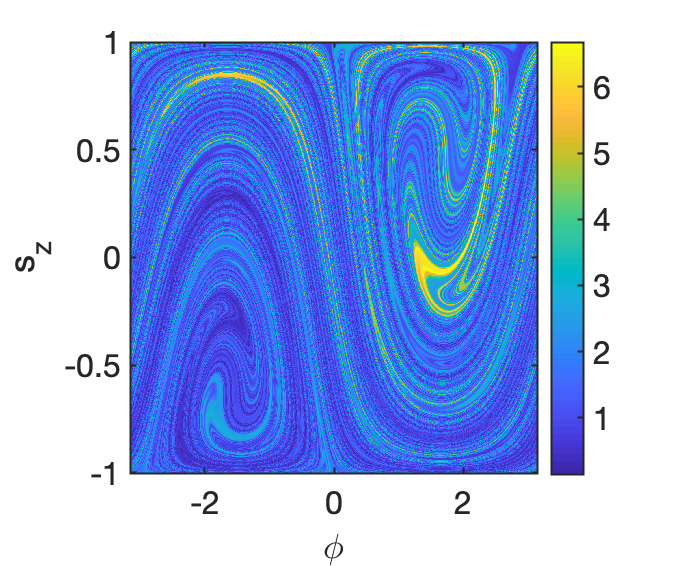}
\includegraphics[width=0.32\textwidth]{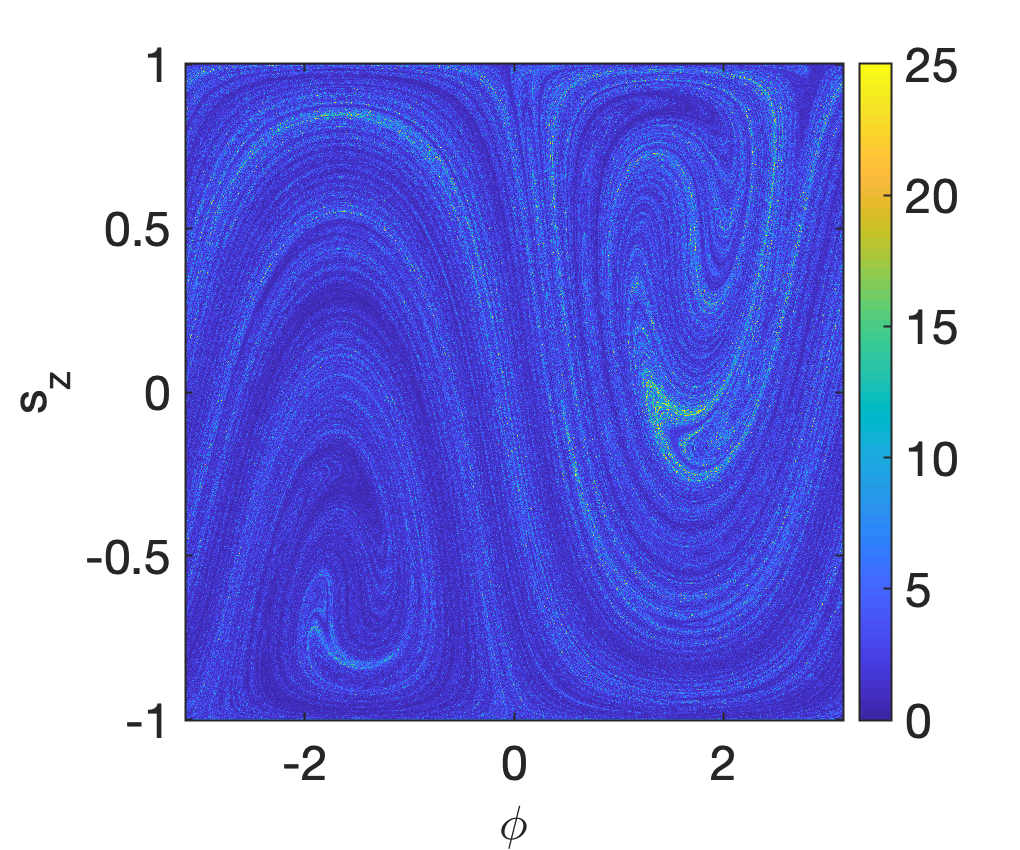}
\caption{Classical intensity behaviour as a function of the initial phase-space position - false-colour plots of the final intensity after $j$ iterations of the classical map. The top row shows the result after $j=5,\, 10$, and $20$ iterations (left to right), respectively, for the parameter values $p=2$, $k=1$, and $\gamma=0.1$. The bottom panel shows the same for $j=2,\, 5$, and $10$ iterations for $p=2$, $k=7$, and $\gamma=0.5$.}
\label{fig_norm_dyn}
\end{figure}

Finally, the classical limit of non-Hermitian systems has an additional variable on top of the phase-space coordinates, that describes the behaviour of the norm of the quantum state in the semiclassical limit,
\begin{equation}
n_{j+1}:=\langle \psi^{(j+1)}|\psi^{(j+1)}\rangle=\langle \psi^{(j)}|\hat F^\dagger\hat F|\psi^{(j)}\rangle. 
\end{equation}
We have already encountered the map for this variable in the $2\times 2$ representation as 
\begin{equation}
n_{j+1}=\Gamma(s_x,s_y,s_z) n_{j},
\end{equation}
with $\Gamma$ defined in equation (\ref{eqn_normfactor}). 
This additional parameter can be viewed as an \textit{intensity} of the classical phase-space trajectory, which growth and decays locally exponentially under the free dynamics proportionally to the momentary $z-$component, but is not affected by the (unitary) kick. In general the intensity is a complicated function of the iteration time for most initial states. To provide an overview of its behaviour over time, we show false-colour plots of the final intensity as a function of the initial phase-space variable for different parameter values and different numbers of iterations in figure \ref{fig_norm_dyn}. We observe that as the number of iterations is increased finer and finer structures emerge. There is an obvious correspondence to the phase-space portraits. In particular in the bottom panel, belonging to the case of the strange attractor, the fractal structure is very clearly visible in the final intensity after increasing numbers of iterations. Note however, that the structure visible in the norm dynamics does not correspond to the attractor, but the repeller of the motion, that would be obtained from a backwards evolution in time.

We further find that for large values of $k$ for which the unitary map is chaotic, even for arbitrarily small values of $\gamma$ the fractal repeller is clearly visible in the iterated intensity, while it is not visible in the dynamics, with the bare eye. This can be seen for example in the middle panel of figure \ref{fig_husimi_norm_dyn} where the classical intensity after three iterations as a function of the initial state is depicted for $p=2$, $k=10$, and $\gamma=0.01$.

After having uncovered some of the rich behaviour of the classical dynamics of this $PT$-symmetric chaotic system, let us now turn our attention to the quantum system.

\section{The quantum system}
\label{sec-quantum}

\begin{figure}[htb]
\centering
\includegraphics[width=0.49\textwidth]{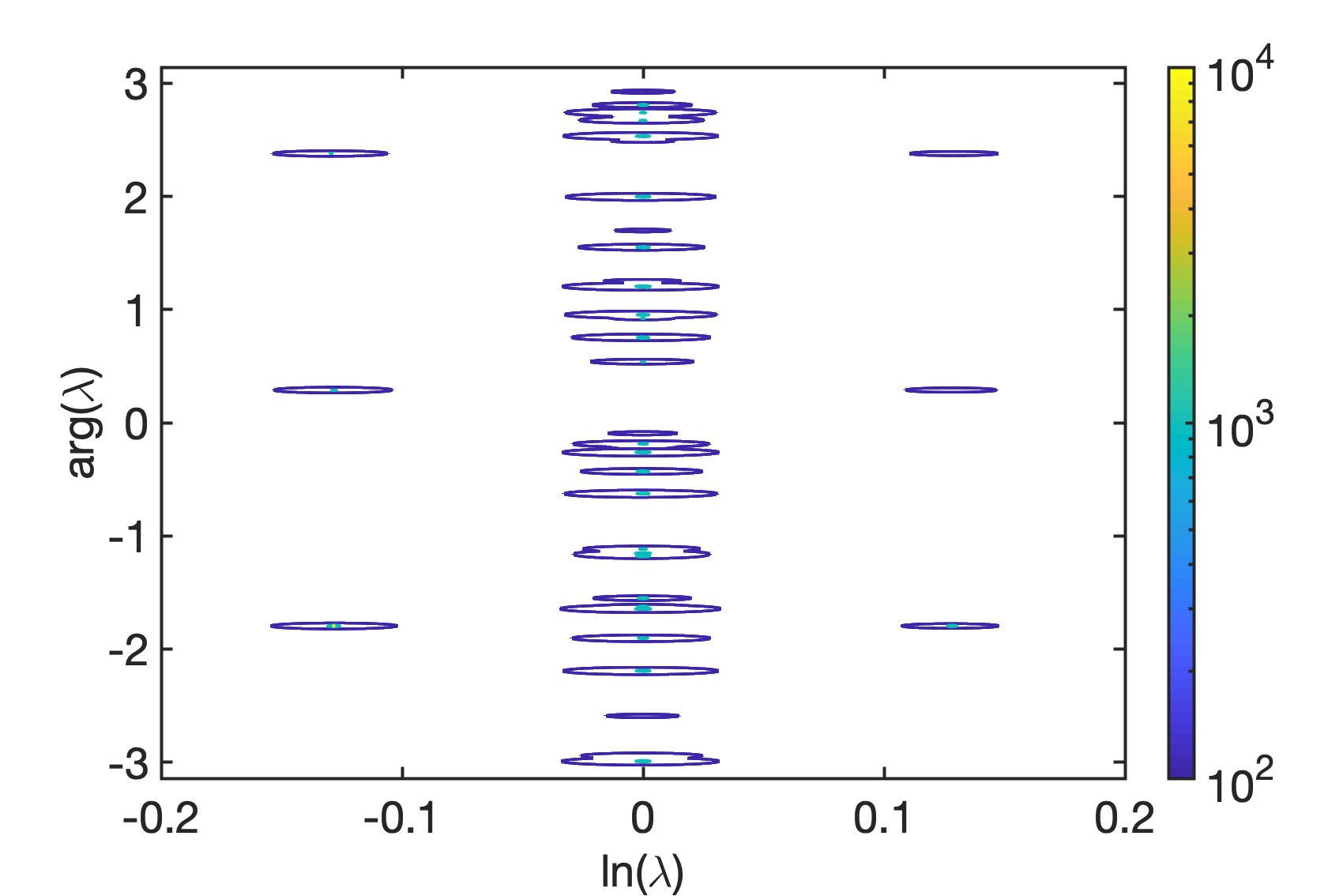}
\includegraphics[width=0.49\textwidth]{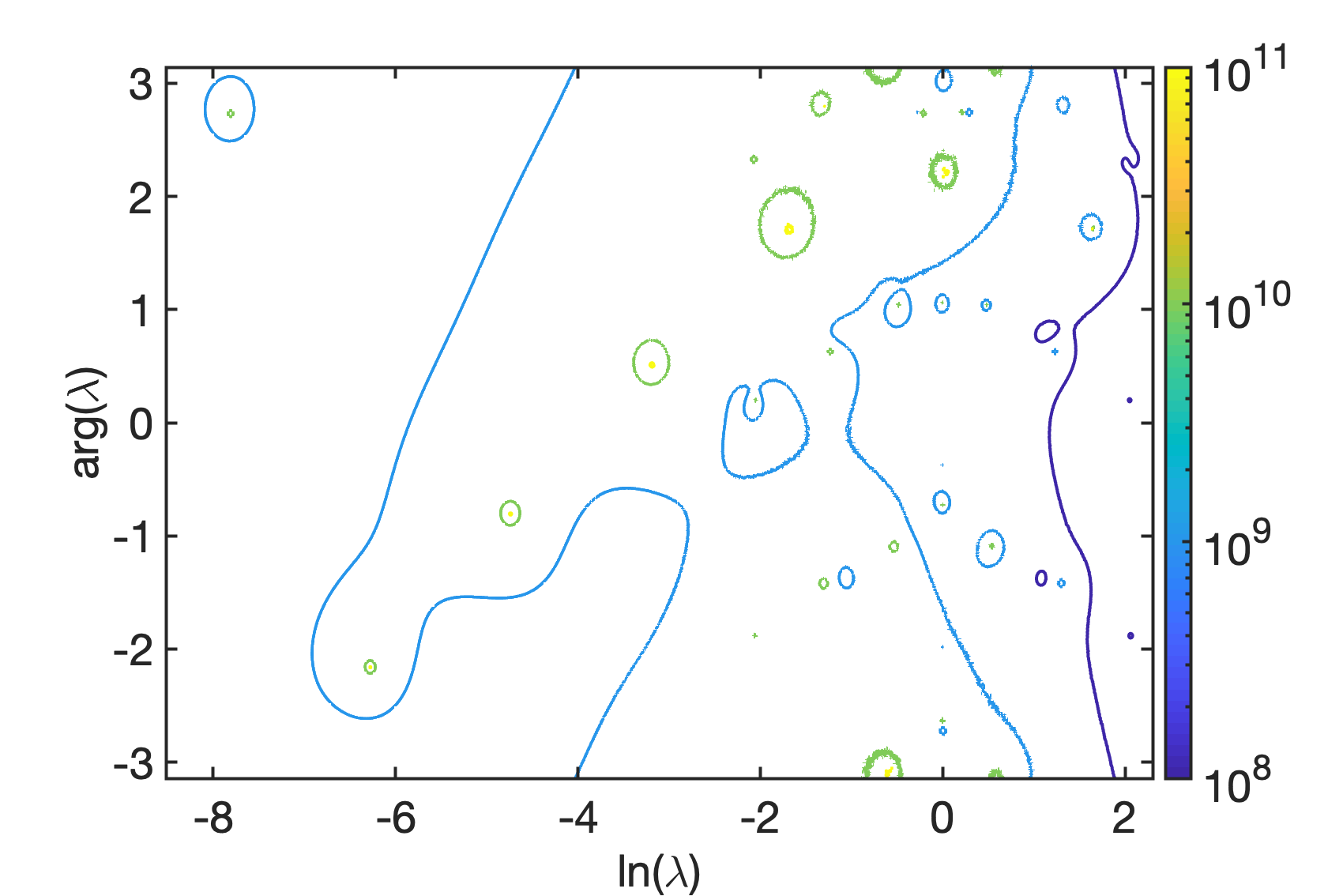}
\caption{Selected contours of the resolvent norm of the Floquet operator (\ref{eqn_Floq}) for $L=20$, $p=2$, $k=1$ and two different values of $\gamma$ in the plane of complex eigenvalues $\lambda$. The left plot shows the level curves at which the resolvent norm equals $10^2$ and $10^3$, respectively, for $\gamma=0.1$. The right plot shows  parts of the level curves of the values $10^ 8$, $10^9$, $10^{10}$, and $10^{11}$ for $\gamma=1$.}
\label{fig_pseudo}
\end{figure}

Let us now consider how the classical features observed in the previous section manifest themselves in the full quantum description. In particular, we will investigate the spectral behaviour of the Floquet operator and the phase-space structures associated to its eigenspaces. The quantum system is very sensitive to non-Hermitian perturbations, and an accurate numerical diagonalisation is problematic for medium sized values of $L$ if the gain-loss parameter is too large. This sensitivity to perturbations is related to the presence of high-order exceptional points in the non-kicked system \cite{Grae08}, and is indeed typical for non-Hermitian and non-unitary (more precisely \textit{non-normal}) operators. To illustrate the issue we show two examples of selected contours of the resolvent norm (where we take the two-norm, computed using the singular value decomposition) of the quantum Floquet operator for a relatively small matrix size of $41$ ($L=20$) in figure \ref{fig_pseudo}. 

The eigenvalues of an operator are defined as the singularities of the resolvent norm, that is, points in the complex plane at which the resolvent norm diverges, i.e., values $\lambda$ at which $||\hat A-\lambda\mathds{I}||=0$. The theory of pseudospectra \cite{Tref05} deals with the concept of values $\lambda_\epsilon$ which are \textit{almost} eigenvalues in the sense that $||\hat A-\lambda_\epsilon\mathds{I}||= \epsilon$ for equivalently $||(\hat A-\lambda_\epsilon\mathds{I})^{-1}||= \epsilon$ or arbitrarily small $\epsilon$, that is, points at which the resolvent norm is large but finite. For normal operators the resolvent norm is very sharply peaked around eigenvalues and the points for which the resolvent norm is larger than $\frac{1}{\epsilon}$ only differ from the actual eigenvalues on the order of $\epsilon$, which confirms the intuition that a pseudo-eigenvalue $\lambda_\epsilon$ is indeed \textit{close} to an eigenvalue \cite{Tref05}. Yet, this intuition turns out to be misleading for operators that are not normal, such as our Floquet operator. In fact, our system, due to the presence of many exceptional points at which eigenvectors coalesce, is a particular notorious example of this phenomenon. This can be seen in the right plot in figure \ref{fig_pseudo} where the level curves at which the resolvent norm equals $10^8$ to $10^{11}$ are plotted for a relatively small matrix size of $41\times 41$, and moderate values of $p=2$, $k=1$, and $\gamma=1$. The level curves extend over a large range of quasi energy values and we can see that solutions to the equation $||\hat F-\lambda_\epsilon\mathds{I}||=10^{-9}$ can differ from each other, and thus the \textit{true} eigenvalue on the order of $10$. This phenomenon is hugely amplified for larger matrix sizes and quickly leads to perturbations on the order of machine accuracy leading to pseudo eigenvalues that are largely uncorrelated with the \textit{true} eigenvalues. In fact, the right conclusion from this behaviour may be to follow Trefethen's advice  and the not consider eigenvalues at all for our system, as they may say very little about the systems behaviour \cite{Tref05}. However, for the present purpose we stick with the analysis of the spectral behaviour of our quantum system to identify typical features of quantum-classical correspondence, but we are forced to confine ourselves to small values of $\gamma$ for which the spectral problem is better behaved, as illustrated for the same values of $L$, $p$, and $k$, as before, but $\gamma=0.1$ in the left panel of figure \ref{fig_pseudo}.

\begin{figure}[htb]
\centering
\includegraphics[width=0.32\textwidth]{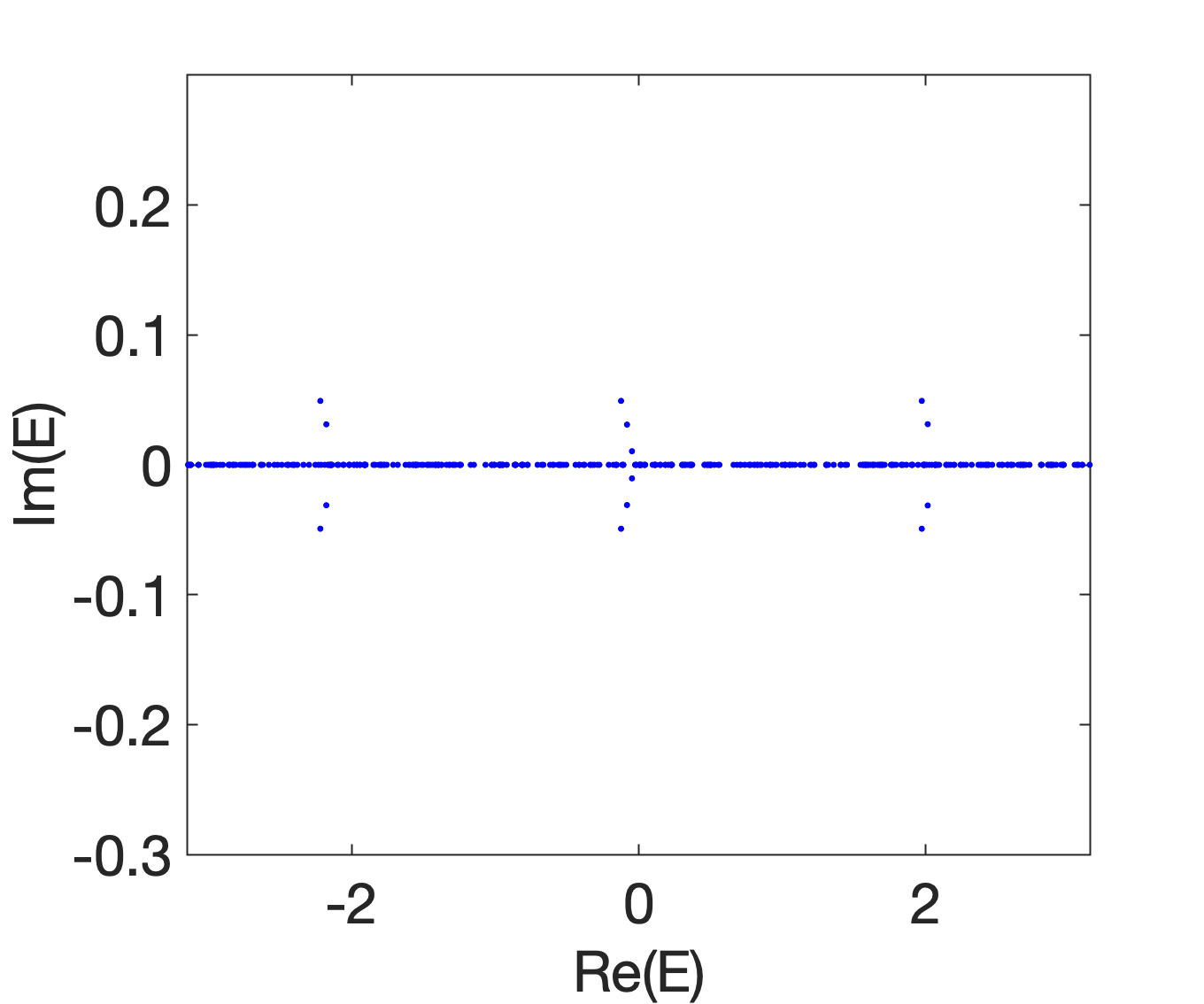}
\includegraphics[width=0.32\textwidth]{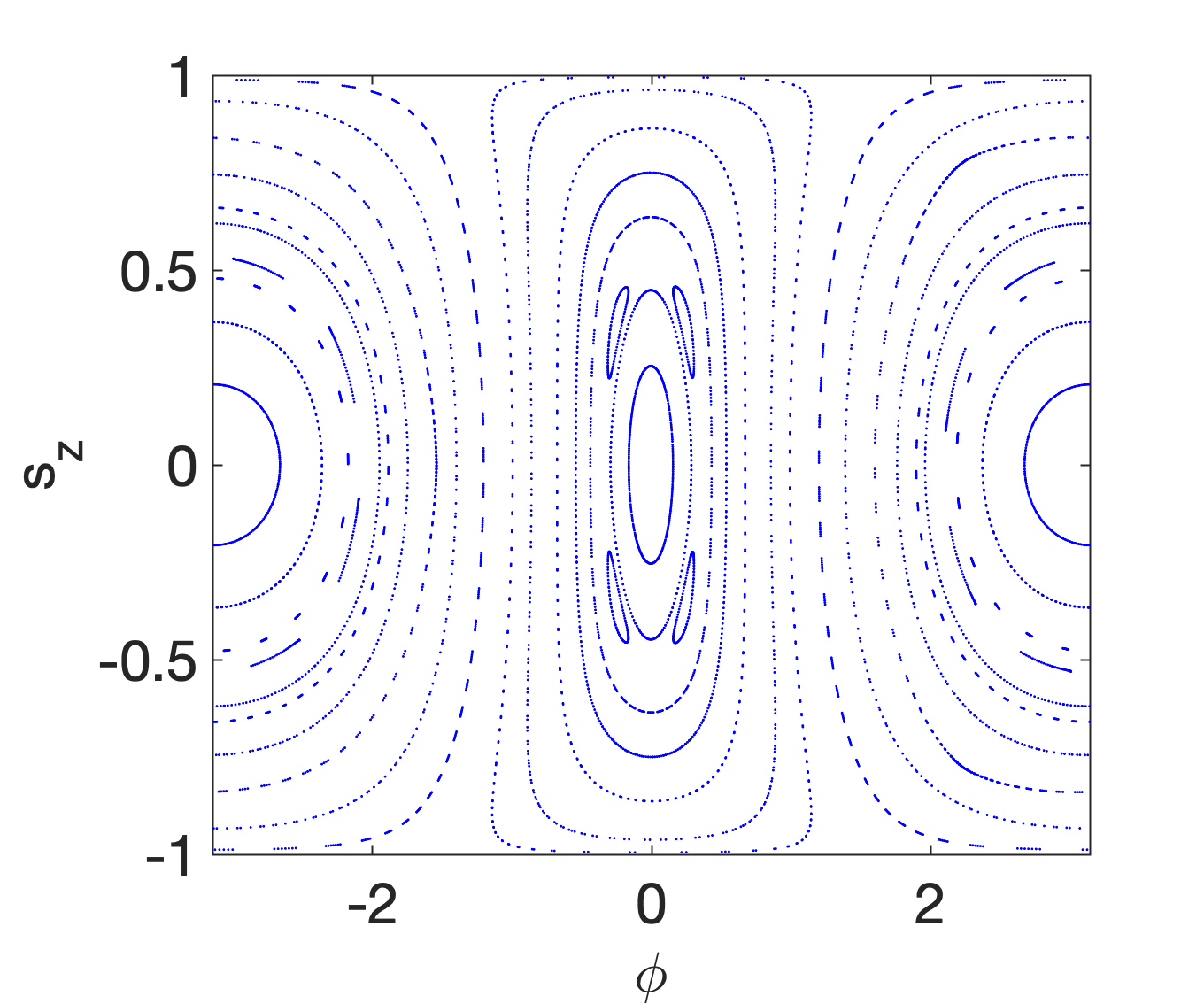}
\includegraphics[width=0.32\textwidth]{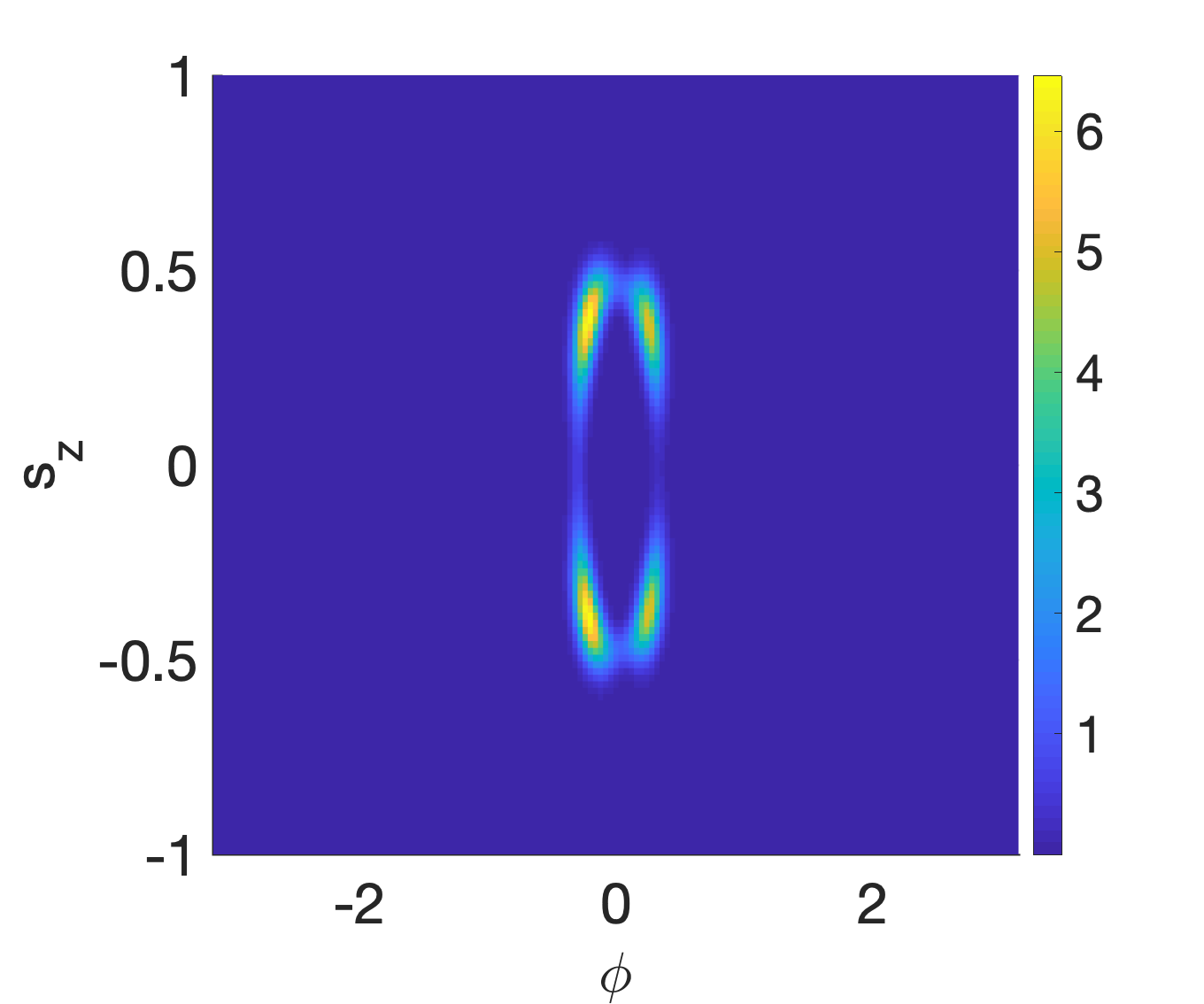}
\includegraphics[width=0.32\textwidth]{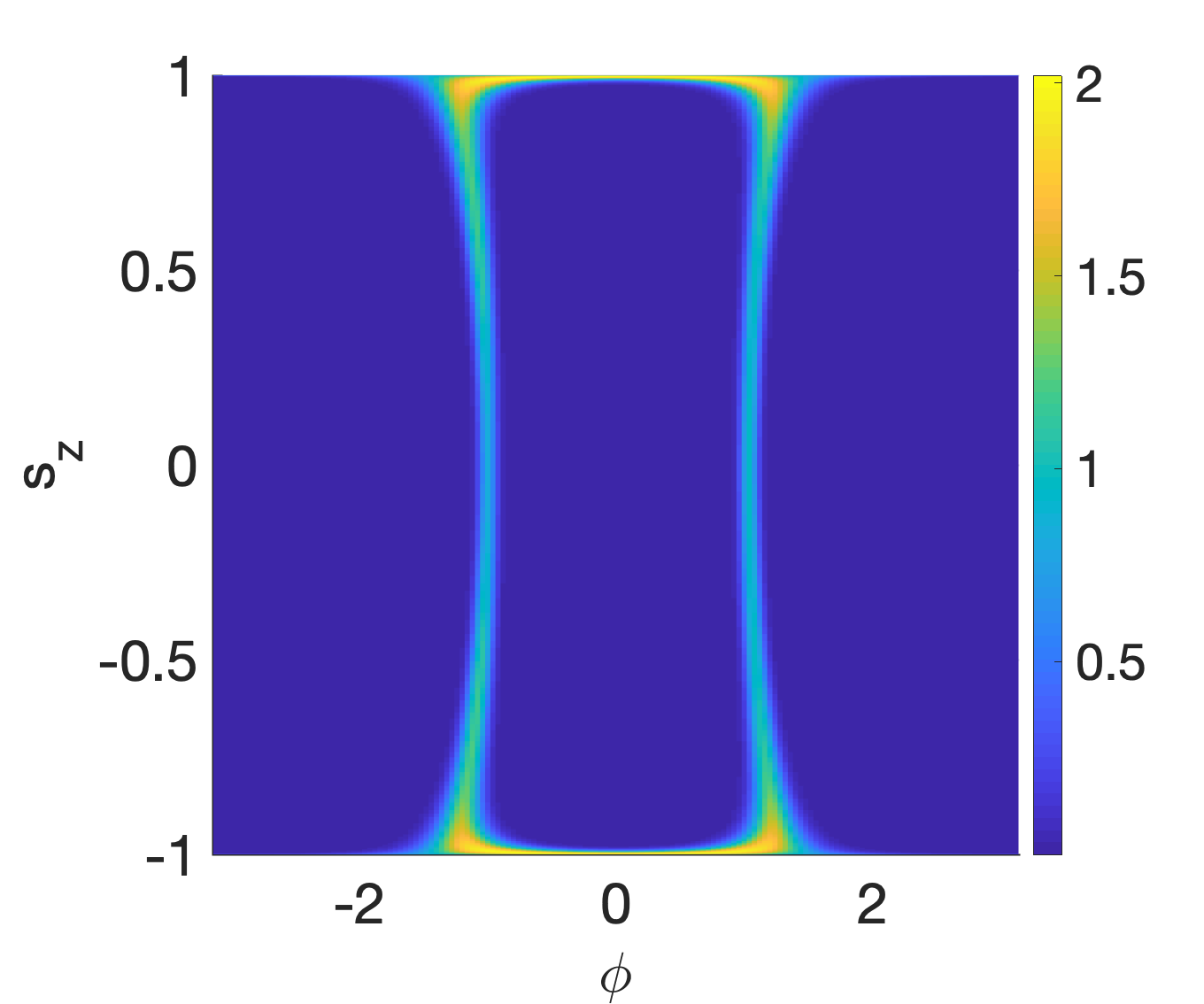}
\includegraphics[width=0.32\textwidth]{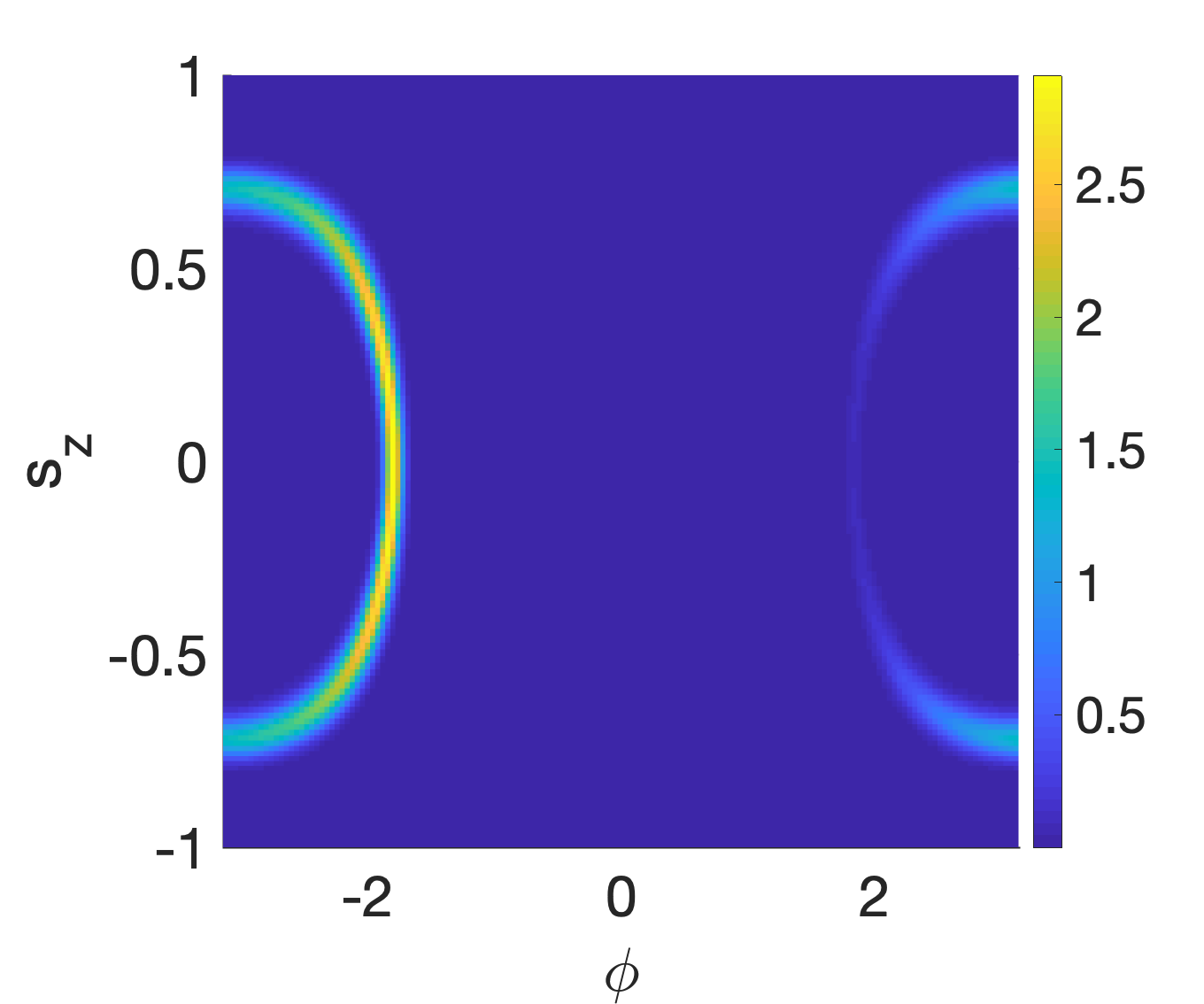}
\includegraphics[width=0.32\textwidth]{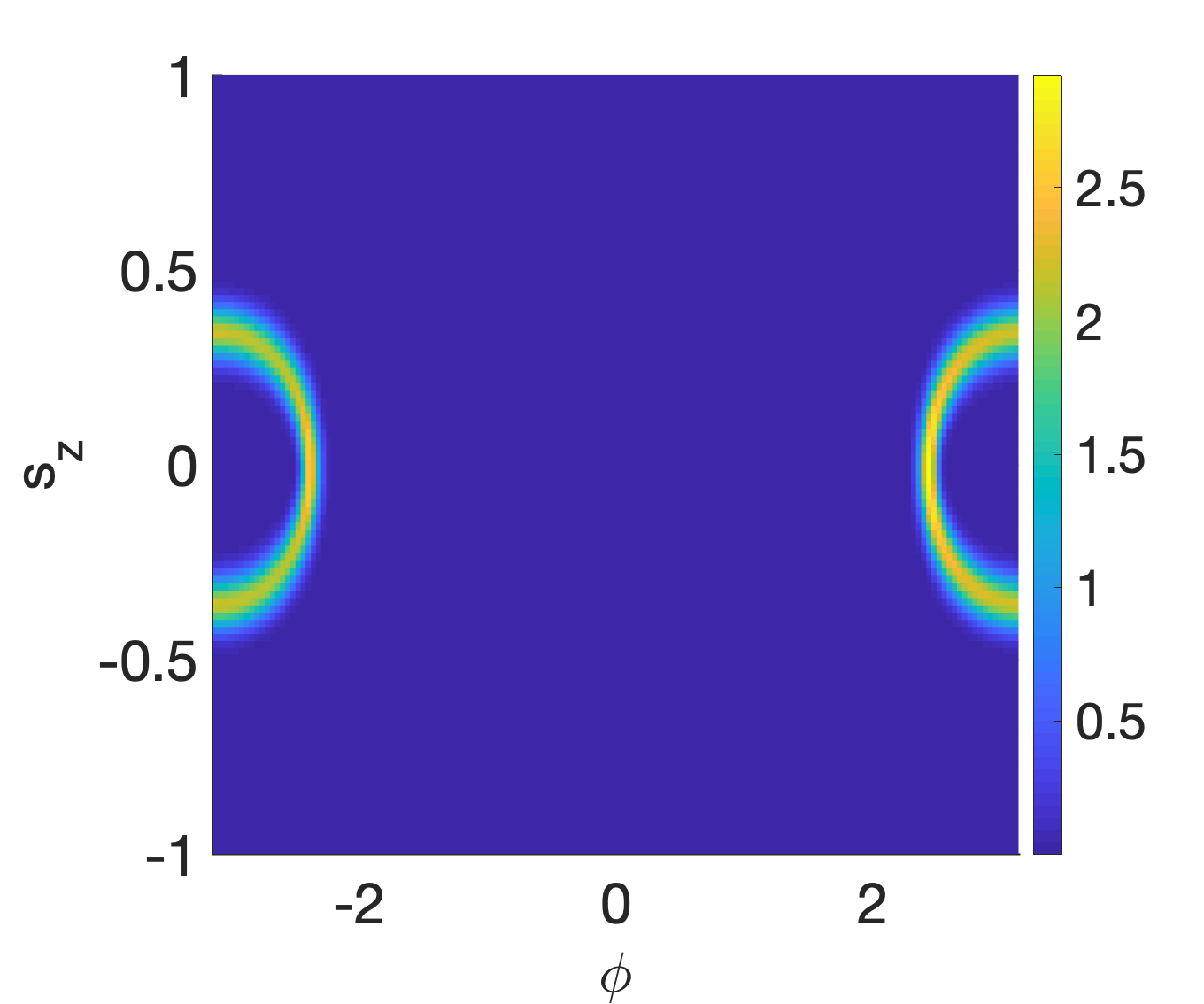}
\caption{Eigenvalues and eigenstates of the quantum Floquet operator in comparison to the classical Poincare section, for $p=2$, $k=0.5$, and $\gamma=0.01$ for $L=200$. The left panel in the top row shows the quasi energies in the complex plane. The middle panel shows the classical Poincare section. The remaining four panels show the Husimi distributions of selected Floquet states.}
\label{fig_husimi_norm_dyn_reg}
\end{figure}  

\begin{figure}[htb]
\centering
\includegraphics[width=0.32\textwidth]{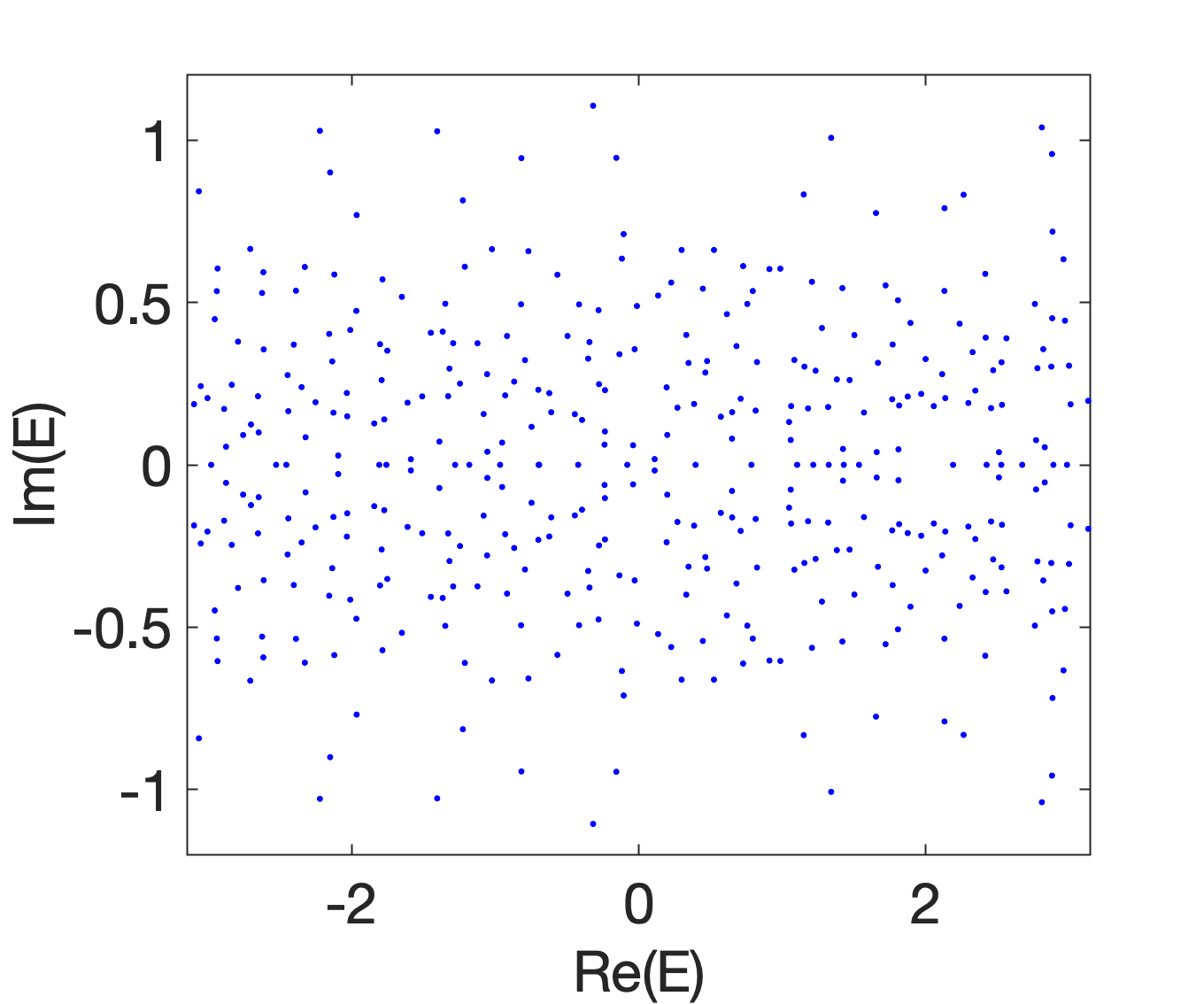}
\includegraphics[width=0.32\textwidth]{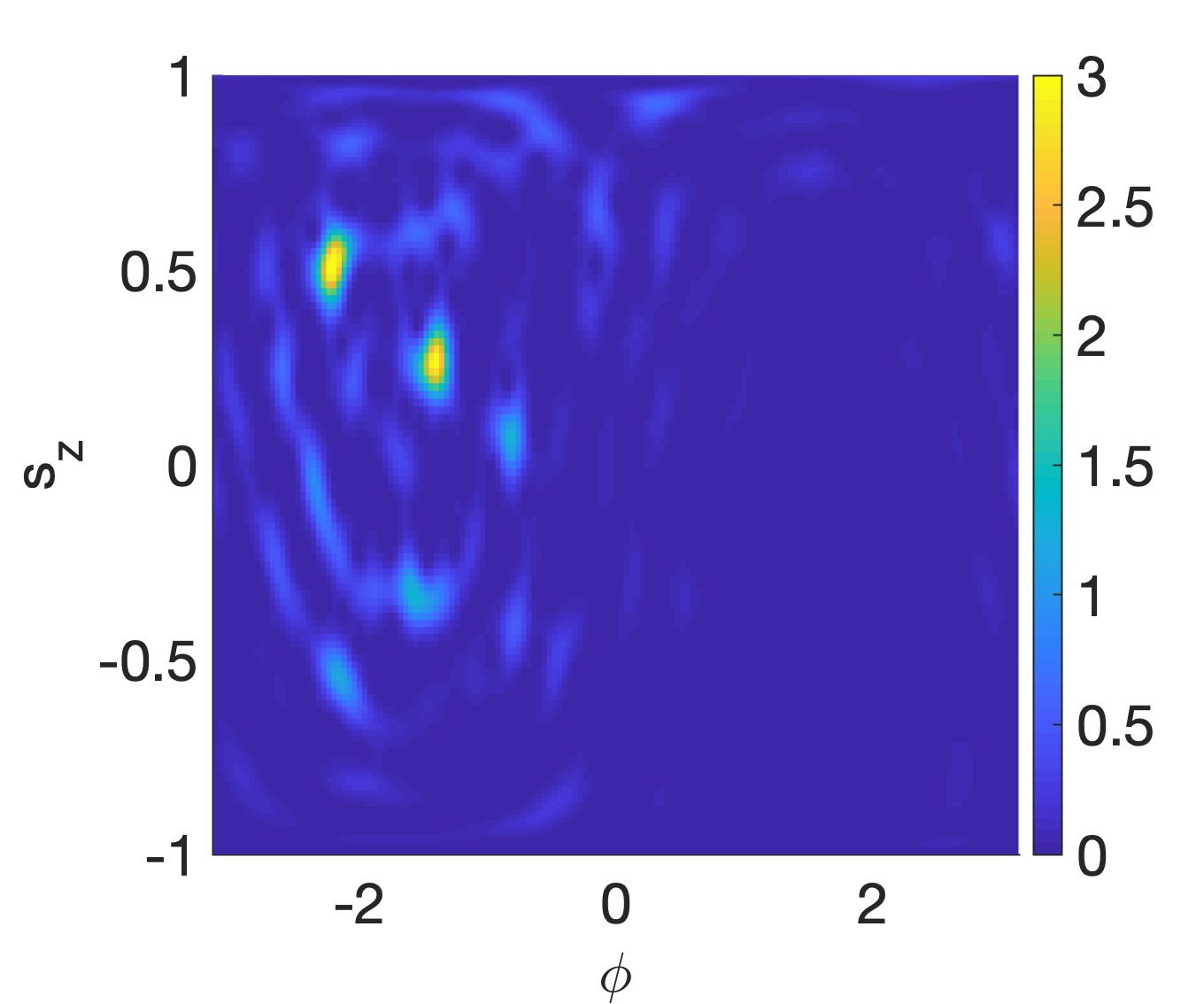}
\includegraphics[width=0.32\textwidth]{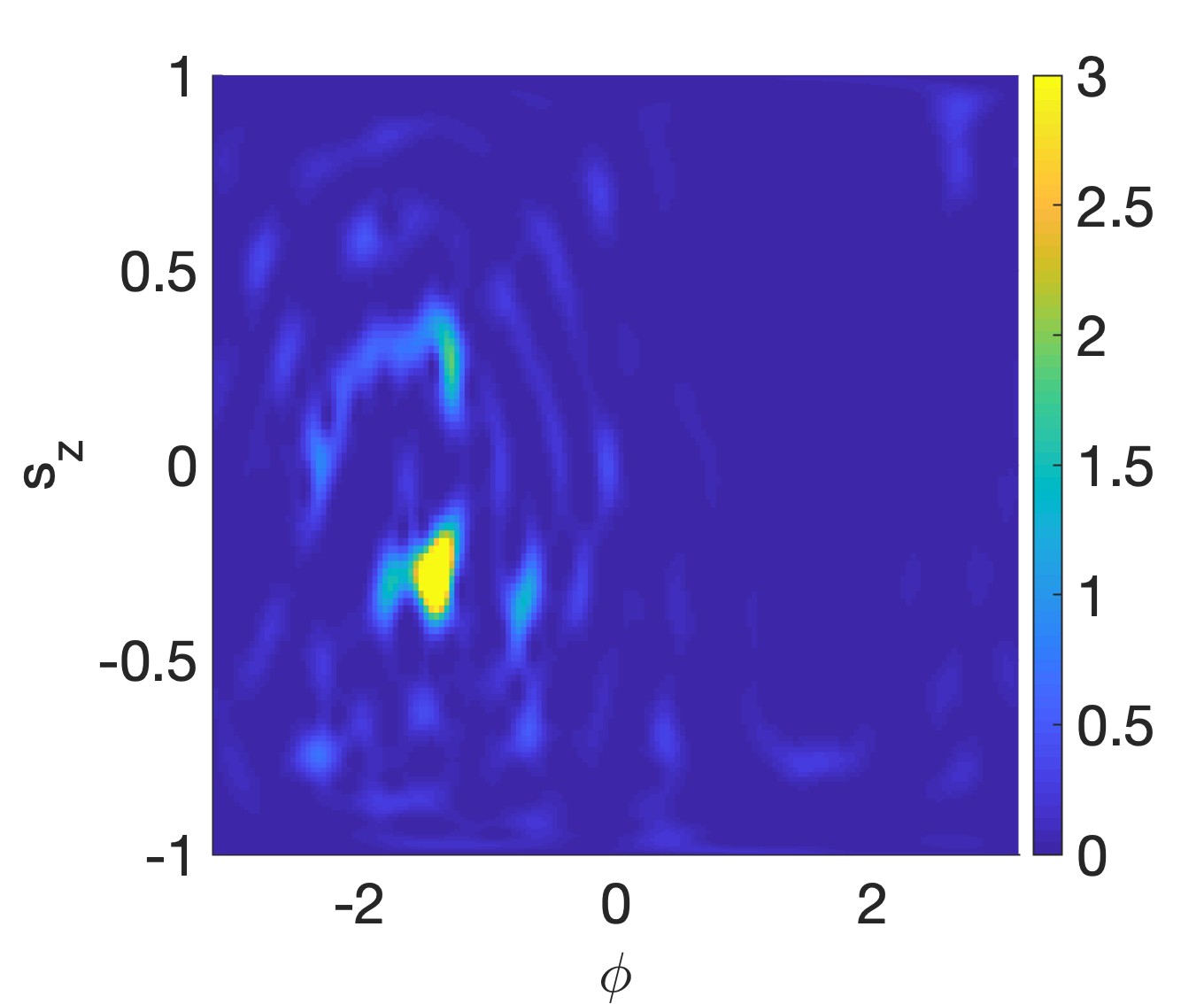}
\includegraphics[width=0.32\textwidth]{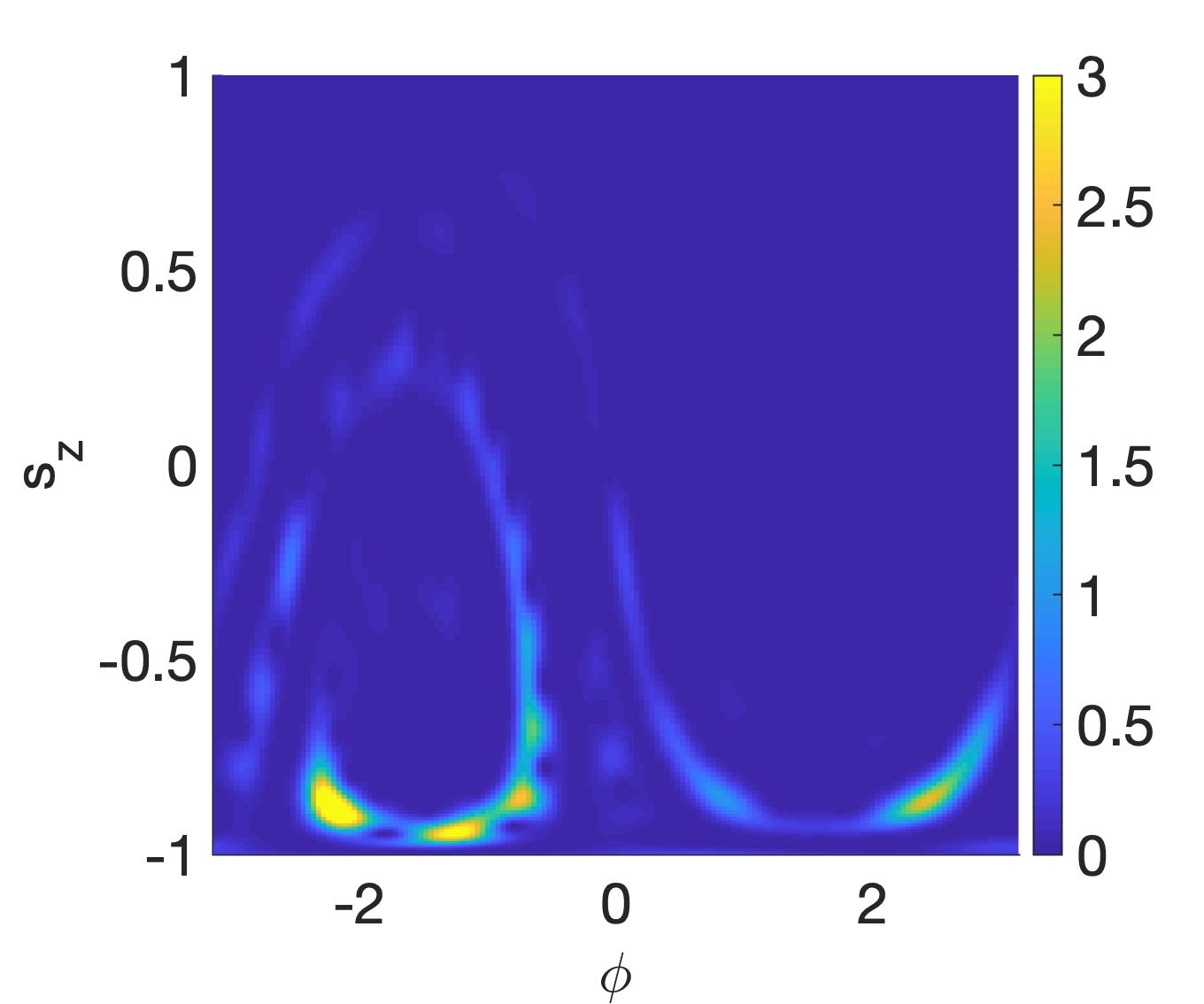}
\includegraphics[width=0.32\textwidth]{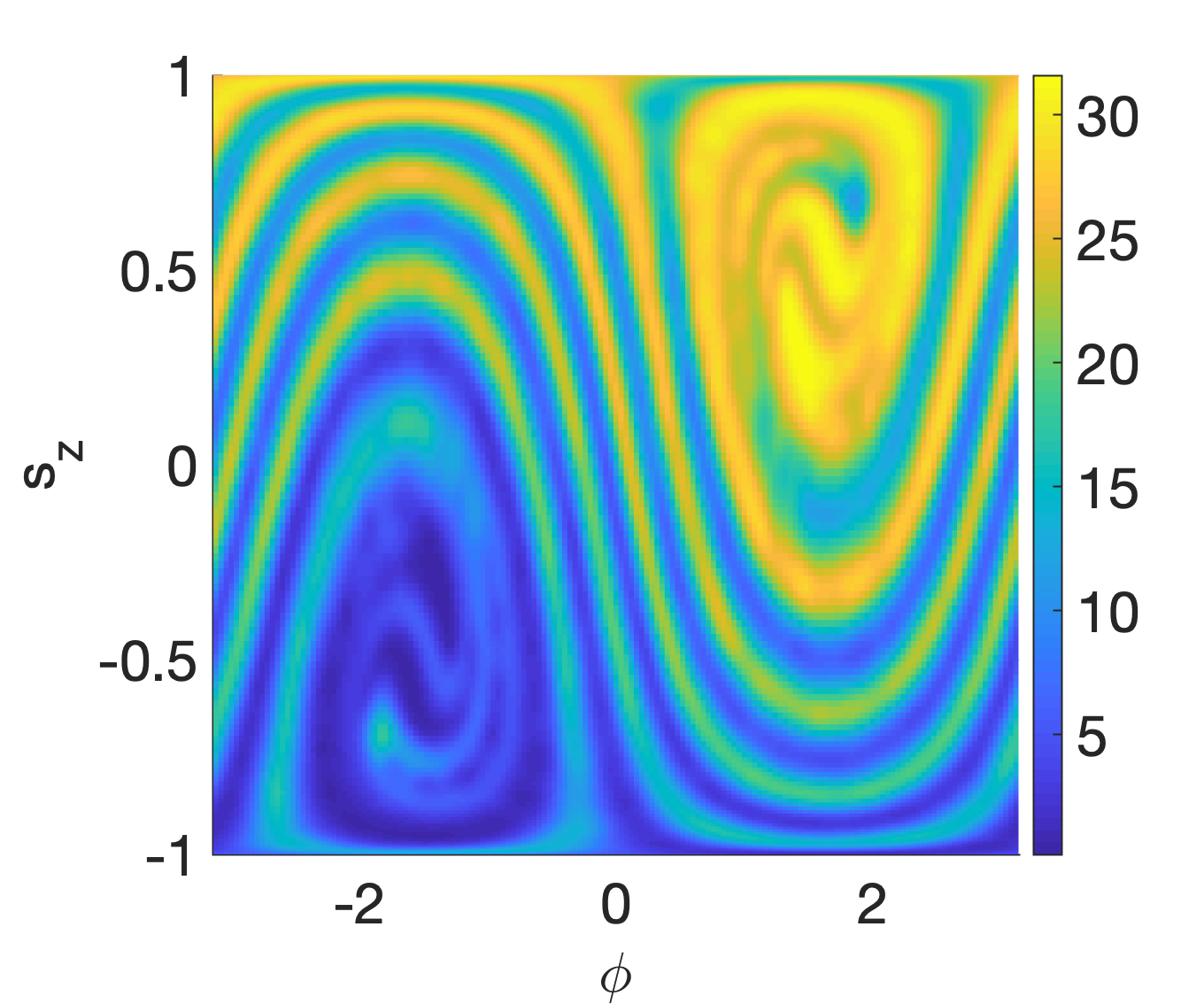}
\includegraphics[width=0.32\textwidth]{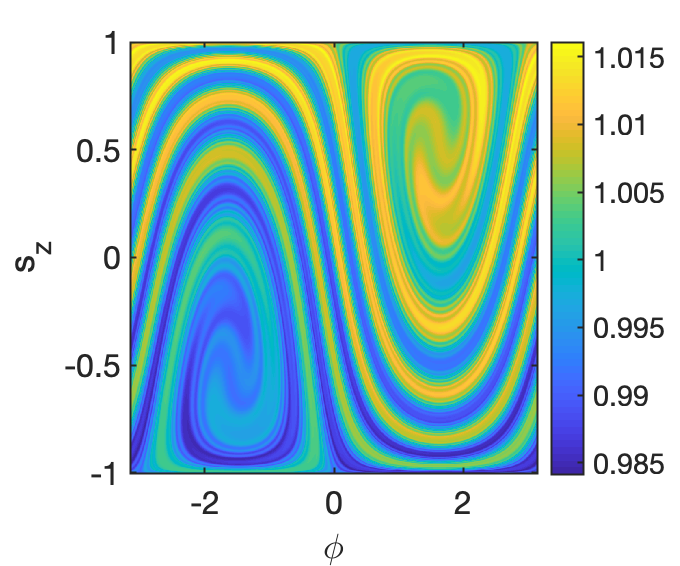}
\caption{Quasi energies and quantum phase-space structures in comparison to the classical intensity patterns, for $p=2$, $k=10$, and $\gamma=0.01$ for $L=200$. The left panel on the top shows the quasi energies in the complex plane. The other two top panels show the Husimi functions of two selected Floquet states. The left panel in the bottom row shows the Husimi function of the fastest growing Floquet state. The middle panel in the bottom row shows the Husimi representation of the subspace of growing states, i.e. the space spanned by those Floquet states with negative imaginary parts of the quasi energies. The right panel shows the classical intensity as a function of the initial phase-space position after three iterations of the classical map.}
\label{fig_husimi_norm_dyn}
\end{figure}

Let us begin with a small value of $k$ for which the classical system is in the regular regime. 
The left panel in the top row of figure \ref{fig_husimi_norm_dyn_reg} shows the quasi energies in the complex plane for $p=2$, and $k=0.5$ and a very small loss-gain parameter $\gamma=0.01$ for $L=200$. The corresponding regular classical dynamics is visualised by a Poincare section in the middle panel of the top row. While the dynamics does not show any sinks and sources, the intensity is not conserved for most initial conditions, but varies slowly, due to the small loss-gain parameter, which introduces an asymmetry in the trajectories. This is reflected in the quantum quasi energies, a handful of which lie visibly off the real axis. The $PT$-symmetry of the system is clearly obvious from the symmetry of the quasi energies with respect to the real axis. Since the system is only mildly non-Hermitian most eigenstates are still approximately orthogonal to each other, and their phase-space distributions are organised according to the structure of the classical phase-space dynamics. This is illustrated in the remaining four panels in figure \ref{fig_husimi_norm_dyn_reg}, which show the Husimi distributions of four different Floquet states.  

In the regime which is chaotic in the unitary case, on the other hand, a small non-Hermitian perturbation has a very different effect. Figure \ref{fig_husimi_norm_dyn} illustrates the case $p=2$, $k=10$ and $\gamma=0.01$, again for $L=200$. In this case almost all the quasi energies (depicted in the complex plane in the left panel of the figure) are complex. Further, the Floquet matrix is far from normal, and most eigenvectors strongly overlap with other eigenvectors. Nevertheless, one can see some structures reminiscent of the classical attractor and repeller in the Husimi functions of individual Floquet states, three of which are depicted in the middle and right panel on the top and the left panel in the bottom row.  To obtain a better insight into the quantum phase-space properties, rather than analysing individual eigenstates, it is useful to consider the Husimi representation of certain invariant subspaces, the \textit{Husimi-Schur representation} as introduced in \cite{Kopp10}. In the middle panel in the bottom row of figure \ref{fig_husimi_norm_dyn} we depict the Husimi representation of the invariant subspace of states with growing norm, that is, we use the Schur vectors of the Floquet operator where the diagonal elements are sorted according to the imaginary part of the quasi energies, and average over the Husimi distributions of the Schur vectors belonging to the positive imaginary parts. The resulting picture shows a striking resemblance to the false-colour plot arising from the classical intensity as a function of the initial phase-space position, which is depicted after 3 iterations in the right panel. This correspondence takes the place of the familiar quantum phase-space localisation on classical Poincare structures in phase space for the unitary case for strongly open quantum systems. Similar observations have been made for other types of openings in chaotic systems, for example in \cite{Kopp10} and \cite{Clau18,Clau19}. It is an interesting open question how the eigenspaces of the quantum system are organised in phase space in regions where the real energy structures and the norm dynamics are competing and neither dominates. This question, however goes beyond the scope of the present paper. 

\begin{figure}[htb]
\centering
\includegraphics[width=0.24\textwidth]{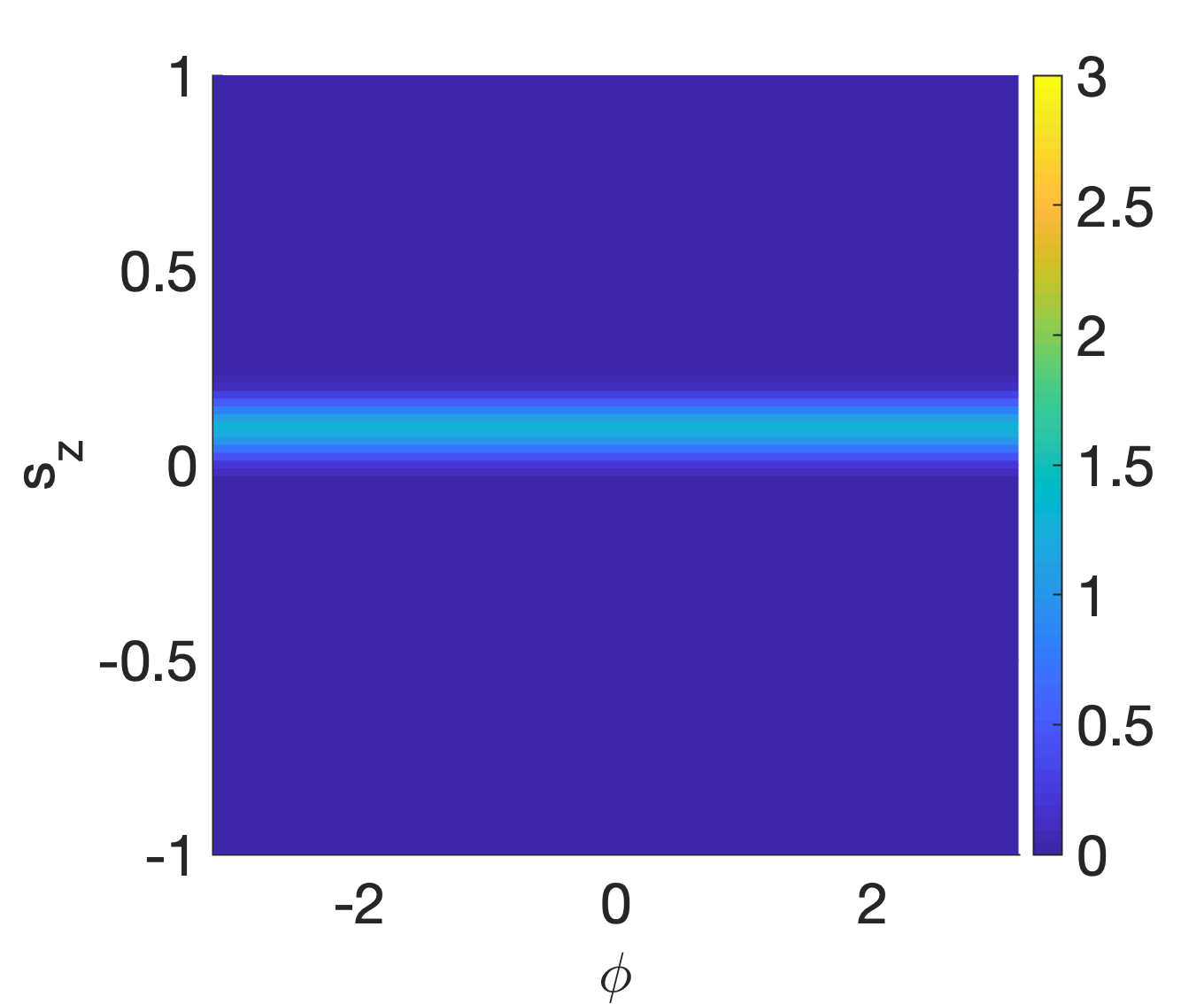}
\includegraphics[width=0.24\textwidth]{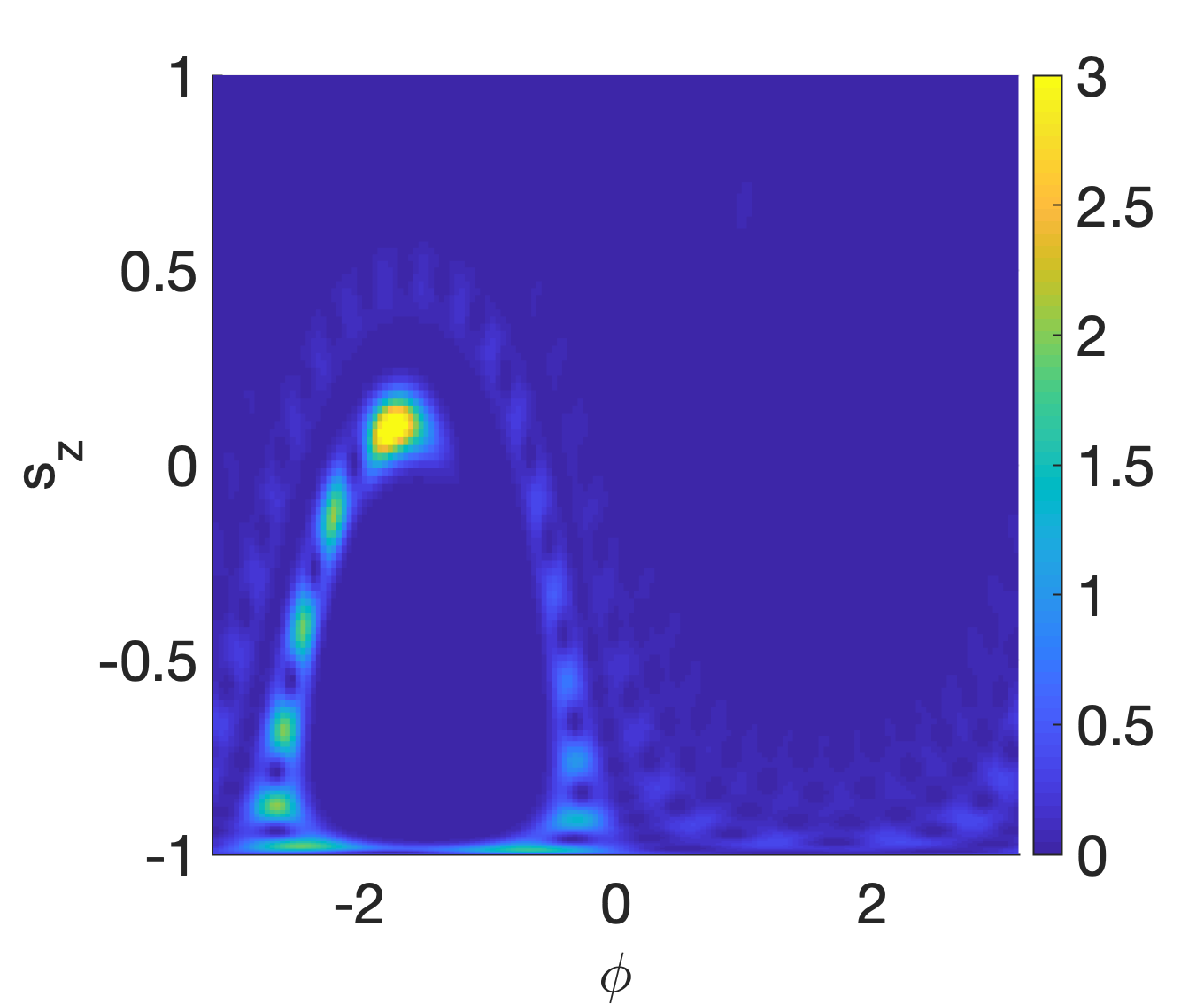}
\includegraphics[width=0.24\textwidth]{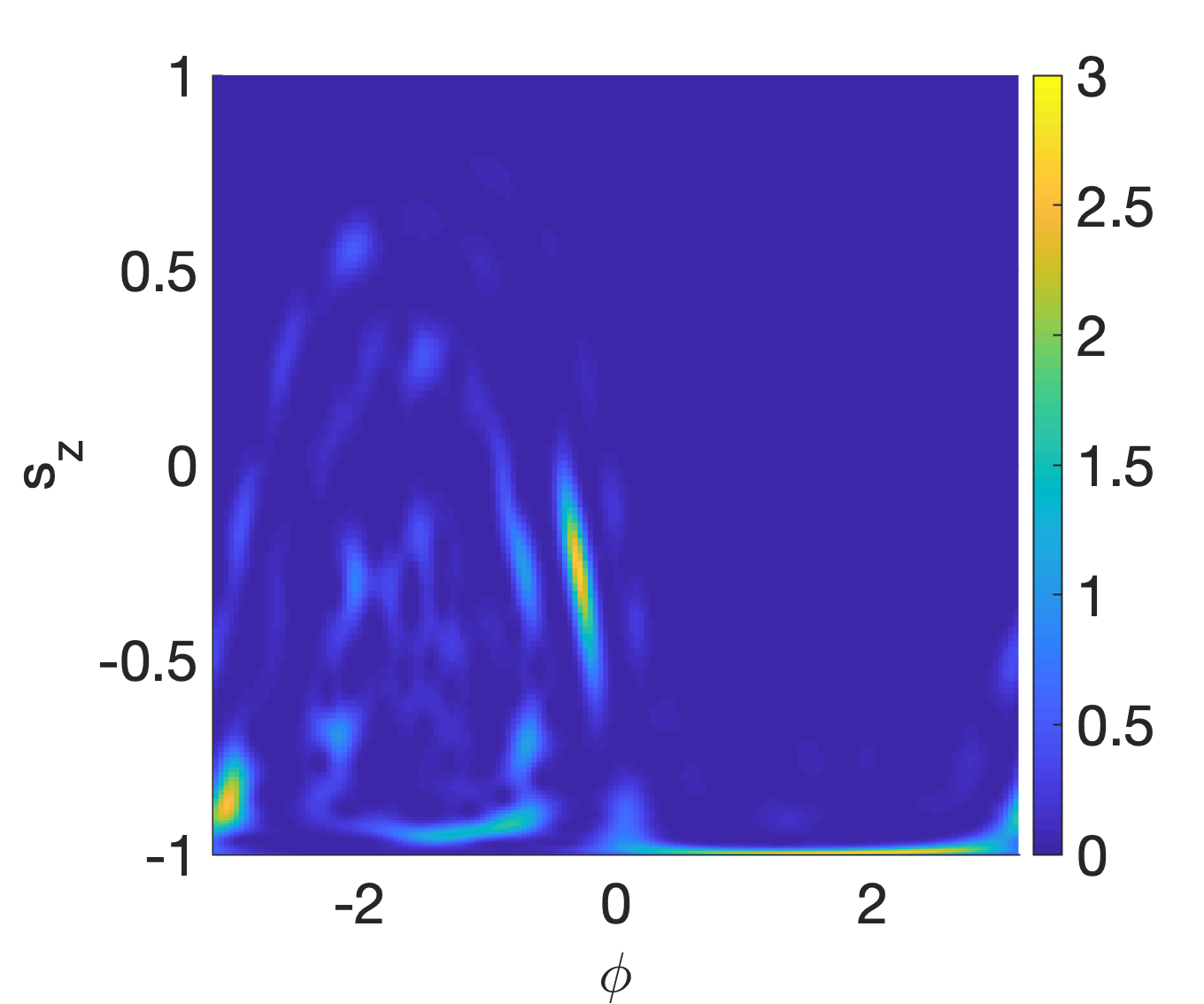}
\includegraphics[width=0.24\textwidth]{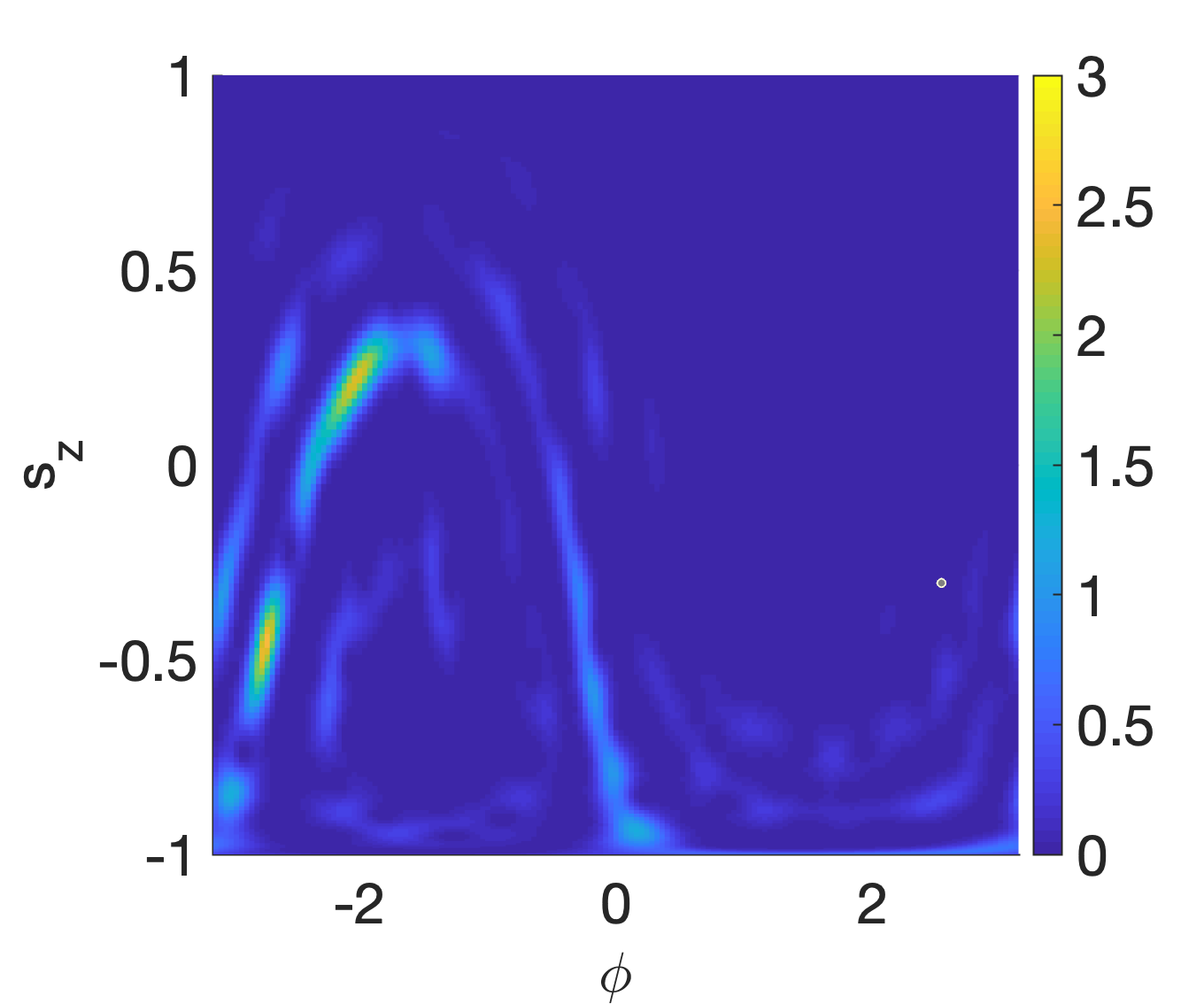}
\includegraphics[width=0.24\textwidth]{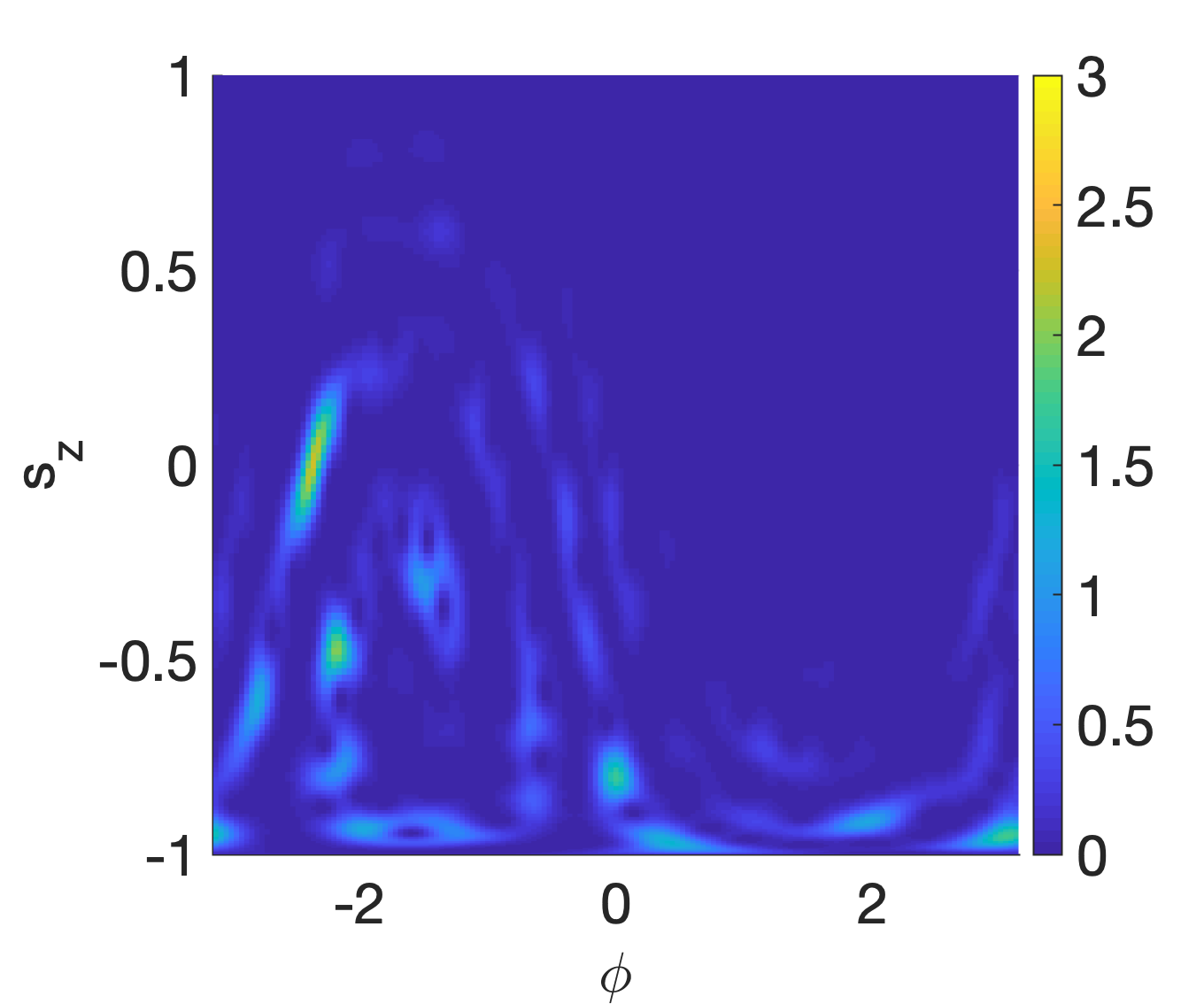}
\includegraphics[width=0.24\textwidth]{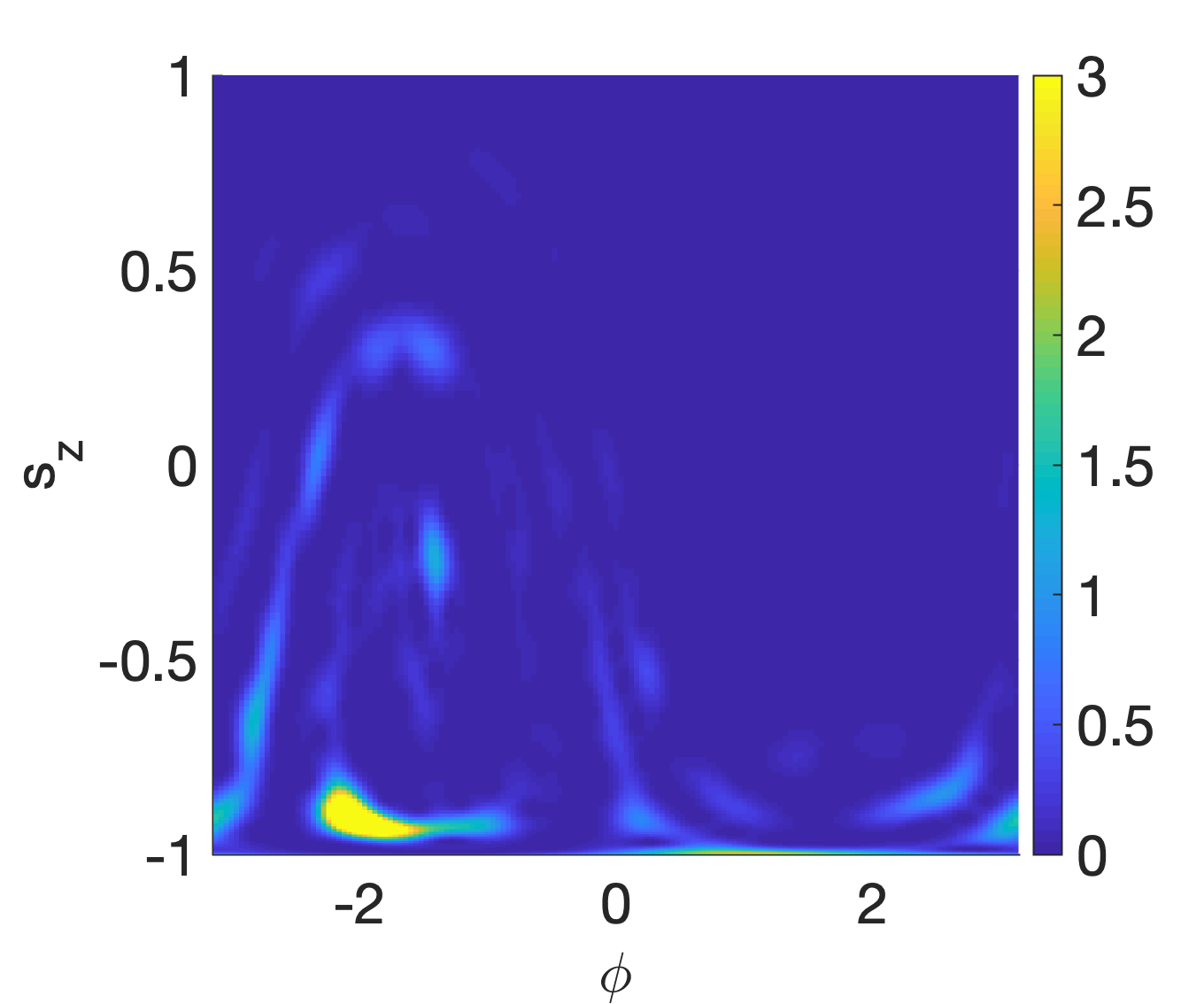}
\includegraphics[width=0.24\textwidth]{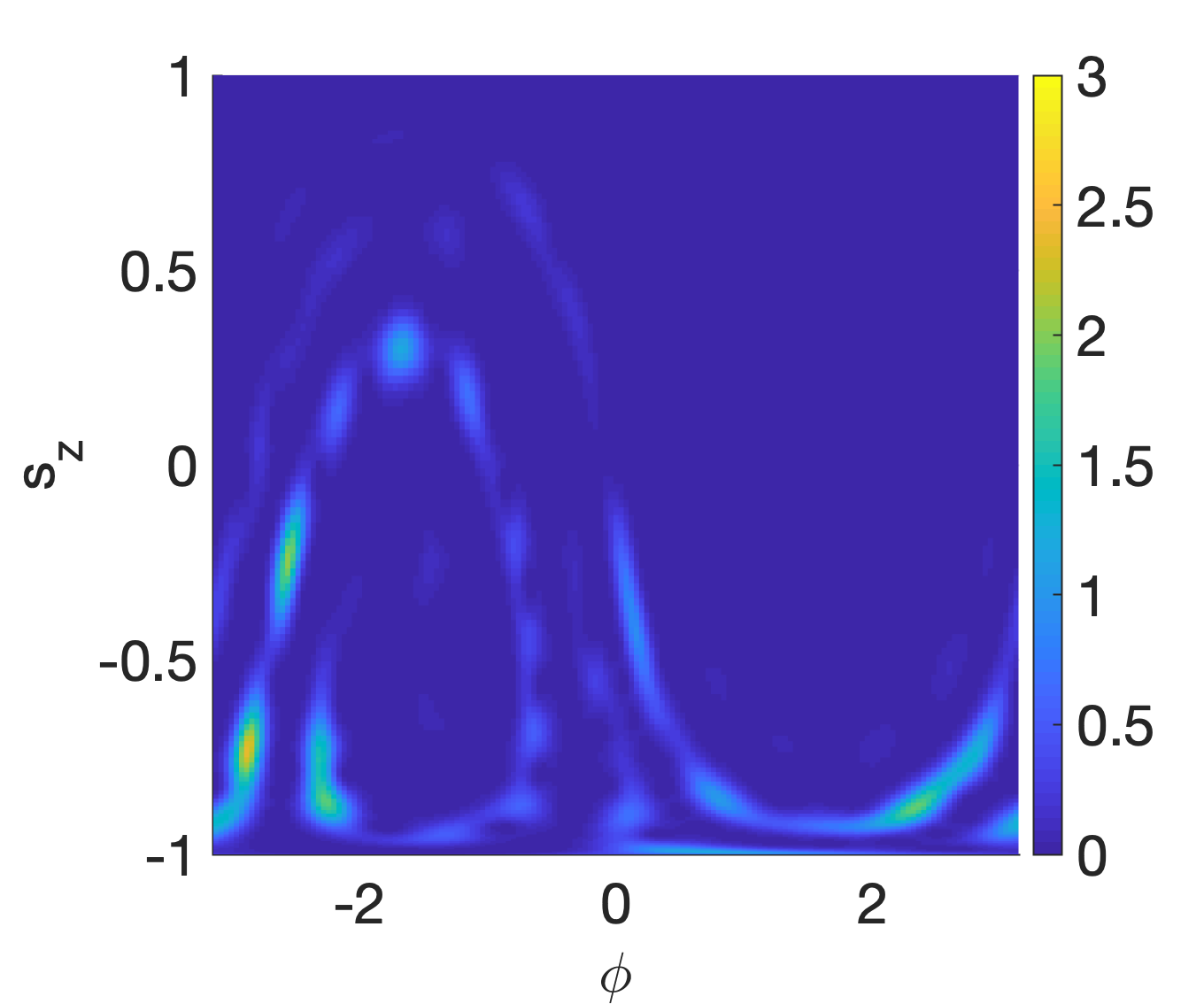}
\includegraphics[width=0.24\textwidth]{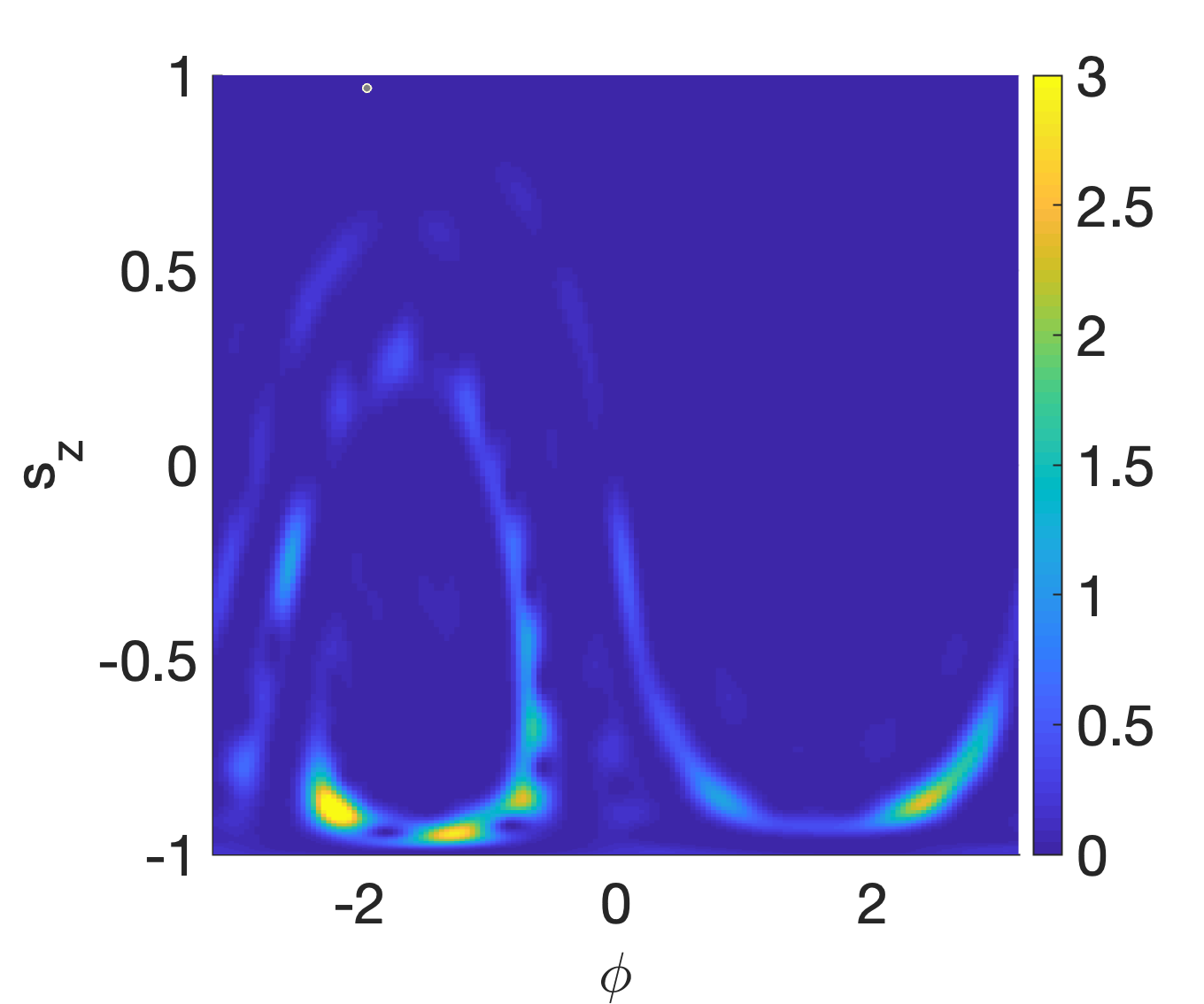}
\caption{Quantum dynamics in phase space for $p=2$, $k=10$, and $\gamma=0.01$ for $L=200$. The figure on the top left shows the Husimi representation of the initial state, the remaining panels show the Husimi representations of the state after $1,2,3,4,5,10$, and $30$ iterations, repsectively, from left to right and top to bottom.}
\label{fig_husimi_dyn}
\end{figure} 

This classical phase-space structure is also visible in the quantum dynamics, as demonstrated in figure \ref{fig_husimi_dyn}, which depicts the Husimi representation of an initial eigenstate of $\hat L_z$ after different numbers of iterations. The spiral structure is clearly visible in the iterated states. After a large number of iterations all that remains is the  fastest growing state. Indeed for the present example after approximately $20$ iterations the state has approached the final configuration. This can be seen from the similarity of the Husimi representation of the state after $30$ iterations, depicted in the lower right panel in figure \ref{fig_husimi_dyn} and the Husimi function of the fastest growing state, depicted in the lower left panel in figure \ref{fig_husimi_norm_dyn}.

\begin{figure}[htb]
\centering
\includegraphics[width=0.49\textwidth]{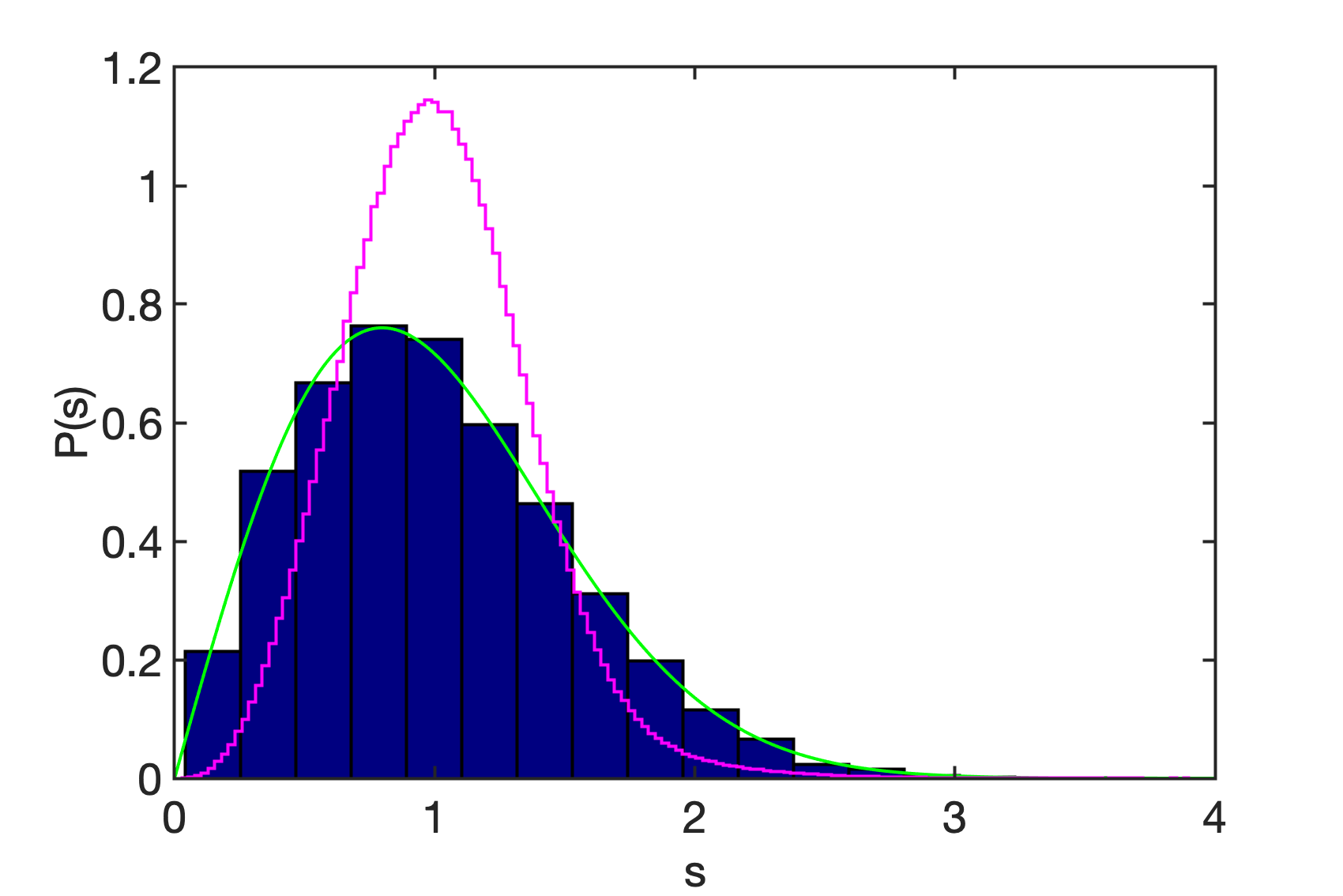}
\includegraphics[width=0.49\textwidth]{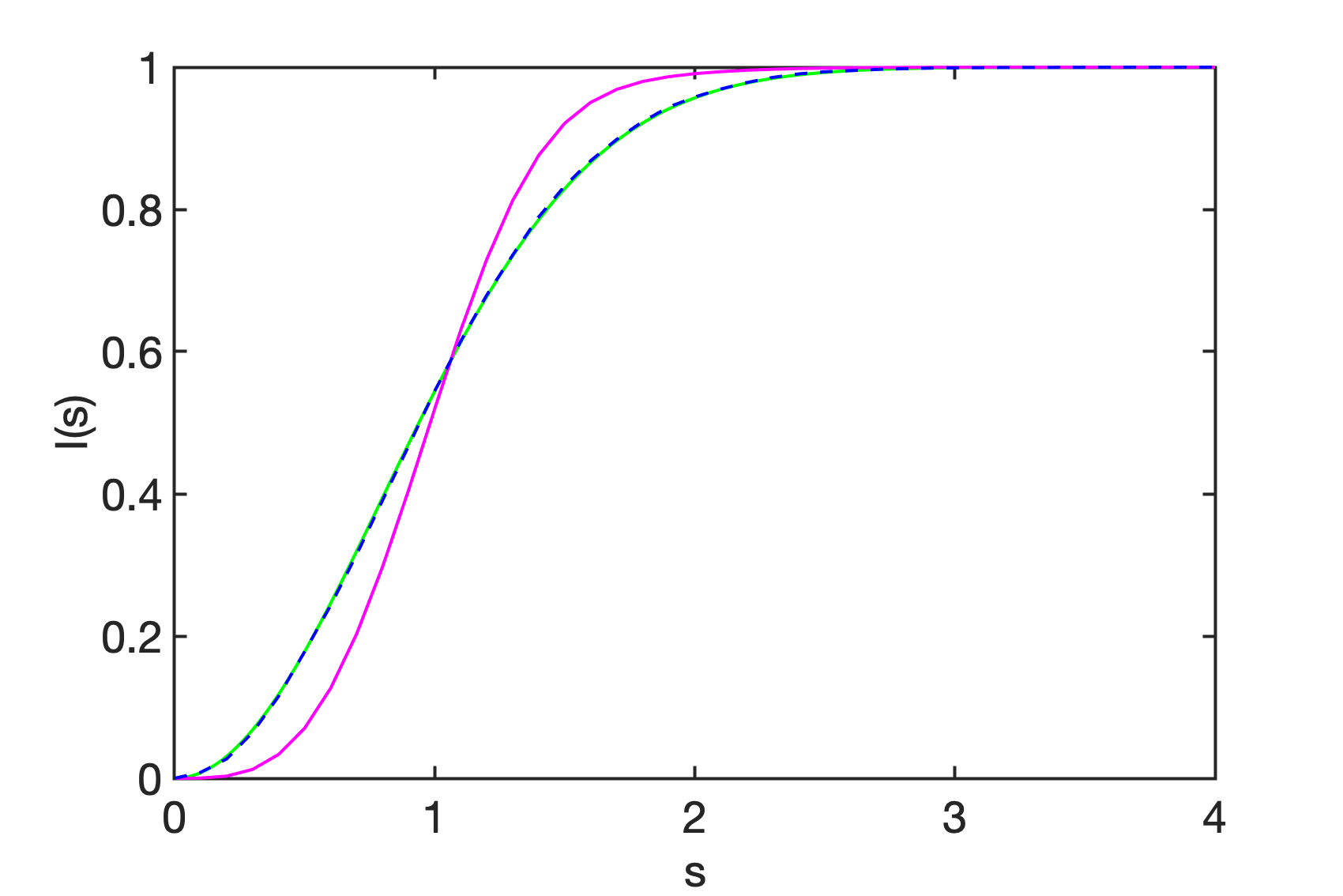}
\caption{Distribution of nearest-neighbour distances for the kicked top with $L = 200, p = 2, \epsilon = 0, \gamma = 0.01$, and $k$ between 0.1 and 1. The left panel shows the distribution of unfolded nearest-neighbour distances in comparison to the two-dimensional Poisson distribution (green) and that of transpose-invariant Gaussian random matrices (magenta). The right panel shows the integrated distributions. }
\label{PTKickedTopSpectralStatisticsRegular}
\end{figure}

\begin{figure}[htb]
\centering
\includegraphics[width=0.49\textwidth]{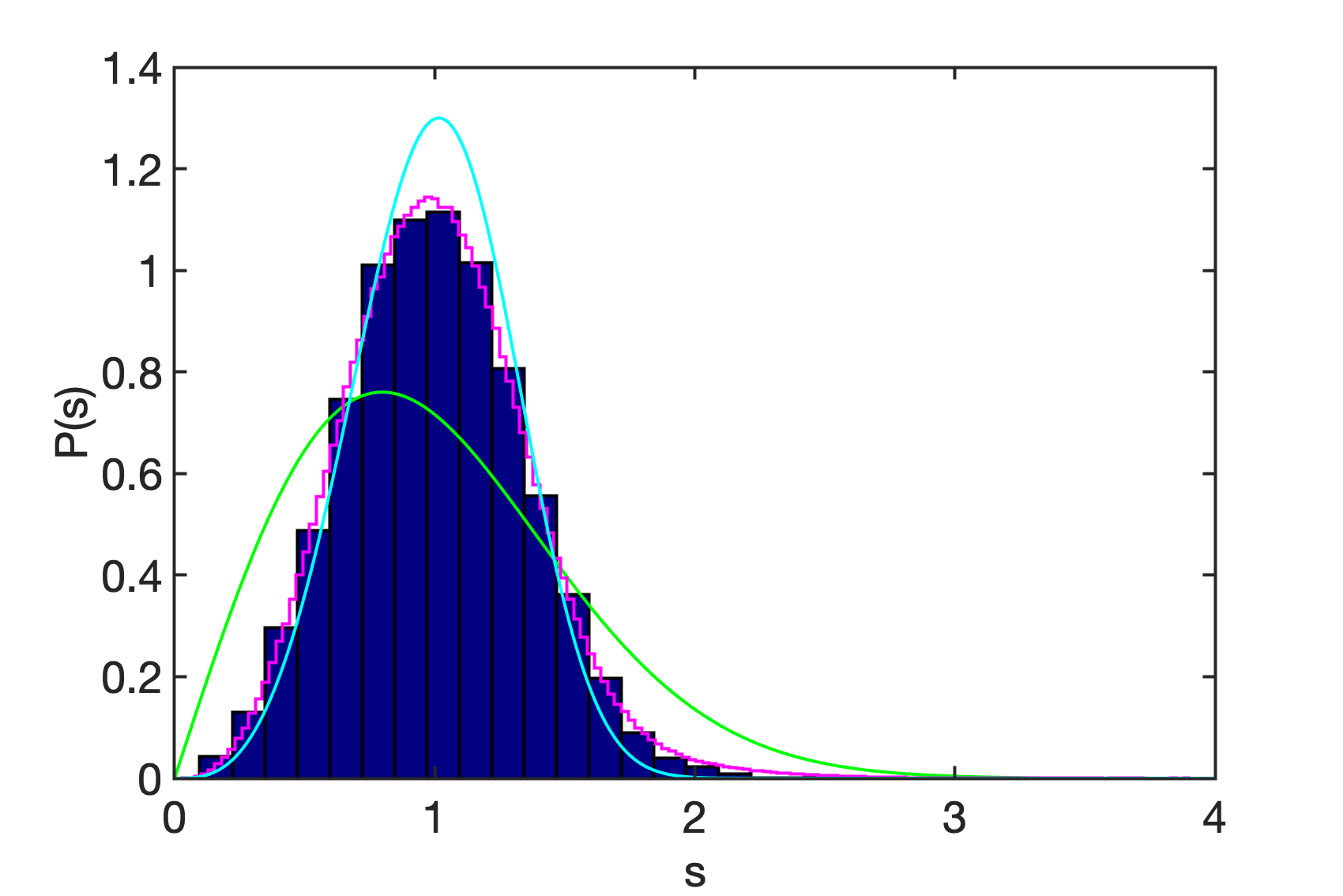}
\includegraphics[width=0.49\textwidth]{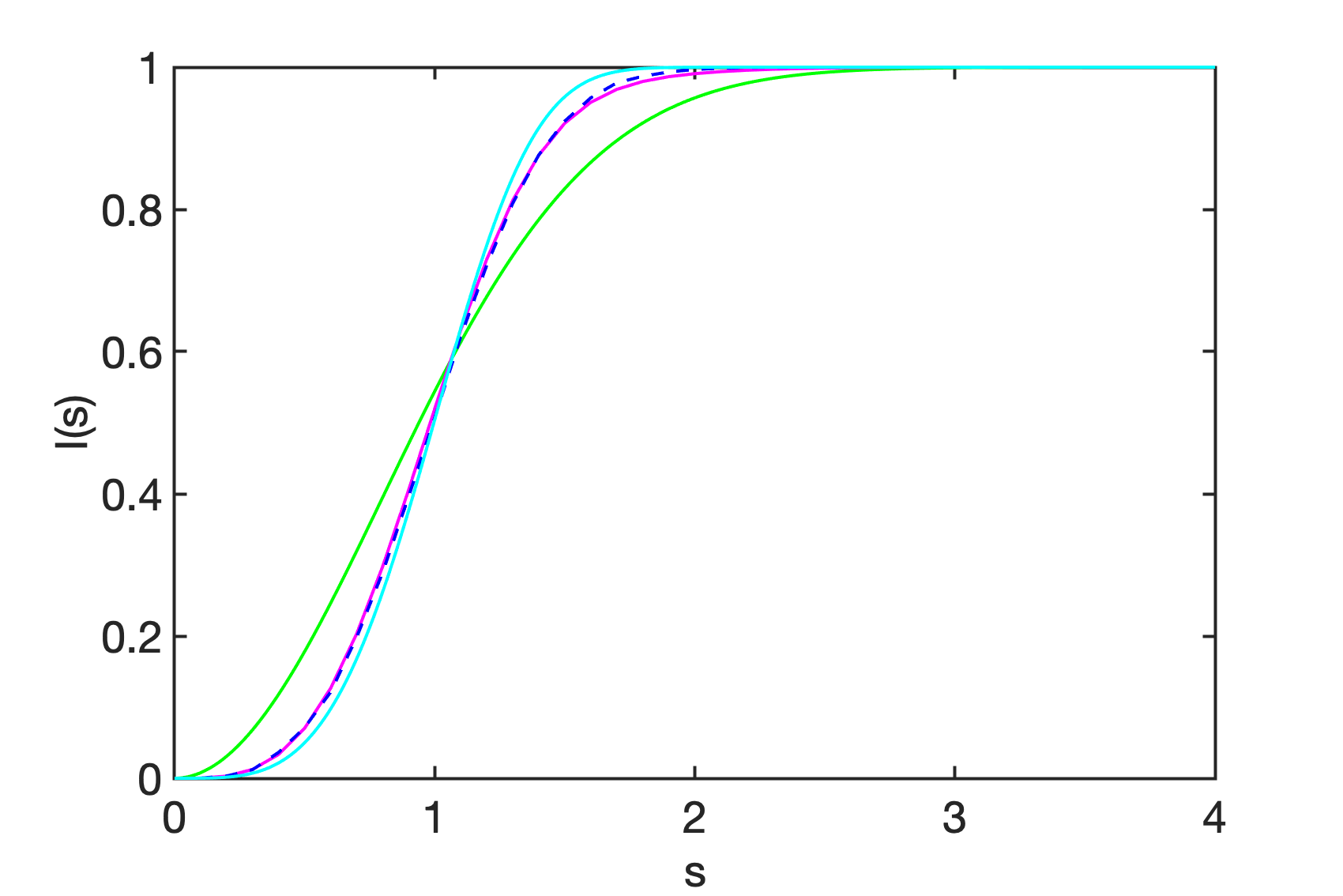}
\includegraphics[width=0.49\textwidth]{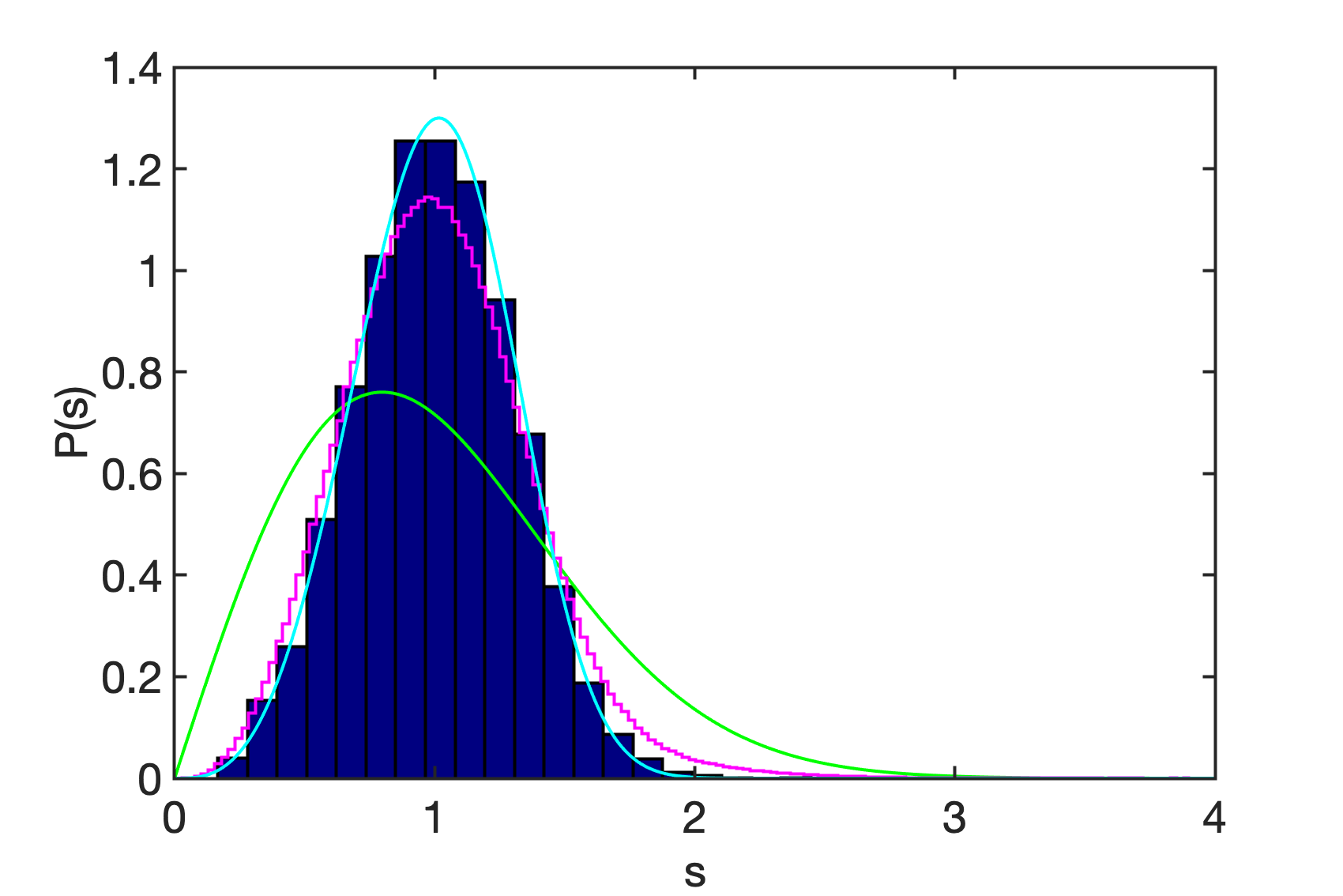}
\includegraphics[width=0.49\textwidth]{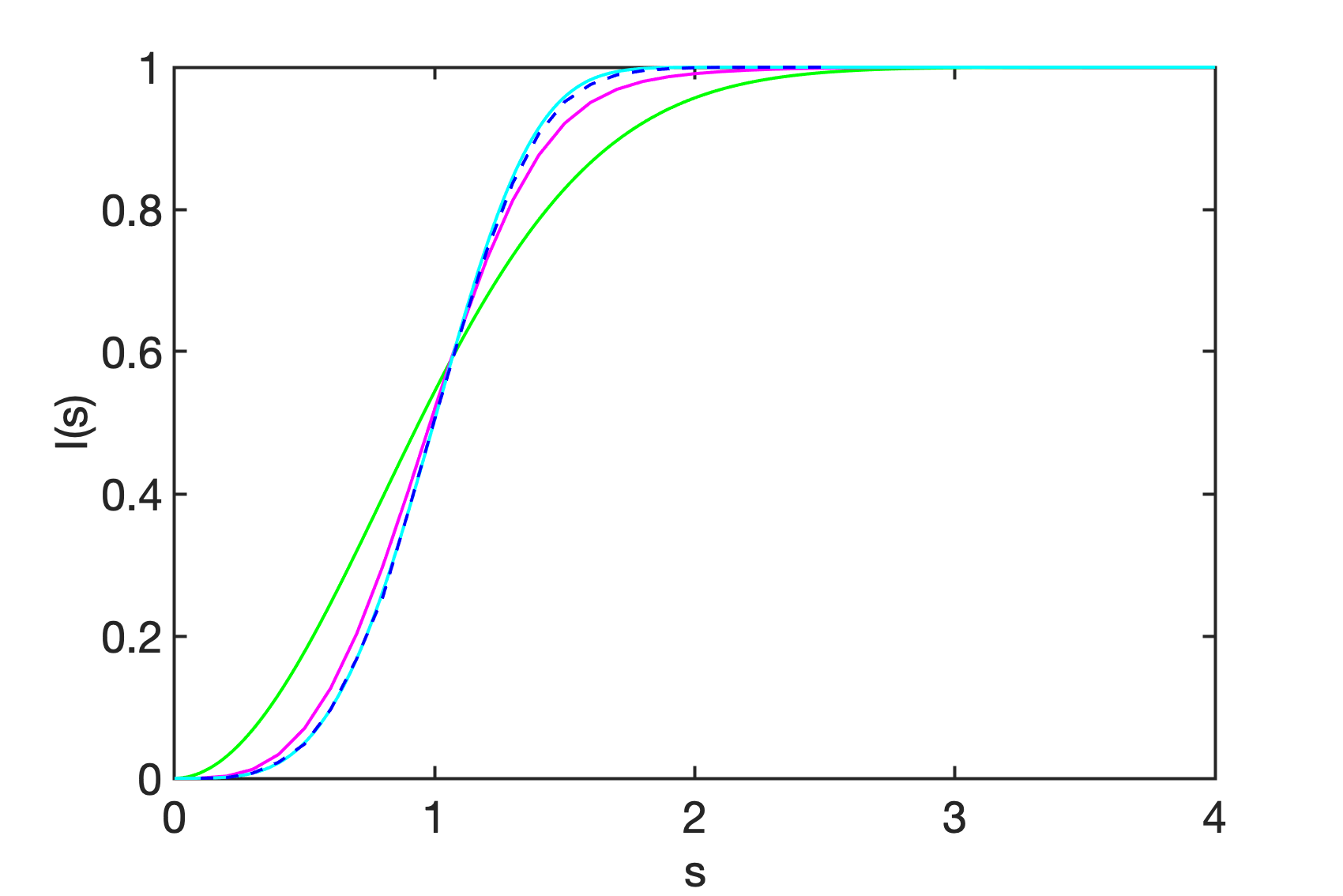}
\caption{Distribution of nearest-neighbour distances for the kicked top with $L = 200, p = 2, \epsilon = 0, \gamma = 0.01$, and $k$ between 9 and 10 (top) and the modified system (\ref{eqn_Floq_y}) with $\eta=1$ (bottom). The left panels show the distributions of unfolded nearest-neighbour distances in comparison to the two-dimensional Poisson distribution (green), that of transpose-invariant Gaussian random matrices (magenta), and the distribution resulting from the Ginibre ensembles (cyan). The right panel shows the integrated distributions.}
\label{PTKickedTopSpectralStatisticsChaotic}
\end{figure}

Let us finally turn to the spectral statistics of our $PT$-symmetric kicked top. In the Hermitian case the level spacing statistics of the quasi energies of the kicked top follow GOE statistics in the chaotic regime, and show a Poisson distribution in the regular regime, in line with the Bohigas-Giannoni-Schmit conjecture. In the non-Hermitian case, the quasi-energies become complex, and a generalisation of level spacings needs to be considered. Here we follow the approach in \cite{Grob88,Grob89,Haak92c} and analyse the distributions of Euclidean nearest-neighbour distances in the complex plane, after unfolding the spectra. For the unfolding we make use of the fact that the quasi energy spectrum is approximately uniform in the real parts, and use the integrated staircase function of the imaginary parts to approximate the underlying smooth distribution and use this to unfold the imaginary parts, while leaving the real parts unscaled. To obtain a suitable amount of quasienergies we combine the data from a number of realisations with slightly altered parameters, similar to the common practice in the Hermitian case. In figure \ref{PTKickedTopSpectralStatisticsRegular} we depict the nearest-neighbour distance distributions in the regular regime, for k varying in $0.05$ increments between $k=0.1$ and $k=1$, for $L=200$ and a small $\gamma=0.01$, where we disregarded quasi energies close to the boundary to avoid boundary effects. The nearest-neighbour distance distribution in the chaotic case, for values of $k=9$ to $k=10$ (in increments of $0.02$) and $\gamma=0.01$ is depicted in  the top row of figure \ref{PTKickedTopSpectralStatisticsChaotic}. For comparison we also plot the two-dimensional Poisson distribution curve (green) that is expected for independent random numbers in the plane, and associated to regular behaviour for complex eigenvalues \cite{Grob88,Grob89,Haak92c}. This distribution is identical to the GOE distribution
\begin{equation}
\label{eqn-2Poisson}
P(s)=\frac{\pi}{2}s \rme^{-\frac{\pi}{4}s^2}.
\end{equation}
The second curve (magenta), shows the nearest-neighbour distribution for eigenvalues of Gaussian random matrices with the constraint $\hat A=\hat A^{T}$, which we numerically obtained from the diagonalisation of $10000$ matrices of the same size as for the kicked top ($401\times401$). This has been identified as one of three universality classes for non-Hermitian Hamiltonians in \cite{Hama20}, which are speculated to replace the three universality classes of Hermitian systems given by Dyson's Gaussian ensembles. Indeed, our Hamiltonian and Floquet operator are invariant under matrix transposition and we observe an excellent agreement of the data in the regular regime with the two-dimensional Poisson curve, and with the Gaussian transpose invariant random matrix model in the chaotic regime. The more familiar universal nearest-neighbour distance distribution class for non-Hermitian matrices that has been identified in the spectra of various physical systems in the past \cite{Grob88,Grob89,Haak92c,Akem19} is depicted by the cyan curves in figure \ref{PTKickedTopSpectralStatisticsChaotic}. The nearest-neighbour spacing distribution is given by that of the eigenvalues of the Ginibre ensembles in the large $N$ limit as \cite{Haak92c}
\begin{equation}
\label{eqn_Ginibre}
P(s)=C\lim_{N\to\infty}\left(\prod_{n=1}^{N-1}e_n((Cs)^2\rme^{-(Cs)^2})\right)\sum_{n=1}^{N-1}\frac{2(Cs)^{2n+1}}{n! e_n((Cs)^2)},
\end{equation}
with
\begin{equation}
e_n(x)=\sum_{m=0}^{n}\frac{x^m}{m!},
\end{equation}
and the numerical constant (arising from demanding normalisation and that $\langle s\rangle=1$)
\begin{equation}
C\approx1.1429\cdots.
\end{equation}
This behaviour can be recovered if we introduce an extra kicking term proportional to $\hat L_y^2$ in our model to yield 
\begin{equation}
\label{eqn_Floq_y}
\hat F =  \ue^{-\ui(p \hat{L}_x +(\epsilon+ \ui \gamma) \hat{L}_z)\frac{\tau}{2}} \ue^{-\ui \frac{k}{L}\hat{L}_z^2}\ue^{-\ui \frac{\eta}{L}\hat{L}_y^2} \ue^{-\ui(p \hat{L}_x +(\epsilon+ \ui \gamma) \hat{L}_z)\frac{\tau}{2}},
\end{equation}
much as in the unitary case, where this extra factor changes the level spacing behaviour from GOE to GUE \cite{Kus88}. The level spacing distribution for the same values of $L$, $p$, $k$, and $\gamma$ and with $\eta=1$ are depicted in the bottom row of figure \ref{PTKickedTopSpectralStatisticsChaotic}, where we observe an excellent agreement between the data and the prediction (\ref{eqn_Ginibre}) from the Ginibre ensemble.

We have further analysed the spectra of a $PT$-symmetric generalisation of another quantum chaotic models showing GOE statistics in the Hermitian case, a triadic Baker map. There are multiple quantisations of the Baker map, and many investigations of open versions (see e.g. \cite{Keat06,Nonn08,Carl16}). Here we use a standard quantisation in which the unitary triadic Baker map is given by the Floquet operator (see, e.g. \cite{Keat06} and references therein)
\begin{equation}
\hat{\mathcal{B}} = F_N^{-1}\begin{pmatrix}F_{N/3} & 0 & 0 \\ 0 & F_{N/3} & 0 \\ 0 & 0 & F_{N/3}\end{pmatrix},
\end{equation}
where $F_N$ is a discrete Fourier transform,
\begin{equation}
(F_N)_{mn} = \frac{1}{\sqrt{N}}\ue^{-\frac{2\pi\ui}{N}\big(m+\frac{1}{2}\big)\big(n+\frac{1}{2}\big)}; \hspace{0.5cm}m, n \in \{0,1,\ldots, N-1\}.
\end{equation}
This map is parameter free, the classical counterpart is fully chaotic, and the quantum quasi energies display random matrix statistics.
We generalise this to a non-unitary $PT$-symmetric map (with $\T^2=\mathbb{I}$) by introducing a parameter $\gamma \in (0, 1)$ that amplifies phase-space trajectories in one third of the phase space, leaves the central third unchanged, and introduces a loss (of equal strength to the gain) in the remaining third, yielding the map
\begin{equation}\hat{\mathcal{B}}_{PT} = F_N^{-1}\mathrm{diag}(\gamma F_{N/3}, F_{N/3}, \frac{1}{\gamma}F_{N/3}).\end{equation}
\begin{figure}[htb]
\centering
\includegraphics[width=0.49\textwidth]{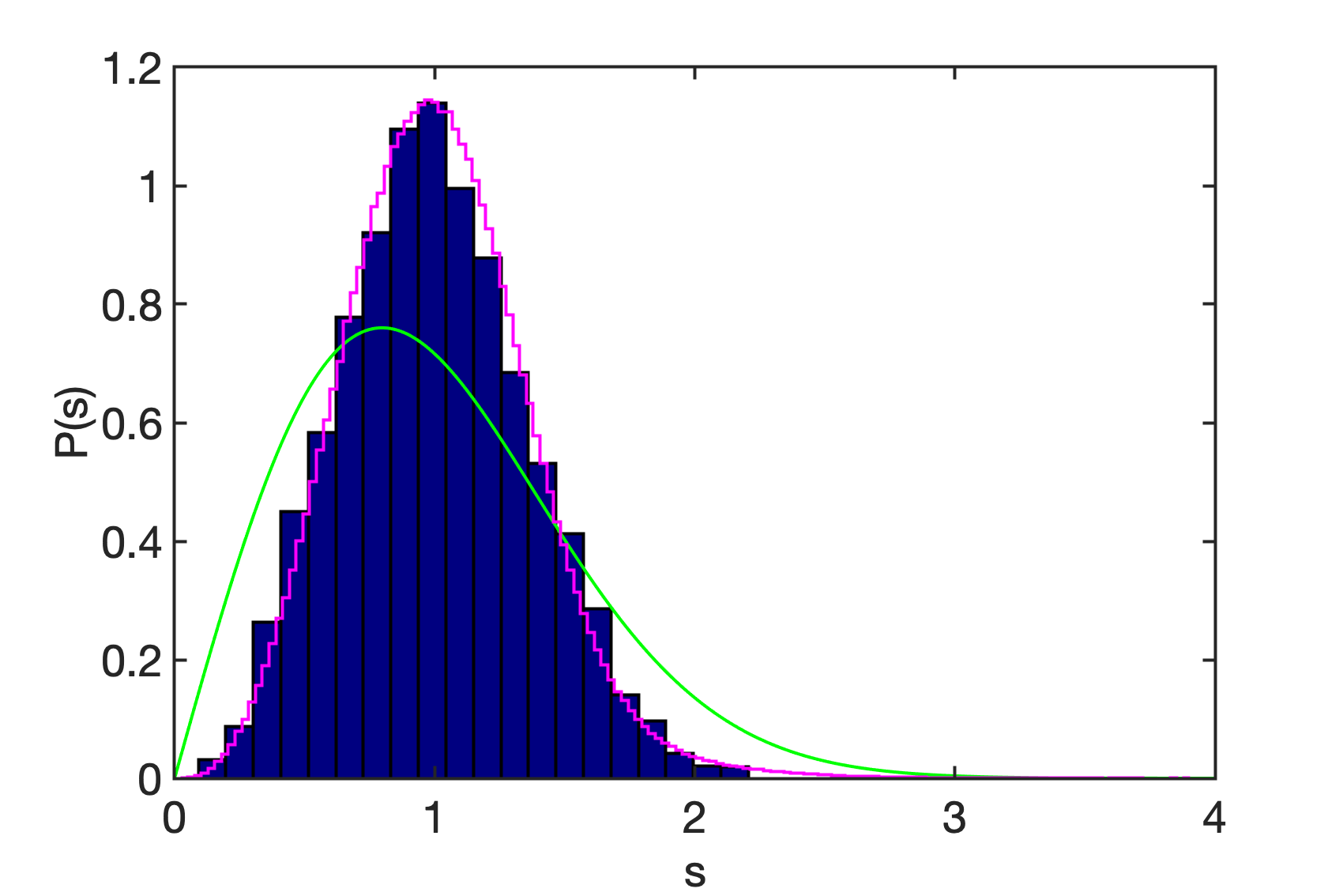}
\includegraphics[width=0.49\textwidth]{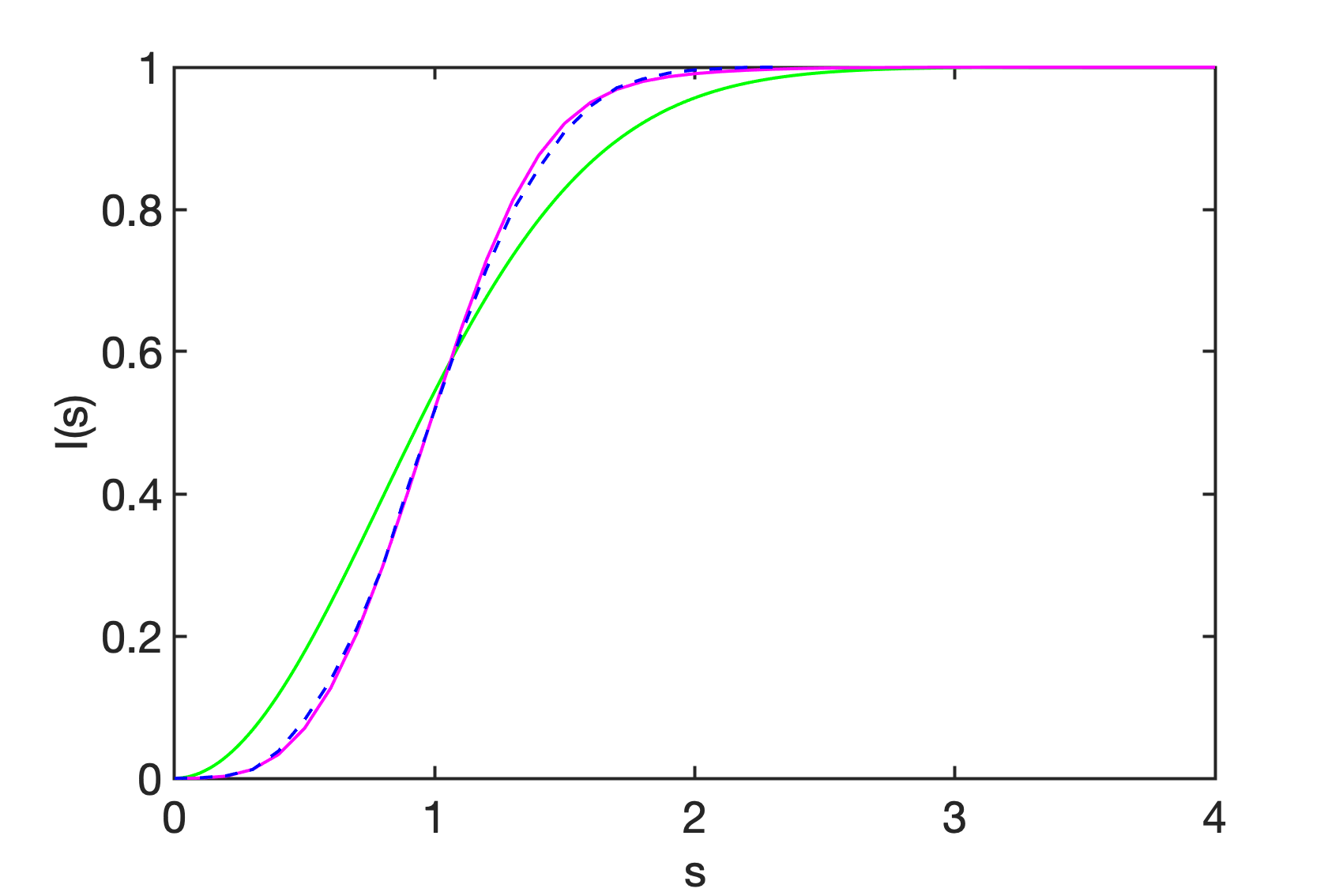}
\caption{Distribution of nearest-neighbour distances for the Baker map with $N = 3000$, and $0.50 \le \gamma \le 0.55$. The left panel shows the distribution of unfolded nearest-neighbour distances in comparison to the two-dimensional Poisson distribution (green)  and that of transpose-invariant Gaussian random matrices (magenta). }
\label{PTBakerMapSpectralStatistics}
\end{figure}
Figure \eqref{PTBakerMapSpectralStatistics} shows the statistics of the nearest-neighbour distances of the quasi energies for the $PT$-symmetric Baker map with matrix size $3000$ averaged over different values of $\gamma$ ranging in $0.01$ steps from $\gamma=0.5$ to $\gamma=0.55$. The Baker map also falls in the class of transpose invariant matrices since the similarity transformation $F_N^{\frac{1}{2}}\hat{\mathcal{B}}_{PT}F_N^{-\frac{1}{2}}$ indeed brings it to a transpose-invariant form, and we observe a close agreement with the corresponding random matrix predictions.  

Preliminary studies of a $PT$-symmetric generalisation of an M-mode Bose-Hubbard model with intermediate interaction strength and unit filling factor also show similar behaviour. This system is known to show the typical quantum classical correspondence of chaotic systems in its many-body and mean-field correspondence (see, e.g. \cite{Kolo04c,Kolo16,Dube16}). This offers an interesting opportunity to observe features of dissipative quantum chaos in many-body quantum systems, as losses in cold atoms in optical lattices are an active area of research and can be well controlled experimentally (see, e.g. \cite{Witt08,Wurt09,Barm11}).

Thus, we conclude that the statistics of the nearest-neighbour spacings in the complex plane for a number of $PT$-symmetric models are indeed universal. This particular statistics, however in the bulk of the spectrum is independent of the $PT$-symmetry of the system. It would be interesting to investigate whether other spectral statistics differ between $PT$-symmetric and more general non-Hermitian models. A natural starting point for this would be the investigation of the statistical features of the real eigenvalues alone, that do not have a counterpart in generic non-Hermitian systems. This, however, would require a much larger data set, and goes beyond the scope of the present paper. 

\section{Summary and Conclusions}
\label{sec_conclusion}
In summary, we have introduced a simple $PT$-symmetric generalisation of a quantum kicked top. We have derived the classical analogue of the dynamics and found that it leads to chaotic phase-space dynamics without sinks or sources for a wide parameter range, and a mixture of stable orbits and dissipative behaviour for larger gain-loss parameters, and strange attractors similar to conventional dissipative chaotic systems when the corresponding unitary system is deep in the chaotic regime. Similar behaviour has been discussed before in the theory of classically reversible dynamical systems, which indeed are the natural counterpart of $PT$-symmetric quantum systems. The classical dynamics arising here has an additional degree of freedom, which can be interpreted as an intensity, that shows intricate dynamical behaviour. In the full quantum system, the nature of the classical counterpart is reflected in the eigenvalue behaviour as well as in quantum phase-space structures. Finally, we have considered the spectral statistics of the quasi energies, and found universal behaviour. We have verified the same behaviour for a $PT$-symmetric generalisation of a triadic Baker map. 
Much remains to be explored. To give just two examples, a detailed characterisation of the rich features of the classical dynamics as well as the investigation of possible universal spectral features that are due to the $PT$-symmetry are interesting tasks for future studies.

\section*{Acknowledgments}
The authors thank Martin Kratky for useful input into the classical dynamics, and Joseph Hall and Simon Malzard for kindly providing a code for the calculation of the Lyapunov exponent, which has been adapted for the present purpose. Further thanks goes to the anonymous referees who have helped to improve this study considerably with their exceptionally insightful questions. E.M.G.  acknowledges  support  from  the Royal Society (Grant. No. UF130339) and from the European Research Council (ERC) under the European Union's Horizon 2020 research and innovation programme (grant agreement No 758453). S.M.N. acknowledges support from an EPSRC DTA grant (Grant No. EP/M506345/1).

\end{document}